\documentstyle{article}

\begin{document}

\title{Calculus of Sea-displacement Operators}
\author{Girish S. \ Setlur(1), D.S. Citrin(2)\\
(1)Condensed Matter Theory Unit,\\
Jawarharlal Nehru Centre for Advanced Scientific Research \\
Jakkur 560064, India. \\
(2)Department of Electrical and Computer Engineering \\
Georgia Institute of Technology, Atlanta, Georgia. }

\maketitle


\begin{abstract}
\noindent Sea-displacement operators for fermions are defined in
terms of the Fermi fields in a one-component Fermi system. The
main conclusions of this article fully corroborate the conjectures
made in our earlier works, and  provide a mathematically rigorous
foundation for these earlier works. These ideas are generalized to
electron-hole systems where we are able to explore clearly the
nature of exciton-exciton interactions. We find that
exciton-exciton interactions in an ideal model of GaAs are not
adequately treated simply as of the two-body type; rather the
interactions are mediated by the exchange of other bosons that are
present in this system. These bosons are identified explicitly and
the exciton Green function is calculated.
This exercise is also intended to be a precursor to a systematic
nonperturbative treatment of gauge theories.
\end{abstract}



\section{Introduction}

This article is meant to address some technical issues that
prevented our earlier work \cite{Setlur} from being universally
embraced, and to place on a firm mathematical foundation some of
the conjectures that appeared there. Indeed as Cune and Apostol
\cite{Cune} pointed out in their very pertinent critique, there
were significant technical drawbacks relating to the meaning of
the square-root of the number operator in the denominator that
cast doubt on the soundness of the physical conclusions. This
article lays to rest once and for all such doubts, and as an added
benefit, we are able to study the precise nature of
exciton-exciton interactions. It is interesting to note that the
rigorous formulation presented here is indispensable for a correct
treatment of charge-conserving two-component Fermi systems such as
excitons and relativistic electrons and positrons. This is
somewhat surprising since in our earlier work\cite{Setlur} we
showed that the lack of a proper definition of the
sea-displacement operator does not invalidate the physical
conclusions in the one-component Fermi system. Here we find that a
naive approach underestimates the nature and strength of
 interactions between excitons but a
more careful treatment brings out many subtle features that are
easy to overlook. Many authors who study exciton-exciton
interactions assume that excitons are bosons
 that interact via two-body interactions where the exciton-number is
 conserved. Kavoulakis and Baym\cite{Kavo} have pointed out the need to
 include Auger-like processes where the exciton-number is not conserved.
 Although assuming that excitons
 are bosons is totally acceptable(it is a matter of definition),
  there are some differences
 between a system such as a hydrogen atom (which interacts via
  two-body forces with other hydrogen atoms) and an exciton.
 The main difference is that in a hydrogen atom the proton and
 electron do not recombine leaving behind photons, where as an
 electron can recombine with a hole. In fact, when two excitons
 scatter off each other it is conceivable that nonradiative
 recombination processes take place in addition to radiative recombination.
 But in place of that, we find
 that an exciton can recombine with a special kind of electron-hole
 pair, which we call a {\em solitron} (to be made clear in the main text)
 and can create an intraband electron-hole pair
 (in other words, the usual kind of sea-displacement bosons found
 in one-component Fermi systems) called a {\em valeron} or
{\em conductron} depending
  upon the nature of the band. Thus we find that the
 two-component charge-conserving undoped Fermi system may be
 thought of as consisting of several kinds of elementary
 excitations (which are postulated to be exact bosons).
 The usual excitons are just one class of such excitations.
 Excitons can be in a bound state or in a scattering state just
 like a hydrogen atom. Furthermore, the excitons can possess a net
 center-of-mass momentum.
 These excitons possess an energy
 dispersion that is slightly different from the zero-center of
 mass excitons and are somewhat important in the analysis.
 Next in order of importance we have in our system a solitron. A
 {\em solitron} is an electron-hole pair that resides at the bottom of
 the conduction and valence bands. This pair is unbound and its
 existence is needed in order that excitons interact with one
 another, especially when we are dealing with Auger-like processes where
 exciton number is not conserved.
 In addition to the solitron, we find the need to invoke two other
 kinds of bosons, valerons and conductrons. {\em Valerons} are
 intravalence band electron-hole excitations. They are analogs of
 the usual sea-bosons \cite{Setlur} in one-component Fermi systems.
 Conductrons similarly are particle-hole excitations in the
 conduction band. All these bosons interact with one another and
 the resulting system is completely equivalent to the interacting
 Fermi system. Only the exciton couples to external radiation
 fields. Therefore, we have to consider excitons as the primary
 objects of interest and the other bosons in the system are like
 gauge-bosons. Material particles (excitons) interact by
 exchanging these other bosons. Furthermore, these gauge bosons
 interact amongst themselves, suggesting an analogy with
 nonabelian gauge theories. Towards the end of this article, we
 point out ways in which our approach can be used to study gauge
 theories. We also compute the exciton Green function using the interaction terms
 that correspond to inelastic scattering of the excitons off the other bosons.
  In future publications, we
 intend to investigate more thoroughly the practical aspects of
 this formalism--specifically, the biexciton Green function and
 nonlinear optical susceptibilities.

\section{Condensate Displacement Operators}

In this section, we provide some details regarding the
correspondence between the canonical bosons and their condensate
displacement analogs. These details are not found in our earlier
works \cite{Setlur}. They are important since they provide a
springboard from which we may write down the analogous statements
for fermions. The correspondence between the sea-displacement
operators and the canonical fermions was not made sufficiently
clear in our earlier works. Indeed, there were inconsistencies in
the technical definition, that even though had no impact on the
physical conclusions (as we shall see in the present article once
and for all) they did leave some room for doubting the soundness
of the framework. Let us therefore start off with the familiar
Bose systems. Let $b^{\dagger}_{ {\bf{k}} }$ and $b_{ {\bf{k}} } $
be canonical boson creation and annihilation operators,
respectively. They obey the commutation rules $[b_{ {\bf{k}} },b_{
{\bf{k}}^{'} }] = 0$ and $[b_{ {\bf{k}} },b^{\dagger}_{
{\bf{k}}^{'} }] = \delta_{ {\bf{k}}, {\bf{k}}^{'} }$. Let us now
introduce the following new objects known as condensate
displacement operators,
\begin{eqnarray}
d_{ {\bf{q}}/2 }({\bf{q}})& =& \frac{1}{\sqrt{{\hat{N}}_{0}}}
b^{\dagger}_{ {\bf{0}} } b_{ {\bf{q}} },
\; {\bf{q}} \neq 0 , \\
d_{ {\bf{0}} }({\bf{0}})& =& 0 , \label{CONDIS}
\end{eqnarray}
where ${\hat{N}}_{0}=b^{\dagger}_{ {\bf{0}} }b_{ {\bf{0}} }$ is
the number operator for the zero-momentum state.
 A word of caution regarding notation. The symbol $ N^{0} $ without a
 hat refers to a positive integer (a c-number) that
 corresponds to the total number of particles in the system, in both
 Bose as well as Fermi systems.
 The symbol $ {\hat{N}} $ refers to the operator that corresponds to
 the total number of particles in the system, be it Bose or Fermi.
 The object $ {\hat{ {\bf{1}} }} = {\hat{N}}/N^{0} $
 is therefore the unit operator.
 The symbol $ {\hat{N}}_{0} $ refers to the operator that
 corresponds to the total number of bosons in the
 zero-momentum state. This object does not appear in the next section where
 we deal with fermions.
 The square root of the operator in the denominator of Eq.\ (\ref{CONDIS})
 deserves special attention.
 In particular, when we act $ d^{\dagger}_{ {\bf{q}}/2 }({\bf{q}}) $
 on a state containing no particle in the zero-momentum state, we get
 an infinity multiplied by the same state, and when this further is acted on
 by $ b_{ {\bf{0}} } $, we get a factor of zero. Zero multiplied by
 infinity is indeterminate. This tells us that
 the condensate displacement operator
 is an ill-defined operator on the Fock space of bosons.
 This is a technical problem that cannot be
 wished away.
 We will mitigate the severity of this problem by postulating
 that all states of the interacting system
 (both ground state and excited states)
 may be expressed as linear combinations of states from a
 {\em restricted Hilbert space} that contains states of the noninteracting
 system with a fixed total number of particles, but excludes those that contain
 no particles in the zero-momentum state.
 Although we are unable to say precisely when this assumption breaks
 down, it is reasonable to assert that even in the case when the interactions
 are strong, either because of the intrinsic nature of the interaction or
 apparently strong due to the dimensionality of the system, the
 zero-momentum state
 of the interacting system will have at least one boson in it, if not a
 macroscopically large number of them. (The number operator is no
 longer a good quantum number in such systems but perhaps this is true in
 some average sense with small fluctuations around the average.)
The subsequent development will justify this.
 Our assumption enables us
 to write down a polar decomposition
 of the
 operator $ b_{ {\bf{0}} } $ as
$b_{ {\bf{0}} } = \exp(-iX^{r}_{0})\sqrt{ {\hat{N}}_{0} }$ where
$X^{r}_{0} = (1/2)(X_{0} + X^{\dagger}_{0})$ is an operator that
is strictly self-adjoint but it is not exactly but almost a
canonical conjugate to the number operator ${\hat{N}}_{0} =
b^{\dagger}_{0}b_{0} \geq 0 $(in Appendix A we see that an
appropriate interpretation of the definition of the conjugate
allows us to circumvent these issues, for fermions we have to be
more careful). On the other hand, $ X_{0} $ is almost self-adjoint
but is strictly a conjugate to the number operator $ {\hat{N}}_{0}
$ (however, in the next section when we deal with fermions,
$X_{0}$ refers to the canonical conjugate of the {\it{total}}
number of {\it{fermions}}). If a definition of $ X^{r}_{0} $ is
desired, we may claim that it is given by the manifestly
self-adjoint formula,
\begin{equation}
X^{r}_{0} = \frac{i}{2}\ln \left( b_{ {\bf{0}} }
 \frac{1}{ \sqrt{ {\hat{N}}_{0}} }
 \right)
 - \frac{i}{2} \ln \left(
 \frac{1}{ \sqrt{ {\hat{N}}_{0}} } b^{\dagger}_{ {\bf{0}} }
 \right) .
\end{equation}
Besides the square root of the number operator in the denominator
that we have already made sense of, we have to make sense of the
logarithm. It is defined to be the power-series expansion
suggested by rewriting the above formula as
\begin{equation}
X^{r}_{0} = \frac{i}{2}\ln \left[ 1 + \left(b_{ {\bf{0}} } -
\sqrt{ {\hat{N}}_{0} }\right) \frac{1}{ \sqrt{ {\hat{N}}_{0} } }
 \right]
 - \frac{i}{2}\ln \left[ 1 +
\frac{1}{ \sqrt{ {\hat{N}}_{0} } } \left(b^{\dagger}_{ {\bf{0}} }
- \sqrt{ {\hat{N}}_{0} }\right)
 \right] .
\label{DEFNX0}
\end{equation}
 The proof of the fact that $ X_{0} $
 is canonically conjugate to the number
 operator $ {\hat{N}}_{0} $
 in the restricted Hilbert space is relegated to
 Appendix A. Therefore we may write
$[X_{0},{\hat{N}}_{0}] = i$. Armed with these facts we now are
able to prove that $ d_{ {\bf{q}}/2 } ({\bf{q}}) $ is a canonical
boson annihilation operator. We may now write $d_{ {\bf{q}}/2
}({\bf{q}}) = \exp(i X^{r}_{0})b_{ {\bf{q}} }$. Since from the
definition, $\exp(iX^{r}_{0})$ depends on neither $b_{ {\bf{q}} }$
nor $b^{\dagger}_{ {\bf{q}} }$, we see that as far as commutation
rules of $ d_{ {\bf{q}}/2 }({\bf{q}}) $ go, they are identical to
those of $ b_{ {\bf{q}} } $ since the two differ by a trivial
phase that commutes with both these objects and their Hermitian
conjugates. Next, we reproduce some facts that have been proved
satisfactorily \cite{Setlur} elsewhere, and these will be used as
the point of departure for a rigorous treatment of fermions.
Earlier we proved\cite{Setlur} (for both $ {\bf{q}} = 0 $ and $
{\bf{q}} \neq 0 $), the following combined formula,
\begin{eqnarray}
 b^{\dagger}_{ {\bf{k}}+{\bf{q}}/2 }b_{ {\bf{k}} - {\bf{q}}/2 }
 &=& {\hat{N}}_{0}\delta_{ {\bf{k}}, 0 }\delta_{ {\bf{q}}, 0 }
 + \sqrt{ {\hat{N}}_{0} }d_{ -{\bf{q}}/2 }(-{\bf{q}})
\delta_{ {\bf{k}}+{\bf{q}}/2, 0 }
 +d^{\dagger}_{ {\bf{q}}/2 }({\bf{q}})
\sqrt{ {\hat{N}}_{0} } \delta_{
{\bf{k}}-{\bf{q}}/2, 0 } \nonumber \\
& &+  d^{\dagger}_{ (1/2)({\bf{k}} + {\bf{q}}/2)
}({\bf{k}}+{\bf{q}}/2) d_{ (1/2)({\bf{k}} - {\bf{q}}/2)
}({\bf{k}}-{\bf{q}}/2) , \\
{\hat{N}}_{0}& =& {\hat{N}} - \sum_{ {\bf{q}} \neq 0 }
d^{\dagger}_{ {\bf{q}}/2 }({\bf{q}})d_{ {\bf{q}}/2 }({\bf{q}}) .
\end{eqnarray}
This equation may be rewritten more elaborately and in a manner
that is conducive to generalization as (${\bf{q}}\neq 0$)
\begin{eqnarray}
b^{\dagger}_{ {\bf{k}} + {\bf{q}}/2 }b_{ {\bf{k}} - {\bf{q}}/2 } &
=& \sqrt{ n_{ {\bf{k}} + {\bf{q}}/2 } }A_{ {\bf{k}} }(-{\bf{q}})
 + A^{\dagger}_{ {\bf{k}} }({\bf{q}})
 \sqrt{ n_{ {\bf{k}}- {\bf{q}}/2 } } \nonumber  \\
 & & + \sum_{ {\bf{q}}_{1} \neq {\bf{q}}, {\bf{0}} }
 A^{\dagger}_{ {\bf{k}} + {\bf{q}}/2 - {\bf{q}}_{1}/2 }({\bf{q}}_{1})
A_{ {\bf{k}} -  {\bf{q}}_{1}/2 }(-{\bf{q}} + {\bf{q}}_{1}) \\
 & & - \sum_{ {\bf{q}}_{1} \neq {\bf{q}}, {\bf{0}}  }
  A^{\dagger}_{ {\bf{k}} - {\bf{q}}/2 + {\bf{q}}_{1}/2 }({\bf{q}}_{1})
A_{ {\bf{k}} + {\bf{q}}_{1}/2 }(-{\bf{q}} + {\bf{q}}_{1}) , \nonumber \\
n_{ {\bf{k}} }& =& b^{\dagger}_{ {\bf{k}} }b_{ {\bf{k}} }
 = n_{B}({\bf{k}})\frac{ {\hat{N}} }{N^{0}}
 + \sum_{ {\bf{q}}_{1} \neq 0 }
A^{\dagger}_{ {\bf{k}} - {\bf{q}}_{1}/2 }({\bf{q}}_{1}) A_{
{\bf{k}} - {\bf{q}}_{1}/2 }({\bf{q}}_{1}) \nonumber \\
 & & - \sum_{ {\bf{q}}_{1} \neq 0 }
A^{\dagger}_{ {\bf{k}} + {\bf{q}}_{1}/2 }({\bf{q}}_{1}) A_{
{\bf{k}} + {\bf{q}}_{1}/2 }({\bf{q}}_{1}) .
\end{eqnarray}
Here $ n_{B}({\bf{k}}) $ is the momentum distribution of
noninteracting bosons at zero temperature, $n_{B}({\bf{k}}) =
\delta_{ {\bf{k}}, 0 }N^{0}$. Also, $A_{ {\bf{k}} }({\bf{q}}) =
\delta_{ {\bf{k}} - {\bf{q}}/2, 0 } d_{ {\bf{q}}/2 }({\bf{q}})$.
It is worthwhile to consider an alternative scheme for making
sense of the definition in Eq.\ (\ref{CONDIS}). It involves
writing the number operator as
\begin{equation}
{\hat{N}}_{0} = N^{0} \left( {\hat{ {\bf{1}} }}
 - \frac{1}{N^{0}}\sum_{ {\bf{q}} \neq 0 }d^{\dagger}_{ {\bf{q}}/2 }
({\bf{q}})d_{ {\bf{q}}/2 }({\bf{q}}) \right) .
\end{equation}
Furthermore, we may write for the occupation number not in the
zero-momentum state ($ {\bf{k}} \neq 0 $) as $n_{ {\bf{k}} } =
d^{\dagger}_{ (1/2){\bf{k}} }({\bf{k}}) d_{ (1/2){\bf{k}}
}({\bf{k}})$. With these identifications, we may rewrite Eq.\
(\ref{CONDIS}) as ($ {\bf{q}} \neq 0$),
\begin{equation}
d_{ {\bf{q}}/2 }({\bf{q}}) = \frac{1}{ \sqrt{ N^{0} } }
 \left( {\hat{ {\bf{1}} }}
 - \frac{1}{N^{0}}\sum_{ {\bf{q}} \neq 0 }d^{\dagger}_{ {\bf{q}}/2 }
({\bf{q}})d_{ {\bf{q}}/2 }({\bf{q}}) \right)^{-\frac{1}{2}}
  b^{\dagger}_{ {\bf{0}} }b_{ {\bf{q}} } .
\label{EQNN2}
\end{equation}
If Eq.\ (\ref{EQNN2}) is interpreted as a power-series expansion
around unity, we can construct an iterative procedure to solve for
$ d_{ {\bf{q}}/2 }({\bf{q}}) $. This may seem redundant given the
fact that we have already made an elegant argument that pins down
the meaning of $ d_{ {\bf{q}}/2 }({\bf{q}}) $ in terms of $ X_{0}
$. The reason for this new approach is that in the case of
fermions we will not have the luxury of introducing an object
similar to $ X_{0} $ for reasons that will become clear in the
next section. Thus we are forced to seek alternatives that are
more fermion-friendly. Unfortunately, these alternatives do not
allow us to venture very far from the noninteracting case, as we
shall soon see. Nevertheless, this exercise is very instructive
since it tells us that the correspondence that we write down for
fermions in the next section has exactly the same features and are
therefore correct, pending the resolution of the interpretation of
the ubiquitous square root in the denominator. Retaining only the
lowest order in the series expansion gives us
\begin{eqnarray}
d_{ {\bf{q}}/2 }({\bf{q}})& =& \frac{1}{\sqrt{N^{0}}}
b^{\dagger}_{0}b_{ {\bf{q}} } , \\
{\hat{N}}_{0}& =& N^{0} {\hat{ {\bf{1}} }} .
\end{eqnarray}
\label{EQNNRR}
Further we have ($ {\bf{k}} \neq 0 $) $n_{ {\bf{k}} } =
d^{\dagger}_{ (1/2){\bf{k}} }({\bf{k}}) d_{ (1/2){\bf{k}}
}({\bf{k}})
 = \frac{1}{N^{0}}( {\hat{ {\bf{1}} }} + {\hat{N}}_{0} )
b^{\dagger}_{ {\bf{k}} }b_{ {\bf{k}} }$. But we also know that
$n_{ {\bf{k}} } =  b^{\dagger}_{ {\bf{k}} }b_{ {\bf{k}} }$. This
seems to suggest one of two things, namely $ n_{ {\bf{k}} } = 0$
or we should restrict ourselves to cases when $ {\hat{ N }}_{0} =
N^{0} {\hat{ {\bf{1}} }} >> {\hat{ {\bf{1}} }}  $. The latter
possibility seems the most attractive until one realizes that the
restricted Hilbert space is not so restrictive as to exclude such
systems where the number of particles is small and finite. This
will become clear in Appendix A where we need to only assume that
the restricted Hilbert space contains no states that have zero
particles in the zero-momentum state. Therefore we are left with
the other possibility, $ n_{ {\bf{k}} } = 0 $. This result is not
as alarming as it seems since Eq.\ (\ref{EQNNRR}) is consistent
only with the state where all the bosons are in the zero-momentum
state and no bosons have higher momenta. Further iterations do not
change this picture. That is, if we interpret the square root as a
power-series expansion, we obtain the first order correction
\begin{equation}
d_{ {\bf{q}}/2 }({\bf{q}}) \approx \left(
 {\hat{ {\bf{1}} }}
+ \frac{1}{2N^{0}}\sum_{ {\bf{q}}_{1} \neq 0 } d^{\dagger}_{
{\bf{q}}_{1}/2 }({\bf{q}}_{1})d_{ {\bf{q}}_{1}/2 }({\bf{q}}_{1})
 \right) \frac{1}{\sqrt{N^{0}}}b^{\dagger}_{ {\bf{0}} }b_{ {\bf{q}} } .
\end{equation}
In order to conform to the iterative scheme, we are obliged to
replace the the $d$'s on the right side by the zeroth-order $d$'s.
But we know that in the zeroth order, $ d^{\dagger}_{
{\bf{q}}_{1}/2 }({\bf{q}}_{1}) d_{ {\bf{q}}_{1}/2 }({\bf{q}}_{1})
= 0 $ and $ {\hat{N}} = N^{0}\hat{\bf{1}} $. Therefore the
first-order $d$ is the same as in the zeroth order, $d_{
{\bf{q}}/2 }({\bf{q}})
  = \frac{1}{\sqrt{N^{0}}}b^{\dagger}_{ {\bf{0}} }b_{ {\bf{q}} }$.
All this points to the futility of interpreting the square root as
a power-series expansion around the noninteracting ground state.
All further iterations lead to precisely the same result. This
should tell us that one should solve the system self-consistently.
Perhaps we should expand around an expectation value [i.e.,
${\hat{N}}_{0} = \langle {\hat{N}}_{0} \rangle {\hat{ {\bf{1}} }}
+ ({\hat{N}}_{0} - \langle {\hat{N}}_{0} \rangle {\hat{ {\bf{1}}
}} )$] with the expectation value determined self-consistently.
This expectation value will be different from $ N^{0} $ even when
interactions are absent, for example, if we consider finite
temperature. Let us now try and compute the finite-temperature
momentum distribution of noninteracting bosons. From elementary
considerations we know that $\langle {\hat{N}}_{0} \rangle
 = N^{0} - \sum_{ {\bf{k}} \neq 0 }\langle b^{\dagger}_{ {\bf{k}} }
b_{ {\bf{k}} } \rangle$ and $\langle n_{ {\bf{k}} } \rangle =
 \langle b^{\dagger}_{ {\bf{k}} }
b_{ {\bf{k}} } \rangle$. The thermodynamic expectation values
involve the chemical potential $ \mu $:
\begin{equation}
\langle n_{ {\bf{k}} } \rangle =
 \frac{1}{exp(\beta(\epsilon_{ {\bf{k}} } - \mu)) - 1} .
\end{equation}
The reason this appears is that in the grand canonical ensemble we
have to compute the trace with the Boltzmann weight $\langle
b^{\dagger}_{ {\bf{k}} } b_{ {\bf{k}} } \rangle = \frac{1}{Z} Tr\{
\exp[-\beta(H-\mu N)] b^{\dagger}_{ {\bf{k}} } b_{ {\bf{k}} }\}  $
where $ Z = Tr\{ \exp[-\beta(H-\mu N) ]\} $ is the grand partition
function. Using the cyclic property of the trace, we may write
\begin{eqnarray*}
\langle  b^{\dagger}_{ {\bf{k}} } b_{ {\bf{k}} } \rangle & =&
\frac{1}{Z} Tr( e^{-\beta(H-\mu N) }  b^{\dagger}_{ {\bf{k}} } b_{
{\bf{k}} } ) \\
&=& \frac{1}{Z} Tr(b_{ {\bf{k}} } e^{-\beta(H-\mu N) }
b^{\dagger}_{ {\bf{k}} }) \\
& =& \frac{1}{Z} Tr( e^{-\beta(H-\mu N) }
 e^{\beta(H-\mu N) }
b_{ {\bf{k}} } e^{-\beta(H-\mu N) }  b^{\dagger}_{ {\bf{k}} }) \\
 &=& e^{-\beta(\epsilon_{ {\bf{k}} } - \mu) }
\langle b_{ {\bf{k}} }b^{\dagger}_{ {\bf{k}} } \rangle \\
& =& e^{-\beta(\epsilon_{ {\bf{k}} } - \mu) } ( 1 + \langle
b^{\dagger}_{ {\bf{k}} } b_{ {\bf{k}} } \rangle ) .
\end{eqnarray*}
In other words,
\[
 \langle  b^{\dagger}_{ {\bf{k}} }
b_{ {\bf{k}} } \rangle
 = \frac{1}{exp(\beta(\epsilon_{ {\bf{k}} } - \mu)) - 1 } .
\]
Let us now try to evaluate this quantity using the
condensate-displacement language:
\begin{eqnarray*}
 \langle  b^{\dagger}_{ {\bf{k}} }
b_{ {\bf{k}} } \rangle& =&
 \langle  d^{\dagger}_{ {\bf{k}}/2 }({\bf{k}})
 d_{ {\bf{k}}/2 }({\bf{k}})
 \rangle \\
& =& \frac{1}{Z} Tr(e^{ -\beta(H-\mu N) } d^{\dagger}_{ {\bf{k}}/2
}({\bf{k}})
 d_{ {\bf{k}}/2 }({\bf{k}}) ) , \\
H& = &\sum_{ {\bf{k}} \neq 0 } \epsilon_{ {\bf{k}} }d^{\dagger}_{
(1/2){\bf{k}} }({\bf{k}}) d_{ (1/2){\bf{k}} }({\bf{k}}) .
\end{eqnarray*}
We also know that $ [{\hat{N}}, d_{ (1/2){\bf{k}} }({\bf{k}})] =
0$:
\begin{eqnarray*}
 \langle  d^{\dagger}_{ {\bf{k}}/2 }({\bf{k}})
 d_{ {\bf{k}}/2 }({\bf{k}})
 \rangle \\
 &= &\frac{1}{Z} Tr(e^{ -\beta(H-\mu N) } d^{\dagger}_{ {\bf{k}}/2 }({\bf{k}})
 d_{ {\bf{k}}/2 }({\bf{k}}) ) \\
 &=& \frac{1}{Z} Tr( e^{ -\beta(H-\mu N) }
e^{\beta(H-\mu N) } d_{ {\bf{k}}/2 }({\bf{k}}) e^{ -\beta(H-\mu N)
} d^{\dagger}_{ {\bf{k}}/2 }({\bf{k}}) ) , \\
 \langle  d^{\dagger}_{ {\bf{k}}/2 }({\bf{k}})
 d_{ {\bf{k}}/2 }({\bf{k}})
 \rangle
 &=& e^{-\beta \epsilon_{ {\bf{k}} }}
\frac{1}{Z} Tr( e^{ -\beta(H-\mu N) } d_{ {\bf{k}}/2 }({\bf{k}})
d^{\dagger}_{ {\bf{k}}/2 }({\bf{k}}) ) \\
 &=& e^{-\beta \epsilon_{ {\bf{k}} }}
(1 + \langle d^{\dagger}_{ {\bf{k}}/2 }({\bf{k}})
 d_{ {\bf{k}}/2 }({\bf{k}})
 \rangle) , \\
 \langle  d^{\dagger}_{ {\bf{k}}/2 }({\bf{k}})
 d_{ {\bf{k}}/2 }({\bf{k}})
 \rangle
 &=& \frac{1}{exp(\beta \epsilon_{ {\bf{k}} }) - 1 } .
\end{eqnarray*}
The crucial $ \mu $ seems to be missing. The reason is somewhat
subtle \cite{footnote}. It has to do with the fact that the trace
in the original case using the parent bosons spans  all states
including those with $ {\hat{N}}_{0} = 0 $. However, in the
condensate-displacement language, the trace is over all states
{\em except} those that have $ {\hat{N}}_{0} = 0 $. There are
quite a number of states that have $ {\hat{N}}_{0} = 0 $ and an
arbitrary total number of particles. It is therefore not
surprising that we have encountered a discrepancy. The best way to
resolve this is to introduce a Lagrange multiplier that allows us
to control how many bosons there are in states with zero momentum.
Thus when it comes to taking the trace over states in the
condensate-displacement language, we have to be careful to include
a new chemical potential that couples to $ {\hat{N}}_{0} $ and not
just to $ {\hat{N}} $. When this is done we can easily show
\[
 \langle  d^{\dagger}_{ {\bf{k}}/2 }({\bf{k}})
 d_{ {\bf{k}}/2 }({\bf{k}})
 \rangle
 = \frac{1}{\exp[\beta (\epsilon_{ {\bf{k}} } - \mu ) ] - 1 } .
\]
These considerations carry over to fermions where the difficulties
are understandably far more severe. After all, trying to describe
fermions using Bose-like objects, and to do it exactly, is a
daunting task. We shall now use this insight to write down a
correspondence for fermions.

\section{The Nature of Sea-Displacement Operators}

In this section, we provide details of the correspondence between
the sea-displacement operators and canonical fermions. The
sea-displacement operators have been introduced
elsewhere\cite{Setlur} but an explicit formula for the
sea-displacement operator in terms of the Fermi fields was
lacking. In this section, we once and for all pin down the
definition of the sea-displacement operator and show that they
obey commutation rules that are somewhat complicated but in the
limit of a generalized random-phase approximation (RPA) (to be
made clear in Appendix B) they obey canonical boson commutation
rules. This exercise hopefully is the final word as far as the
technicalities of the correspondence goes. Furthermore, as we
shall see in the conclusion, the rigorous formulation reinforces
the rather startling claim\cite{Setlur} that in 1D, so long as the
interaction between the fermions is purely repulsive and possesses
a Fourier transform, and is sufficiently weak so that all states
of the interacting system (both ground state as well as excited
states) may be expressed as linear combinations of low lying
excited states of the noninteracting system and its ground state,
then the momentum distribution of the interacting system possesses
a sharp Fermi surface at precisely the same place as the
noninteracting system. Let us now proceed to the main task at
hand.

Let $ c_{ {\bf{k}} } $ and $ c^{\dagger}_{ {\bf{k}} } $ be
canonical fermion annihilation and creation operators. We may
write $\{c_{ {\bf{k}} }, c_{ {\bf{k}}^{'} }\} = 0$ and $\{c_{
{\bf{k}} }, c^{\dagger}_{ {\bf{k}}^{'} }\} = \delta_{ {\bf{k}},
{\bf{k}}^{'} }$. Let $ n_{ {\bf{k}} } = c^{\dagger}_{ {\bf{k}}
}c_{ {\bf{k}} } $ denote the number operator. The sea-displacement
operators for fermions are defined as ($ {\bf{q}} \neq 0 $)
\begin{eqnarray}
A_{ {\bf{k}} }({\bf{q}})& =& n_{F}({\bf{k}}-{\bf{q}}/2) [1 -
n_{F}({\bf{k}}+{\bf{q}}/2)] \frac{1}{\sqrt{n_{ {\bf{k}}-{\bf{q}}/2
} }} c^{\dagger}_{ {\bf{k}} - {\bf{q}}/2 }c_{ {\bf{k}} +
{\bf{q}}/2 }\label{AKQ} \\
A_{ {\bf{k}} }({\bf{0}}) &=& 0
\end{eqnarray}
where $n_{F}({\bf{k}}) = \theta(k_{F} - |{\bf{k}}|) $ is the
zero-temperature Fermi distribution. As before, the square root of
the number operator in the denominator in the definition requires
clarification, a point noted  by Cune and Apostol\cite{Cune}.
Their approach for dealing with this problem is unfortunately, not
adequate\cite{Cunecrit}. Neither it seems is the approach used for
dealing with this problem in the case of bosons, namely that we be
able to interpret the object ${\hat{U}}({\bf{k}}) =
\frac{1}{\sqrt{n_{ {\bf{k}} } }} c^{\dagger}_{ {\bf{k}} }$ as
being a unitary operator. This reason for this additional
complication probably stems from the fact that in order for this
object to be unitary, we should be able to find a self-adjoint
canonical conjugate $ P_{ {\bf{k}} } $ of the number operator such
that $ [P_{ {\bf{k}} }, n_{ {\bf{k}}^{'} }] = i\delta_{ {\bf{k}},
{\bf{k}}^{'} } $. Such an object is not likely to exist not only
because of the positivity of the number operator $n_{ {\bf{k}} }$,
but also because of idempotence \cite{idem}. In fact, it is a
simple matter to convince ourselves that idempotence is
inconsistent with the existence of $ P_{ {\bf{k}} } $. On the one
hand, we have $ [P_{ {\bf{k}} },n^{2}_{ {\bf{k}} }] =
 [P_{ {\bf{k}} },n_{ {\bf{k}} }]n_{ {\bf{k}} } +
n_{ {\bf{k}} }[P_{ {\bf{k}} },n_{ {\bf{k}} }] = 2in_{ {\bf{k}} }$.
On the other hand, $[P_{ {\bf{k}} },n^{2}_{ {\bf{k}} }] = [P_{
{\bf{k}} },n_{ {\bf{k}} }] = i $. This suggests that $n_{ {\bf{k}}
} = (1/2) {\hat{ {\bf{1}} }}$--a state of affairs seldom realized
if at all. Some readers of this work familiar with the more
traditional bosonization approaches may point to the importance of
`point-splitting'--`a procedure that does not allow us to write $
n_{ {\bf{k}} } = c^{\dagger}_{ {\bf{k}} }c_{ {\bf{k}} } $. They
may wish to suggest this as the main reason for all these
difficulties. It is likely that point-splitting is a necessary
technical consideration only in systems that have $ k_{F}
\rightarrow \infty $ and $ m \rightarrow \infty $ but $ v_{F} =
k_{F}/m < \infty $, in other words, a system where both the mass
of the particle and the density are infinite such that the Fermi
velocity is finite \cite{Stone}. We deal with systems that are
more physical (finite $ k_{F} $ and $ m $) and are, by and large,
immune to these considerations. While the metaphorical definition
of Eq.(~\ref{AKQ}) is quite adequate for most practical purposes,
it is desirable to have a better understanding of the meaning of
the square root.

 Let us divert our attention for
some time to the more mechanical aspects of this program in the
hope that soon we will be able to address this delicate technical
problem of the square root in the denominator. In a manner
entirely analogous to that of bosons we may write
\begin{eqnarray}
c^{\dagger}_{ {\bf{k}} + {\bf{q}}/2 }c_{ {\bf{k}} - {\bf{q}}/2 } &
=& \sqrt{ n_{ {\bf{k}} + {\bf{q}}/2 } }A_{ {\bf{k}} }(-{\bf{q}})
 + A^{\dagger}_{ {\bf{k}} }({\bf{q}})
 \sqrt{ n_{ {\bf{k}}- {\bf{q}}/2 } }\nonumber \\
& & + \lambda \sum_{ {\bf{q}}_{1} \neq {\bf{q}}, {\bf{0}} }
 A^{\dagger}_{ {\bf{k}} + {\bf{q}}/2 - {\bf{q}}_{1}/2 }({\bf{q}}_{1})
A_{ {\bf{k}} -  {\bf{q}}_{1}/2 }(-{\bf{q}} + {\bf{q}}_{1})
\label{OFFDIAG} \\
& & - \lambda \sum_{ {\bf{q}}_{1} \neq {\bf{q}}, {\bf{0}}  }
  A^{\dagger}_{ {\bf{k}} - {\bf{q}}/2 + {\bf{q}}_{1}/2 }({\bf{q}}_{1})
A_{ {\bf{k}} + {\bf{q}}_{1}/2 }(-{\bf{q}} + {\bf{q}}_{1})  .
\end{eqnarray}
Furthermore,
\begin{equation}
n_{ {\bf{k}} } = c^{\dagger}_{ {\bf{k}} }c_{ {\bf{k}} }
 =  n_{F}({\bf{k}}) \frac{ {\hat{N}} }{ N^{0} }
+ \lambda \sum_{ {\bf{q}}_{1} \neq {\bf{0}} } A^{\dagger}_{
{\bf{k}} - {\bf{q}}_{1}/2 }({\bf{q}}_{1}) A_{ {\bf{k}} -
{\bf{q}}_{1}/2 }({\bf{q}}_{1})
 -  \lambda \sum_{ {\bf{q}}_{1} \neq {\bf{0}} }
 A^{\dagger}_{ {\bf{k}}+ {\bf{q}}_{1}/2 }({\bf{q}}_{1})
A_{ {\bf{k}} + {\bf{q}}_{1}/2 }({\bf{q}}_{1})
 \label{DIAG}
\end{equation}
where $ n_{F}({\bf{k}}) = \theta(k_{F} - |{\bf{k}}|) $ is the
zero-temperature Fermi distribution function. The parameter $
\lambda $
 is a book-keeping device, familiar from perturbation theory, where in the
 end we set $ \lambda = 1 $. This is a very useful device as we shall soon find
 out, one that partially addresses the problem of the square root in the
 denominator. The proof of Eqs.\ (\ref{OFFDIAG}) and (\ref{DIAG}) rests
 crucially on the meaning of the definition of $ A_{ {\bf{k}} }({\bf{q}}) $
 shown in Eq.\ (\ref{AKQ}). Let us now consider the problem of
 ascertaining the meaning of $A_{ {\bf{k}} }({\bf{q}}) $ in a `perturbative'
 manner. That is, we shall interpret the square root of the number operator
 as being the square root of the Eq.\ (\ref{DIAG}) where we find a natural
 unit operator about which we may expand the square root. Thus if we consider
 the lowest order in $ \lambda $ we have
$n_{ {\bf{k}} } = n_{F}({\bf{k}}) {\hat{ {\bf{1}} }}$. Since $ n_{
{\bf{k}}-{\bf{q}}/2 } = {\bf{1}} $ when
 $ n_{F}({\bf{k}}-{\bf{q}}/2) = 1 $, we may write,
\begin{equation}
A_{ {\bf{k}} }({\bf{q}}) = n_{F}({\bf{k}}-{\bf{q}}/2) [1 -
n_{F}({\bf{k}}+{\bf{q}}/2) ]
 c^{\dagger}_{ {\bf{k}} - {\bf{q}}/2 }
c_{ {\bf{k}} + {\bf{q}}/2 } . \label{EQCUNE}
\end{equation}
These equations obviously solve Eqs.\ (\ref{OFFDIAG}) and
(\ref{DIAG}) in the lowest order $ \lambda^{0} = 1 $ except that
we seem to have the additional curiosity $c^{\dagger}_{ {\bf{k}} +
{\bf{q}}/2 }c_{ {\bf{k}} - {\bf{q}}/2 } = 0$ if $ n_{F}({\bf{k}} +
{\bf{q}}/2) = 1 $ and $ n_{F}({\bf{k}} - {\bf{q}}/2) = 1 $ or if $
n_{F}({\bf{k}} + {\bf{q}}/2) = 0 $ and $ n_{F}({\bf{k}} -
{\bf{q}}/2) = 0 $. This is hardly surprising given the fact that
 $ n_{ {\bf{k}} } = n_{F}({\bf{k}}){\hat{ {\bf{1}} }} $
  is consistent with our restricted Hilbert
space having exactly one element, namely the ground state of the
noninteracting Fermi sea $ | FS \rangle $. This means
$c^{\dagger}_{ {\bf{k}} + {\bf{q}}/2 }c_{ {\bf{k}} - {\bf{q}}/2 }
| FS \rangle   = 0$ when $ n_{F}({\bf{k}} + {\bf{q}}/2) = 1 $ and
$ n_{F}({\bf{k}} - {\bf{q}}/2) = 1 $ or if $ n_{F}({\bf{k}} +
{\bf{q}}/2) = 0 $ and $ n_{F}({\bf{k}} - {\bf{q}}/2) = 0 $, a fact
easily verified. Parenthetically, we note that Eq.\ (\ref{EQCUNE})
was anticipated in the comment by Cune and Apostol \cite{Cune}.
Just as in the case of bosons, further iterations do not affect
these conclusions. Therefore, if one starts with the
noninteracting system and expands around the noninteracting ground
state, one is able to treat only the noninteracting system. Just
for the sake of completeness, let us see where the first order
terms lead us. When $ n_{F}({\bf{k}} + {\bf{q}}/2) = 1 $ and $
n_{F}({\bf{k}} - {\bf{q}}/2) = 0 $, we have from the definition
Eq.\ (\ref{AKQ}) a result that is independent of $\lambda$ and
therefore valid to all orders in $ \lambda $, $\sqrt{n_{
{\bf{k}}+{\bf{q}}/2 } }A_{ {\bf{k}} }(-{\bf{q}})
 = c^{\dagger}_{ {\bf{k}} + {\bf{q}}/2 }c_{ {\bf{k}}- {\bf{q}}/2 }$.
In addition,  we have $A^{\dagger}_{ {\bf{k}} }({\bf{q}}) \sqrt{
n_{ {\bf{k}} - {\bf{q}}/2 } }
 = c^{\dagger}_{ {\bf{k}} + {\bf{q}}/2 }c_{ {\bf{k}}- {\bf{q}}/2 }$
when $n_{F}({\bf{k}} + {\bf{q}}/2) = 0 $ and $ n_{F}({\bf{k}} -
{\bf{q}}/2) = 1 $. These two identifications are entirely
consistent with Eq.\ (\ref{OFFDIAG}) to all orders in $ \lambda $.
The other cases namely, $ n_{F}({\bf{k}} \pm {\bf{q}}/2) = 0 $ and
$ n_{F}({\bf{k}} \pm {\bf{q}}/2) = 1 $ are given below. They have
to be proven iteratively. When $ n_{F}({\bf{k}}+{\bf{q}}/2) = 0 $
and $  n_{F}({\bf{k}}-{\bf{q}}/2) = 0 $, we can see from Eq.\
(\ref{OFFDIAG}) that $c^{\dagger}_{ {\bf{k}} + {\bf{q}}/2 }c_{
{\bf{k}} - {\bf{q}}/2 }
 = \lambda \sum_{ {\bf{q}}_{1} \neq {\bf{q}}, {\bf{0}} }
A^{\dagger}_{ {\bf{k}} + {\bf{q}}/2 - {\bf{q}}_{1}/2
}({\bf{q}}_{1}) A_{ {\bf{k}} - {\bf{q}}_{1}/2
}(-{\bf{q}}+{\bf{q}}_{1})$. We now show that this result is
identically zero. This is consistent with our earlier claim that
further iterations do not change the zeroth order results. Since,
now if we replace the $A$'s by the zeroth order ones, we get a
first order result for $ c^{\dagger}_{ {\bf{k}} + {\bf{q}}/2 }c_{
{\bf{k}} - {\bf{q}}/2 } $ equal to zero:
\begin{eqnarray*}
c^{\dagger}_{ {\bf{k}} + {\bf{q}}/2 }c_{ {\bf{k}} - {\bf{q}}/2 } &
=& \lambda \sum_{ {\bf{q}}_{1} \neq {\bf{q}}, {\bf{0}} }
A^{\dagger}_{ {\bf{k}} + {\bf{q}}/2 - {\bf{q}}_{1}/2
}({\bf{q}}_{1}) A_{ {\bf{k}} - {\bf{q}}_{1}/2
}(-{\bf{q}}+{\bf{q}}_{1}) \\
 &=& \lambda \sum_{ {\bf{q}}_{1} \neq {\bf{q}}, {\bf{0}} }
n_{F}({\bf{k}} + {\bf{q}}/2 - {\bf{q}}_{1}) c^{\dagger}_{ {\bf{k}}
+ {\bf{q}}/2 }c_{ {\bf{k}} + {\bf{q}}/2 - {\bf{q}}_{1} }
c^{\dagger}_{ {\bf{k}} + {\bf{q}}/2 - {\bf{q}}_{1} } c_{ {\bf{k}}
- {\bf{q}}/2 } \\
 &=& \lambda \sum_{ {\bf{q}}_{1} \neq {\bf{q}}, {\bf{0}} }
n_{F}({\bf{k}} + {\bf{q}}/2 - {\bf{q}}_{1}) ({\hat{ {\bf{1}} }} -
n_{ {\bf{k}} + {\bf{q}}/2 - {\bf{q}}_{1} } ) c^{\dagger}_{
{\bf{k}} + {\bf{q}}/2 }c_{ {\bf{k}} - {\bf{q}}/2 }.
\end{eqnarray*}
The number operator $ n_{ {\bf{k}} + {\bf{q}}/2 - {\bf{q}}_{1} } =
n_{F}({\bf{k}} + {\bf{q}}/2 - {\bf{q}}_{1}) {\hat{ {\bf{1}} }} $,
and therefore the first order result is zero as well. Similarly,
we find the same answer when $ n_{F}({\bf{k}}+{\bf{q}}/2) = 1 $
and $ n_{F}({\bf{k}}-{\bf{q}}/2) = 1 $. To put it another way,
iterations around $ n_{ {\bf{k}} } = n_{F}({\bf{k}}){\hat{
{\bf{1}} }} $ do not change the form of $ n_{ {\bf{k}} } $. That
is, $ n_{ {\bf{k}} } $ remains frozen at the value $
n_{F}({\bf{k}}){\hat{ {\bf{1}} }} $. This means that the only
state consistent with such an identity is the noninteracting
ground state. Using this, we can convince ourselves that the
program is consistent, just as the program was consistent in the
Bose case which has been proved rigorously by other means.

In order to do better than just remain at the noninteracting
ground state, one must, just as in the Bose case replace $ n_{
{\bf{k}} } =  \langle n_{ {\bf{k}} } \rangle {\hat{ {\bf{1}} }} +
(n_{ {\bf{k}} } - \langle n_{ {\bf{k}} } \rangle {\hat{ {\bf{1}}
}} ) $, and expand around an expectation value
 $  \langle n_{ {\bf{k}} } \rangle \neq 0 $ for all $ {\bf{k}} $.
Thus we are obliged to consider a system with interactions and at
a finite temperature in order for this scheme to be of practical
significance. The philosophy is that we solve for $ \langle n_{
{\bf{k}} } \rangle $ self-consistently and then pass to the limit
of weak interactions or low temperature if one wants to study the
ideal case. Therefore, we may write
\begin{equation}
\frac{1}{\sqrt{n_{ {\bf{k}} - {\bf{q}}/2 }}}
 = \frac{1}{\sqrt{ \langle n_{ {\bf{k}} - {\bf{q}}/2 } \rangle }}
\left[ {\hat{ {\bf{1}} }} + \frac{ n_{ {\bf{k}} - {\bf{q}}/2 } -
 \langle n_{ {\bf{k}} - {\bf{q}}/2 } \rangle {\hat{ {\bf{1}} }} }
{ \langle n_{ {\bf{k}} - {\bf{q}}/2 } \rangle } \right]
^{-\frac{1}{2}} \label{SQRT} .
\end{equation}
Let us now use the insight obtained in the case of bosons to write
down the correct commutation rules obeyed by the sea-displacement
operators. This is a nontrivial task given the subtleties
involved. The Bose case was soluble exactly via a polar
decomposition which we do not have here. We write down here the
final answers, details of which may be found in Appendix B. The
exact commutation rules obeyed by the sea-displacement operators
it seems are rather complicated. The sea-displacement operators
do, however, obey {\em exact closed} commutation rules as we shall
see in Appendix B. There is a natural and simple regime where the
{\em approximate} commutation rules are those of canonical bosons.
It is the regime of the RPA of Bohm and Pines \cite{Bohm}. They
are
\begin{eqnarray}
\left[ A_{\bf k}({\bf q}), A_{{\bf k}^{'}}({\bf
q}^{'})\right]_{RPA}
& =& 0 ,  \label{COMM1} \\
\left[ A_{\bf k}({\bf q}), A^{\dagger}_{{\bf k}'}({\bf q}')
\right]_{RPA} &=&
 n_{F}({\bf{k}}-{\bf{q}}/2)[1-n_{F}({\bf{k}}+{\bf{q}}/2)]
\delta_{{\bf k}, {\bf k}^{'}} \delta_{{\bf q}, {\bf q}^{'}} .
\label{COMM2}
\end{eqnarray}
We saw in the Bose case that the RPA-commutation rules may be
taken to be the exact rules the only care that we had to exercise
was to couple the condensate bosons to their own chemical
potential. Here too we have to couple the sea-bosons to a momentum
dependent chemical potential as we will soon show. We find that it
is quite appropriate to view the objects $ A_{ {\bf{k}}
}({\bf{q}}) $ as being exact bosons, although for special values
of the indices they behave rather strangely. These simple-looking
results belie the notoriously difficult and technical problem of
the square root in the denominator, which incidentally, we have
only partially resolved [see Eq.\ (\ref{SQRT})]. One potential
criticism of this is to claim that a scheme such as Eq.\
(\ref{SQRT}) is inconsistent with the exact commutation rules
presented in Appendix B. The rebuttal to such a critique is that
even Eq.\ (\ref{SQRT}) is approximate, no matter how many orders
are summed--the reason being that this assumes that fluctuations
in the number operator are small compared with the mean. This is
inconsistent since we may show for example that
 $ \langle n^{2}_{ {\bf{k}} } \rangle -  \langle n_{ {\bf{k}} } \rangle^{2}
=  \langle n_{ {\bf{k}} } \rangle(1 - \langle n_{ {\bf{k}} }
\rangle)$. Therefore, the more noideal the momentum distribution
is, the more it fluctuates. In fact, a momentum distribution that
is highly nonideal $  \langle n_{ {\bf{k}} } \rangle \approx 1/2 $
for most momenta, has the largest fluctuation equal to the mean
itself. There is another important point that should be mentioned.
The definition in Eq.\ (\ref{AKQ}) closely resembles the
definition of condensate displacement operators for bosons. This
is certainly a plus. Further, the factor $
n_{F}({\bf{k}}-{\bf{q}}/2)[1 - n_{F}({\bf{k}}+{\bf{q}}/2)]$ serves
as a cutoff function. It fulfills a very important role. It
ensures that the kinetic energy operator is positive-definite. The
authors initially tried a number of alternatives that attempted to
include terms beyond the RPA-like term, for instance one with $[1
-n_{F}({\bf{k}} + {\bf{q}}/2)] [1 -n_{F}({\bf{k}} - {\bf{q}}/2)]$.
Such attempts are always unsuccessful for the simple reason that
they lead to non-positive kinetic energy operators. With the
cutoff function we have introduced, we may write the kinetic
energy operator in the sea-displacement language as $K = k_{0}
{\hat{N}} + \sum_{ {\bf{k}},{\bf{q}} \neq 0 } \omega_{ {\bf{k}}
}({\bf{q}}) A^{\dagger}_{ {\bf{k}} }({\bf{q}}) A_{ {\bf{k}}
}({\bf{q}})$. It can be seen that $ \omega_{ {\bf{k}} }({\bf{q}})
= ({\bf{k}\cdot{\bf q}}/m)
n_{F}({\bf{k}}-{\bf{q}}/2)[1-n_{F}({\bf{k}}+{\bf{q}}/2)] \geq 0 $.
Here $k_{0} = \frac{1}{N^{0}} \sum_{ {\bf{k}} }\epsilon_{ {\bf{k}}
}n_{F}({\bf{k}}) $ is the kinetic energy per particle. In order
not to disappoint the attentive reader, we collect here some facts
that are true in the absolute sense. From the exact definition in
Eq.\ (\ref{AKQ}) it is fairly obvious that (independent of the
meaning of the square root)
\begin{equation}
 [A_{ {\bf{k}} }({\bf{q}}), n_{ {\bf{p}} }] =
 A_{ {\bf{k}} }({\bf{q}})
(\delta_{ {\bf{p}}, {\bf{k}} + {\bf{q}}/2 }
 - \delta_{ {\bf{p}}, {\bf{k}} - {\bf{q}}/2 }) .
\label{ACOMMN}
\end{equation}
More  generally,
\begin{equation}
F([\{n_{ {\bf{p}} } +\delta_{ {\bf{p}}, {\bf{k}}+{\bf{q}}/2 }
-  \delta_{ {\bf{p}}, {\bf{k}}-{\bf{q}}/2 } \}  ])
A_{ {\bf{k}} }({\bf{q}}) =
A_{ {\bf{k}} }({\bf{q}})F([n])
\label{EQNFA}
\end{equation}
From the definition in Eq.(~\ref{AKQ}) we may write,
\begin{equation}
A_{ {\bf{k}} }({\bf{q}})A^{\dagger}_{ {\bf{k}} }({\bf{q}})
 = n_{F}({\bf{k}}-{\bf{q}}/2)[1 - n_{F}({\bf{k}}+{\bf{q}}/2)]
 (1 - n_{ {\bf{k}}+ {\bf{q}}/2 })
 \label{AADAG}
\end{equation}
 The other commutation rules are collected in
Appendix B. They are important only for special values of the
indices $ {\bf{k}}, {\bf{k}}^{'}, {\bf{q}} $ and $ {\bf{q}}^{'} $.
In the RPA-sense and whenever we are willing to overlook these
nuances (which is always, practically speaking) the RPA
commutation rules of Eq.(~\ref{COMM1})  and Eq.(~\ref{COMM2}) are
quite sufficient and easy to use. Let us now try and compute the
properties of the free Fermi theory using the machinery that we
have just laid down. For this we need to first ask : How to
express $ A^{\dagger}_{ {\bf{k}} }({\bf{q}})A_{ {\bf{k}}
}({\bf{q}})$? We expect the answer to depend crucially on the
meaning of the square root. We do not have any more insight into
the nature of this object, but we will be needing its expectation
value when we try to compute the finite temperature properties of
noninteracting Fermi system in the sea-displacement language.
Fortunately, the expectation value is obtained by a reasonably
straightforward method and one that is very reminiscent of the
Bose case. We shall have occasion to introduce a chemical
potential that is momentum dependent and scales as the logarithm
of the volume of the system so that one may venture into regions
where the square root in the denominator vanishes (as does the
numerator) where the sea-displacement method breaks down. The
introduction of the chemical potential restores the meaning of the
trace that in the Fermi language was intended to span over all the
states. Let us now evaluate the finite-temperature properties,
specifically, the finite-temperature momentum distribution of
noninteracting fermions in the sea-displacement language. To this
end, let us now compute the following correlation function:
\begin{eqnarray}
\langle A^{\dagger}_{ {\bf{k}} }({\bf{q}})A_{ {\bf{k}} }({\bf{q}})
\rangle& = &\frac{1}{Z({\bf{k}}-{\bf{q}}/2)} Tr
\left(e^{-\beta(H-\mu N - \mu_{ {\bf{k}} - {\bf{q}}/2 } n_{
{\bf{k}} - {\bf{q}}/2 })}
 A^{\dagger}_{ {\bf{k}} }({\bf{q}})A_{ {\bf{k}} }({\bf{q}})
\right) , \\
Z({\bf{k}} - {\bf{q}}/2)& =&  Tr(e^{-\beta(H-\mu N - \mu_{
{\bf{k}} - {\bf{q}}/2 } n_{ {\bf{k}} - {\bf{q}}/2 })} ) .
\end{eqnarray}
Using the cyclic property of the trace,  we have
\begin{eqnarray}
\langle A^{\dagger}_{ {\bf{k}} }({\bf{q}})A_{ {\bf{k}} }({\bf{q}})
\rangle & =& \frac{1}{Z({\bf{k}}-{\bf{q}}/2)} Tr\left[ A_{
{\bf{k}} }({\bf{q}})e^{-\beta(H-\mu N - \mu_{ {\bf{k}} -
{\bf{q}}/2 } n_{ {\bf{k}} - {\bf{q}}/2 })}
 A^{\dagger}_{ {\bf{k}} }({\bf{q}})\right] \nonumber \\
&=& \frac{1}{Z({\bf{k}}-{\bf{q}}/2)} Tr\left[ e^{-\beta(H-\mu N -
\mu_{ {\bf{k}} - {\bf{q}}/2 }n_{ {\bf{k}} - {\bf{q}}/2 })}
e^{\beta(H-\mu N - \mu_{ {\bf{k}} - {\bf{q}}/2 }n_{ {\bf{k}} -
{\bf{q}}/2 })} \right. \\
& & \times \left. A_{ {\bf{k}} }({\bf{q}}) e^{-\beta(H-\mu N -
\mu_{ {\bf{k}} - {\bf{q}}/2 } n_{ {\bf{k}} - {\bf{q}}/2 })}
 A^{\dagger}_{ {\bf{k}} }({\bf{q}}) \right] \nonumber \\
& =&  e^{-\beta\frac{ {\bf k}\cdot{\bf q} }{m}} e^{-\beta \mu_{
{\bf{k}} - {\bf{q}}/2 }} \frac{1}{Z({\bf{k}}-{\bf{q}}/2)} Tr\left[
e^{-\beta(H-\mu N - \mu_{ {\bf{k}} - {\bf{q}}/2 }n_{ {\bf{k}} -
{\bf{q}}/2 }) } A_{ {\bf{k}} }({\bf{q}})
 A^{\dagger}_{ {\bf{k}} }({\bf{q}})
\right] \nonumber \\
&= &e^{-\beta\frac{ {\bf k}\cdot{\bf q} }{m}} e^{-\beta \mu_{
{\bf{k}} - {\bf{q}}/2 }} \langle A_{ {\bf{k}} }({\bf{q}})
 A^{\dagger}_{ {\bf{k}} }({\bf{q}})
\rangle . \nonumber
\end{eqnarray}
But we know from Eq.\ ({\ref{AADAG}) that
\begin{eqnarray}
\langle A_{ {\bf{k}} }({\bf{q}})A^{\dagger}_{ {\bf{k}} }({\bf{q}})
\rangle &=& n_{F}({\bf{k}}-{\bf{q}}/2)[1 -
n_{F}({\bf{k}}+{\bf{q}}/2)][ 1 - n^{\beta}({\bf{k}}+{\bf{q}}/2)]
 \nonumber \\
 & \approx & n_{F}({\bf{k}}-{\bf{q}}/2)[1 - n_{F}({\bf{k}}+{\bf{q}}/2)]
\label{EQNAPPR}
\end{eqnarray}
at low temperatures, since the thermodynamic expectation value
 $ n^{\beta}({\bf{k}}+{\bf{q}}/2) $
vanishes exponentially when $ n_{F}({\bf{k}}+{\bf{q}}/2) = 0 $.
Therefore we may write $\langle A^{\dagger}_{ {\bf{k}}
}({\bf{q}})A_{ {\bf{k}} }({\bf{q}})\rangle
 = n_{F}({\bf{k}}-{\bf{q}}/2)[1 -  n_{F}({\bf{k}}+{\bf{q}}/2)]
\exp(-\beta {\bf k}\cdot{\bf q}/m) \exp(-\beta \mu_{ {\bf{k}} -
{\bf{q}}/2 })$. The chemical potential $ \mu_{ {\bf{k}} } $ has to
be chosen so that the correct thermodynamic expectation values are
recovered. It is clear that in the absence of the chemical
potential, we are bound to obtain answers that are incorrect,
unless one is at absolute zero. It is also clear that in order for
Eq.\ (\ref{SENSE}) to make sense, one must choose a chemical
potential that scales as the logarithm of the volume of the
system. Let us set $\exp (\beta \mu_{ {\bf{k}} } ) =
V\exp(\lambda_{ {\bf{k}} })$ for some $ \lambda_{ {\bf{k}} } $.
This may be determined by
 demanding the self-consistency of
\begin{eqnarray}
n_{F, \beta}({\bf{k}}) &=& n_{F}({\bf{k}})
 + \sum_{ {\bf{q}} \neq 0 } n_{F}({\bf{k}}-{\bf{q}})
[1 - n_{F}({\bf{k}})] e^{-\beta \frac{ {\bf k}\cdot{\bf q} }{m} }
e^{\beta
\epsilon_{ {\bf{q}} }} e^{-\beta \mu_{ {\bf{k}}-{\bf{q}} } } \nonumber \\
& &- \sum_{ {\bf{q}} \neq 0 } n_{F}({\bf{k}}) (1 -
n_{F}({\bf{k}}+{\bf{q}})) e^{-\beta \frac{ {\bf k}\cdot{\bf q}
}{m} } e^{-\beta \epsilon_{ {\bf{q}} }} e^{-\beta \mu_{ {\bf{k}} }
}
\label{SENSE} , \\
n_{F,\beta}(k) &= &n_{F}(k) \left( 1 - \frac{ e^{ -\lambda_{k} }
}{V} e^{\beta \frac{k^{2}}{2m}} \sum_{ {\bf{q}} \neq {\bf{k}} } [1
- n_{F}(q)]e^{-\beta \frac{q^{2}}{2m}} \right) \nonumber \\
& & + [1-n_{F}(k)]e^{-\beta \frac{k^{2}}{2m} } \frac{1}{V} \sum_{
{\bf{q}} } n_{F}(q)e^{\beta \frac{q^{2}}{2m}}e^{-\lambda_{q}} .
\label{NUMSEA}
\end{eqnarray}
Let us now focus on a system in two space dimensions, in which
case
\begin{eqnarray*}
n_{F}(k) e^{-\lambda_{k}}
 &=& \left( \frac{ 4 \pi \beta }{2m} \right)
n_{F}(k)(1 - n_{F,\beta}(k)) e^{-\beta \frac{k^{2}}{2m} } e^{\beta
\frac{k_{F}^{2}}{2m} }\\
&  \approx &  \left( \frac{2 \pi \beta}{m}
\right) n_{F}(k), \\
\frac{1}{V}\sum_{ {\bf{q}} }n_{F}(q) e^{\beta\frac{q^{2}}{2m}}
e^{-\lambda_{q}} &  =& \frac{1}{(2\pi)^{2}} \int^{k_{F}}_{0} 2 \pi
q \, dq \,  \left( \frac{4 \pi \beta }{2m} \right) (1 -
n_{F,\beta}(q)) e^{\beta
\frac{k^{2}_{F} }{2m} }\\
& =&  e^{\beta \frac{ k^{2}_{F} }{2m} } .
\end{eqnarray*}
The last result follows if we assume, as we shall, that we are
working at low enough temperatures $ k_{B}T << E_{F} $. For this
program to be self-consistent,   we have to make this assumption.
Perhaps there is a way out of it, but for now the authors are
unable to find it. In any event, it is this regime that is of
considerable physical significance and the fact that the
sea-displacement scheme is consistent in this regime is very
comforting indeed. Moving on, we write
\begin{equation}
\frac{1}{V}\sum_{ {\bf{q}} } [1 - n_{F}({\bf{q}})] e^{-\beta
\frac{ q^{2} }{2m} }
 = \left( \frac{m}{2 \pi \beta } \right)
e^{-\beta \frac{ k^{2}_{F} }{2m} } .
\end{equation}
Substituting these back into Eq.\ (\ref{NUMSEA}) we find
$n_{F,\beta}({\bf{k}}) =
 n_{F}({\bf{k}})n_{F,\beta}({\bf{k}})
 + [1 - n_{F}({\bf{k}})]
\exp [ -\beta ( k^{2} - k^{2}_{F} )/2m]$. At low temperatures when
$ |{\bf{k}}| > k_{F} $, we know that $ n_{ {F},\beta }({\bf{k}})
\approx \exp[ -\beta ( k^{2} - k^{2}_{F} )/2m]$. Therefore, the
sea-displacement language gives the exact same result as the
original Fermi language. Having successfully fixed the chemical
potential of the sea-displacements, we now turn to the  problem of
computing the correlation functions of the noninteracting system
at finite temperature.  For instance we have the four-point
function, $F({\bf{k}}{\bf{q}};{\bf{k}}^{'}{\bf{q}}^{'}) =
 \langle c^{\dagger}_{ {\bf{k}} + {\bf{q}}/2 }c_{ {\bf{k}} - {\bf{q}}/2 }
c^{\dagger}_{ {\bf{k}}^{'} - {\bf{q}}^{'}/2 } c_{ {\bf{k}}^{'} +
{\bf{q}}^{'}/2 } \rangle$. In the Fermi language, it is simply
given by
\begin{equation}
F({\bf{k}}{\bf{q}};{\bf{k}}^{'}{\bf{q}}^{'}) = \delta_{ {\bf{k}},
{\bf{k}}^{'} }\delta_{ {\bf{q}}, {\bf{q}}^{'} }
n_{F,\beta}({\bf{k}}+{\bf{q}}/2) [1 -
n_{F,\beta}({\bf{k}}-{\bf{q}}/2) ]  \label{FOURFERMI}
\end{equation}
It may be shown that in the sea-displacement language, the same
quantity may be evaluated as
\[
F({\bf{k}}{\bf{q}};{\bf{k}}^{'}{\bf{q}}^{'})  = \]
\[
  \delta_{
{\bf{k}}, {\bf{k}}^{'} } \delta_{ {\bf{q}}, {\bf{q}}^{'} } \left[
\langle \sqrt{ n_{ {\bf{k}} + {\bf{q}}/2 } } A_{ {\bf{k}}
}(-{\bf{q}}) A^{\dagger}_{ {\bf{k}}^{'} }(-{\bf{q}}^{'}) \sqrt{n_{
{\bf{k}}^{'}+{\bf{q}}^{'}/2 }} \rangle \right. \nonumber \\
 + \left \langle
A^{\dagger}_{ {\bf{k}} }({\bf{q}}) \sqrt{ n_{ {\bf{k}}-{\bf{q}}/2
} } \sqrt{n_{ {\bf{k}}^{'}-{\bf{q}}^{'}/2 }} A_{ {\bf{k}}^{'}
}({\bf{q}}^{'})
 \rangle \right]
\]
\[
+ \delta_{ {\bf{k}}, {\bf{k}}^{'} } \delta_{ {\bf{q}},
{\bf{q}}^{'}}
 \sum_{ {\bf{q}}_{1} \neq {\bf{q}}, 0 }
\langle A^{\dagger}_{ {\bf{k}}+{\bf{q}}/2-{\bf{q}}_{1}/2
}({\bf{q}}_{1}) A_{ {\bf{k}}+{\bf{q}}/2-{\bf{q}}_{1}/2
}({\bf{q}}_{1}) \rangle n_{F}({\bf{k}}+{\bf{q}}/2-{\bf{q}}_{1}) [
1 - n_{F}({\bf{k}}-{\bf{q}}/2) ]
\]
\begin{equation}
+  \delta_{ {\bf{k}}, {\bf{k}}^{'} } \delta_{ {\bf{q}},
{\bf{q}}^{'} }
 \sum_{ {\bf{q}}_{1} \neq {\bf{q}}, 0 }
\langle A^{\dagger}_{ {\bf{k}}-{\bf{q}}/2+{\bf{q}}_{1}/2
}({\bf{q}}_{1}) A_{ {\bf{k}}-{\bf{q}}/2+{\bf{q}}_{1}/2
}({\bf{q}}_{1}) \rangle n_{F}({\bf{k}}+{\bf{q}}/2) [ 1 -
n_{F}({\bf{k}}-{\bf{q}}/2 + {\bf{q}}_{1})]  .
 \label{FOURSEA}
\end{equation}
 Equation (\ref{FOURSEA}) is obtained by placing two
expressions in Eq.\ (\ref{OFFDIAG}) next to each other,
contracting the indices and finally using Eq.\ (\ref{EQNAPPR}).
Insofar as the scheme for computing the momentum distribution is
satisfactory, it may shown that formula in Eq.\ (\ref{FOURSEA}) is
in fact identical to (\ref{FOURFERMI}). To show this rigorously,
we have to properly interpret the average $\langle \sqrt{ n_{
{\bf{k}}+{\bf{q}}/2 } }A_{ {\bf{k}} }(-{\bf{q}}) A^{\dagger}_{
{\bf{k}}^{'} }(-{\bf{q}}^{'}) \sqrt{ n_{
{\bf{k}}^{'}+{\bf{q}}^{'}/2 } } \rangle$. In fact it is a simple
matter to evaluate this quantity exactly in the Fermi language.
From Eq.\ (\ref{AKQ}) we have $\sqrt{n_{ {\bf{k}}+{\bf{q}}/2 }
}A_{ {\bf{k}} }(-{\bf{q}})
 = c^{\dagger}_{ {\bf{k}}+ {\bf{q}}/2 }c_{ {\bf{k}} - {\bf{q}}/2 }
n_{F}({\bf{k}}+{\bf{q}}/2 ) [1 - n_{F}({\bf{k}}-{\bf{q}}/2 ) ]$.
Therefore we may write quite unambiguously
\begin{eqnarray}
& & \langle \sqrt{n_{ {\bf{k}}+{\bf{q}}/2 } }A_{ {\bf{k}}
}(-{\bf{q}}) A^{\dagger}_{ {\bf{k}}^{'} }(-{\bf{q}}^{'}) \sqrt{
n_{
{\bf{k}}^{'} + {\bf{q}}^{'}/2 } } \rangle \nonumber \\
 & = &n_{F, \beta}( {\bf{k}}+ {\bf{q}}/2 )
[1 - n_{F,\beta}( {\bf{k}}-{\bf{q}}/2 ) ]
n_{F}({\bf{k}}+{\bf{q}}/2 ) [1 - n_{F}({\bf{k}}-{\bf{q}}/2 )]
\delta_{ {\bf{k}}, {\bf{k}}^{'} } \delta_{ {\bf{q}}, {\bf{q}}^{'}
} .
\end{eqnarray}
Similarly we may write,
\begin{eqnarray}
 & & \langle A^{\dagger}_{ {\bf{k}} }({\bf{q}}) \sqrt{ n_{
{\bf{k}}-{\bf{q}}/2 } } \sqrt{ n_{ {\bf{k}}^{'}-{\bf{q}}^{'}/2 } }
A_{ {\bf{k}}^{'} }({\bf{q}}^{'}) \rangle \nonumber \\
& = & n_{F,\beta}( {\bf{k}}+ {\bf{q}}/2 ) [1 - n_{F,\beta}(
{\bf{k}}-{\bf{q}}/2 ) ] n_{F}({\bf{k}}-{\bf{q}}/2 ) [1 -
n_{F}({\bf{k}}+{\bf{q}}/2 ) ] \delta_{ {\bf{k}}, {\bf{k}}^{'} }
\delta_{ {\bf{q}}, {\bf{q}}^{'} } . \label{EQNCORR}
\end{eqnarray}
The main conclusion of these arguments is that one must be careful
when evaluating these expectation values. In particular, it is
wrong to make the approximation $ n_{ {\bf{k}} + {\bf{q}}/2 }
\approx \langle n_{ {\bf{k}} + {\bf{q}}/2 } \rangle {\hat{
{\bf{1}} }}$, since this implies
\[
\langle A^{\dagger}_{ {\bf{k}} }({\bf{q}})\sqrt{ n_{
{\bf{k}}-{\bf{q}}/2 } } \sqrt{ n_{ {\bf{k}}^{'}-{\bf{q}}^{'}/2 } }
A_{ {\bf{k}}^{'} }({\bf{q}}^{'}) \rangle  =_{?}
n_{F}({\bf{k}}-{\bf{q}}/2) n_{F}({\bf{k}}^{'}-{\bf{q}}^{'}/2)
\langle A^{\dagger}_{ {\bf{k}} }({\bf{q}})A_{ {\bf{k}}^{'}
}({\bf{q}}^{'}) \rangle
\]
\begin{equation}
 =  n_{F}({\bf{k}}-{\bf{q}}/2)[1 - n_{F}({\bf{k}}+{\bf{q}}/2)]
\delta_{ {\bf{k}}, {\bf{k}}^{'} } \delta_{ {\bf{q}}, {\bf{q}}^{'}
} \left( \frac{1}{V} \right) e^{-\beta \frac{ {\bf k}\cdot{\bf q}
}{m} } \left( \frac{ 2 \pi \beta }{m} \right) .  \label{INCORR}
\end{equation}
This result is inconsistent with the correct result in Eq.\
(\ref{EQNCORR}) that does not vanish in the thermodynamic limit.
Having said this, it is still not a poor approximation at low
temperatures ($ k_{B}T << E_{F} $). The reason being that although
the approximation on the right-hand side of Eq.\ (\ref{INCORR})
vanishes in the thermodynamic limit at any finite temperature, the
exact answer for this correlation function as derived above also
vanishes for a different reason. At low temperatures, due to the
presence of a product such as $ n_{F,\beta}( {\bf{k}}+{\bf{q}}/2
)[1 - n_{F}({\bf{k}}+{\bf{q}}/2) ]$, the correlation function
vanishes exponentially fast as $\exp[-(\epsilon_{
{\bf{k}}+{\bf{q}}/2 } - E_{F})/k_{B}T]$. Therefore, while it is
certainly advisable to be cautious, we are allowed some leeway at
low temperatures. Having done all this, it is now a simple matter
to convince ourselves that Eq.\ (\ref{FOURSEA}) is identical to
(\ref{FOURFERMI}). If $ n_{F}({\bf{k}}+{\bf{q}}/2)  = 1 $ and $
n_{F}({\bf{k}}-{\bf{q}}/2) = 0 $, then all but the first term in
Eq.(~\ref{FOURSEA}) vanish and the identity is trivially
satisfied. If $ n_{F}({\bf{k}}+{\bf{q}}/2)  = 0 $ and $
n_{F}({\bf{k}}-{\bf{q}}/2) = 1 $, then all but the second term
vanish and the identity is again trivially satisfied. What remains
now is to compute the terms when $ n_{F}({\bf{k}}+{\bf{q}}/2) = 0
$ and $ n_{F}({\bf{k}}-{\bf{q}}/2) = 0 $ and when $
n_{F}({\bf{k}}+{\bf{q}}/2) = 1 $ and $ n_{F}({\bf{k}}-{\bf{q}}/2)
= 1 $. In the first instance, we have to evaluate the sum $S_{1} =
\sum_{ {\bf{q}}_{1} \neq {\bf{q}}, 0 } \langle A^{\dagger}_{
{\bf{k}}+{\bf{q}}/2-{\bf{q}}_{1}/2 }({\bf{q}}_{1}) A_{
{\bf{k}}+{\bf{q}}/2-{\bf{q}}_{1}/2 }({\bf{q}}_{1}) \rangle
n_{F}({\bf{k}}+{\bf{q}}/2-{\bf{q}}_{1})$. For this we have to make
use of the simplified expression in 2D $\langle A^{\dagger}_{
{\bf{k}} }({\bf{q}}) A_{ {\bf{k}} }({\bf{q}}) \rangle
 = n_{F}({\bf{k}}-{\bf{q}}/2) (1 - n_{F}({\bf{k}}+{\bf{q}}/2))
\exp (-\beta {\bf k}\cdot{\bf q}/m)2 \pi \beta/(mV)$. Therefore,
we may write $ S_{1} $ as
\begin{equation}
S_{1} = \frac{1}{V} \sum_{ {\bf{Q}}_{1} } n_{F}({\bf{Q}}_{1})
e^{\beta \frac{ Q^{2}_{1} }{2m} } \left( \frac{ 2 \pi \beta }{m}
\right) e^{ - \beta \frac{ ({\bf{k}}+{\bf{q}}/2)^{2} }{2m} }
 = e^{\beta \frac{ k_{F}^{2} - ({\bf{k}}+{\bf{q}}/2)^{2} }{2m} }.
\end{equation}
At low temperatures we may approximate Eq.\ (\ref{FOURFERMI}) as
$F({\bf{k}}{\bf{q}};{\bf{k}}{\bf{q}}) = \exp \{ \beta [ k^{2}_{F}
- ({\bf{k}}+{\bf{q}}/2)^{2}]/2m \}$ since when $
n_{F}({\bf{k}}-{\bf{q}}/2) = 0 $ and $ E_{F} >> kT $, and $
n_{F}({\bf{k}}+{\bf{q}}/2) = 0 $ implies $ |{\bf{k}}+{\bf{q}}/2| >
k_{F} $. Therefore we have $1 - n_{F,\beta}({\bf{k}}-{\bf{q}}/2)
\approx 1$ and $n_{F,\beta}({\bf{k}}+{\bf{q}}/2) = \exp\{\beta [
k^{2}_{F}  - ({\bf{k}}+{\bf{q}}/2)^{2}]/2m\}$, and hence the
result follows. Similarly, when $ n_{F}({\bf{k}}+{\bf{q}}/2) = 1 $
and $ n_{F}({\bf{k}}-{\bf{q}}/2) = 1 $ we have
\begin{eqnarray}
& & \sum_{ {\bf{q}}_{1} \neq {\bf{q}}, 0 } \langle A^{\dagger}_{
{\bf{k}}-{\bf{q}}/2 + {\bf{q}}_{1}/2 } ({\bf{q}}_{1})A_{
{\bf{k}}-{\bf{q}}/2 + {\bf{q}}_{1}/2 }({\bf{q}}_{1}) \rangle [1 -
n_{F}({\bf{k}}-{\bf{q}}/2+{\bf{q}}_{1})] \nonumber \\
 &=& \frac{1}{V} \sum_{ {\bf{Q}}_{1} }
(1 - n_{F}({\bf{Q}}_{1})) e^{-\beta \frac{ Q^{2}_{1} }{2m} }
\left( \frac{ 2 \pi \beta }{m} \right) e^{\beta \frac{
({\bf{k}}-{\bf{q}}/2)^{2} }{2m} }
 = e^{\beta \frac{ ({\bf{k}}-{\bf{q}}/2)^{2} - k^{2}_{F} }{2m} } .
\end{eqnarray}
Analogous to the earlier case, we may approximate Eq.\
(\ref{FOURFERMI}) as $F({\bf{k}}{\bf{q}};{\bf{k}}{\bf{q}}) = \exp
\{ \beta [ ({\bf{k}}-{\bf{q}}/2)^{2} - k^{2}_{F} ]/2m\}$.
Therefore, this scheme gives the right answers at finite
temperature for both the momentum distribution and the four-point
functions.

\subsection{ Role of Partial Isometries }

The term partial isometry \cite{Reed} refers to, intuitively, an
almost unitary operator. The absolute unitary nature is spoiled by
an object that is by now ubiquitous in this article, namely the
canonical conjugate $ X_{0} $ of the total number operator. Thus
we would like to interpret the object
\begin{equation}
U({\bf{k}}) = n_{F}({\bf{k}}) \frac{1}{ \sqrt{n_{ {\bf{k}} }} }
c^{\dagger}_{ {\bf{k}} }
\end{equation}
as an almost unitary operator when $ n_{F}({\bf{k}}) = 1 $. The
presence of the square root of the number operator in the
denominator led others \cite{Cune} to conclude that our theory
contains divergences. A closer examination tells us that this is
not the case. The creation operator $ c^{\dagger}_{ {\bf{k}} } $
scales as the square root of the number operator. In other words,
if we write
 \begin{equation}
 c^{\dagger}_{ {\bf{k}} } = \sqrt{ n_{ {\bf{k}} } }
 \exp\{ -i\Theta([n];{\bf{k}})\} \exp (iP^{\dagger}_{ {\bf{k}}
}) \label{CREAT}
 \end{equation},
 the creation operator becomes a nontrivial
complex operator of unit modulus (ideally) times the square root
of the number operator. The creation operator differs from the
square root of the number operator by a phase factor. Analogous to
what we found in our earlier works \cite{Setlur}, this phase
factor consists of a canonical conjugate to the number operator
denoted by $ P_{ {\bf{k}} } $ and a real functional of the number
 operator $ \Theta([n];{\bf{k}}) $. The sea-displacement operator may be
written in the following suggestive form:
\begin{equation}
A_{ {\bf{k}} }({\bf{q}}) = n_{F}({\bf{k}}-{\bf{q}}/2) [1 -
n_{F}({\bf{k}}+{\bf{q}}/2)] e^{-i\Theta([n];{\bf{k}}-{\bf{q}}/2)}
e^{i\mbox{  }P^{\dagger}_{ {\bf{k}}-{\bf{q}}/2 } } c_{ {\bf{k}} +
{\bf{q}}/2 } \label{MANAFORM}
\end{equation}
This form, although more manageable than the definition in Eq.\
(\ref{AKQ}), is ambiguous since we have yet to pin down the
meaning of $P_{ {\bf{k}} }$ and $\Theta([n];{\bf{k}})$. We shall
adopt an approach similar to the one we outlined in our earlier
works\cite{Setlur} where we transform to the Fourier space and
write down a formula for $ P_{ {\bf{k}} } $ in terms of so-called
momentum currents (as opposed to real currents). As before, we
find in our formalism the object $ X_{0} $ that spoils the unitary
nature of these operators. This object though troublesome, is very
important since without it our program would be inconsistent. To
understand the need for this object more clearly let us write (as
we rightly anticipate) $P_{ {\bf{k}} } = X_{0} + {\tilde{P}}_{
{\bf{k}} }$ where $ {\tilde{P}}_{ {\bf{k}} } $ is manifestly
self-adjoint whereas $ X_{0} $ is not self-adjoint and obeys the
relation $ [X_{0},{\hat{N}}] = i $. If $ X_{0} $ is interpreted as
being self adjoint, then the resulting formula for $ A^{\dagger}_{
{\bf{k}} }({\bf{q}})A_{ {\bf{k}} }({\bf{q}}) $ obtained from Eq. \
(\ref{MANAFORM}) would be inconsistent with Eq.\ (\ref{DIAG})
 unless $ n_{ {\bf{k}} } = n_{F}({\bf{k}}) {\hat{ {\bf{1}} }} $,
 a state of affairs
realized only in the noninteracting system at zero temperature.
Therefore, it is imperative that we do not interpret $ X_{0} $ as
being self-adjoint. However, just as in the Bose case we may still
choose the hermitian part of $ X_{0} $ provided we interpret the
polar decomposition in Eq.(~\ref{CREAT}) in the distribution
theoretic sense to be made clear below.
 Furthermore, it is not clear
what the role of idempotence is in this formalism.
In Appendix B, we write down {\it{closed}} commutation rules obeyed
by the sea-displacement operators. These are the {\it{exact}}
rules. The derivation of these do not require the use of
idempotence, but we have to make use of the fact
 $ (c_{ {\bf{k}} })^{2} = 0 $. This and other similar  observations
 such as idempotence require
point-splitting techniques. This procedure is well-known to the
traditional bosonizing community, mainly because it is
indispensable in dealing with relativistic systems in 1+1
dimensions which have a different mathematical structure than the
nonrelativistic systems that we consider here. Perhaps these
techniques can even be made mathematically rigorous in those
systems. We had to consider the field operators in real space to
be operator-valued distributions\cite{Setlur}, which required the
introduction of a separable one-particle Hilbert space, which
amounts to placing the system in a box with periodic boundary
conditions. Similarly, we are now forced, it seems, to do the same
with momentum space, that is, we postulate in {\em addition} a
short distance cut-off that amounts to introducing a lattice
constant. Space is now discretized and finite as well. There is a
large macroscopic length scale $ L $ (dimension of the box) and a
small microscopic length scale $ a $ (the lattice spacing). In the
end we set $ L \rightarrow \infty $ and
 $ a \rightarrow 0 $ and hope that our edifice remains intact.
Let us try and make more sense out of Eq.(~\ref{MANAFORM}). First,
we would like a formula for $ X_{0} $ in terms of the Fermi fields
{\it{without}} involving $ \Theta([n];{\bf{k}}) $. To this end,
let us postulate the existence of a 'super boson'. It is defined
as,
\begin{equation}
{\hat{C}} = e^{-iX_{0}}\sqrt{ {\hat{N}} }
\end{equation}
where $ {\hat{N}} = \sum_{ {\bf{k}} } c^{\dagger}_{ {\bf{k}} }c_{
{\bf{k}} } $ and $ [X_{0}, {\hat{N}}] = i $. If $ X_{0} $ is
 strictly self-adjoint, then $ {\hat{C}} $ is indeed an exact
 boson. Otherwise it is a quasi-boson. We would not like to
 have $ X_{0} $ as self-adjoint for reasons already alluded to.
The claim is that the filled Fermi sea is obtained by creating
 $ N^{0} $ number of these bosons from the vacuum of the fermions.
\begin{equation}
(C^{\dagger})^{N^{0}}|0\rangle
 = |FS \rangle = \left( \prod_{ |{\bf{k}}| < k_{F} }
 c^{\dagger}_{ {\bf{k}} } \right)|0\rangle
\end{equation}
We may rewrite the filled Fermi sea as follows,
\begin{equation}
|FS \rangle = \left( \prod_{ |{\bf{k}}| < k_{F} }
 c^{\dagger}_{ {\bf{k}} } \right)|0\rangle
 = exp \left( \sum_{ {\bf{k}} }n_{F}({\bf{k}})
 ln\left( c^{\dagger}_{ {\bf{k}} } \right) \right)|0\rangle
\end{equation}
From this we may deduce,
\begin{equation}
C^{\dagger}
 = exp \left(
\frac{1}{N^{0}}\sum_{ {\bf{k}} }n_{F}({\bf{k}})
 ln\left( c^{\dagger}_{ {\bf{k}} } \right) \right)
 \approx \sqrt{N^{0}}e^{iX^{\dagger}_{0}}
\end{equation}
upto some phases that we will soon show are unimportant.From this
we may read off a formula for $ X_{0} $.
\begin{equation}
X_{0} = \frac{i}{N^{0}}\sum_{ {\bf{k}} }n_{F}({\bf{k}})
 ln\left( c_{ {\bf{k}} } \right)
\label{DEFNFX0}
\end{equation}
From this it is clear that $ X_{0} $ is not self-adjoint and
further we have to show that it is a conjugate to $ {\hat{N}} $.
For this let us compute,
\begin{equation}
e^{i\lambda {\hat{N}} } X_{0} e^{-i\lambda {\hat{N}} }
 = X_{0} + \lambda
\end{equation}
Using the definition in Eq.(~\ref{DEFNFX0}) we have,
\begin{equation}
e^{i\lambda {\hat{N}} } X_{0} e^{-i\lambda {\hat{N}} } =
\frac{i}{N^{0}}\sum_{ {\bf{k}} }n_{F}({\bf{k}})
 ln\left( e^{i\lambda {\hat{N}} } c_{ {\bf{k}} }
 e^{-i\lambda {\hat{N}} }\right)
  = X_{0} + \lambda
\end{equation}
Therefore the conjugate obeys all the expected properties,
independent of the meaning of the logarithm. We would now like to
ascertain how this object commutes with other objects such as $
n_{ {\bf{p}} } = c^{\dagger}_{ {\bf{p}} }c_{ {\bf{p}} } $.
\begin{equation}
e^{i\lambda n_{ {\bf{p}} } } X_{0} e^{-i\lambda n_{ {\bf{p}} } } =
  \frac{i}{N^{0}}\sum_{ {\bf{k}} }n_{F}({\bf{k}})
 ln\left( e^{i\lambda n_{ {\bf{p}} } } c_{ {\bf{k}} }
 e^{-i\lambda n_{ {\bf{p}} } }\right)
 = X_{0} + \lambda \frac{ n_{F}({\bf{p}}) }{ N^{0} }
\label{X0NN}
\end{equation}
 Now we would like a formula for
 the self-adjoint operator $ {\tilde{P}}_{ {\bf{k}} } $.
 First we would like to point out some obvious pitfalls.
 If one takes the point of view advocated in our earlier
 work\cite{Setlur}, then we have to face up to the fact that
 the number operator obeys idempotence since only the real
 space quantities are distributions whereas the momentum space
 quantities are bona fide operators. Idempotence, unfortunately
 is inconsistent with the existence of a canonical conjugate
 such as $ P_{ {\bf{k}} } $ as has been argued earlier.
 This then means that somehow we have to eschew idempotence in
 favor of retaining the conjugate. It is still unclear
 to the authors why idempotence is such a big hurdle and how to
 overcome it. We take the point of view that any
   $ {\tilde{P}}_{ {\bf{k}} } $ should result in a $ A_{ {\bf{k}}}({\bf{q}}) $
that obeys Eq.(~\ref{DIAG}) and furthermore we also demand as
indicated earlier, the identity in Eq.(~\ref{AADAG}). This matter
has been studied by us quite thoroughly and we are unable to find
a completely satisfactory answer. However, we write down one
possibility. This definition involves the problematic line
integral encountered in our earlier work as well\cite{Setlur}. The
resolution to this may be obtained by transforming to the real
space analogous to the approach used for making sense of the
DPVA\cite{Setlur}.
 The quantity $ {\tilde{P}}_{ {\bf{k}} } $ may
 be written down as follows.
\begin{equation}
{\tilde{P}}_{ {\bf{k}} } =  \int^{ {\bf{k}} }d{\bf{l}}\mbox{ }
(-1/n_{ {\bf{p}} }){\bf{I}}({\bf{p}})
 + \Theta([n];{\bf{k}})
 - \int^{ {\bf{k}} }d{\bf{l}}\mbox{  }[-i\Theta,\nabla {\tilde{P}}]({\bf{p}})
\end{equation}
The line integral is to be interpreted as being carried out after
transforming the quantities to real space analogous to the
DPVA\cite{Setlur}. The quantity
 $ {\bf{I}}({\bf{p}}) = (1/2i)[c^{\dagger}_{ {\bf{p}} }\nabla
 c_{ {\bf{p}} } - (\nabla c^{\dagger}_{ {\bf{p}} }) c_{ {\bf{p}} }
 ]$. Finally, we have to write down a prescription for $ \Theta $.
 Since we have discretised the momenta by placing the system in a
 box, we may use the following natural ansatz. There is no need to
 make contact with the free theory very likely since in the
 number-phase respresentation the free theory is recovered
 automatically. Let $ {\bf{k}}_{ {\bf{n}} } = (2\pi/L)(n_{1},n_{2},n_{3})
 $.
 Define the mapping $ t({\bf{k}}_{ {\bf{n}} }) = 2^{n_{1}} \times
 3^{n_{2}} \times 5^{n_{3}} $. It may be seen that $ t $ is a
 bijection. From this we may write\cite{Leggett},
 \begin{equation}
 \Theta([n];{\bf{k}}_{ {\bf{n}} }) = \pi\sum_{ {\bf{p}}_{ {\bf{m}} }
 }\theta(t({\bf{k}}_{ {\bf{n}} }) - t({\bf{p}}_{ {\bf{m}} }))
 n_{ {\bf{p}}_{ {\bf{m}} } }
 \end{equation}
Here $ \theta(x) = 1 $ if $ x > 0 $ and $ \theta(x) = 0 $ is $ x
\leq 0 $ is the usual Heaviside step function.
 This obeys the required recursion relation below,
necessary for ensuring that Fermi statistics are obeyed.
\begin{equation}
\Theta([n_{ {\bf{p}}_{1} }
 - \delta_{ {\bf{p}}_{1}, {\bf{k}}^{'} } \} ]
;{\bf{k}}) + \Theta([n];{\bf{k}}^{'}) - \Theta([n];{\bf{k}})
-\Theta([\{ n_{ {\bf{p}}_{1} }
 - \delta_{ {\bf{p}}, {\bf{k}} } \} ]
;{\bf{k}}^{'})
 = \pm \pi
\label{recur}
\end{equation}

\subsection{Short-Range Interactions}

We have found a curious feature in our
investigations of the nature of the RPA that is worth mentioning.
It seems that depending upon how one groups the
Fermi field operators, one encounters an enigmatic duality between
repulsion and attraction(see Ref.\ \cite{Mahan}, chapter 4). It is
as follows. Consider the interaction term\cite{Comjel} in jellium
\cite{Mahan}:
\begin{eqnarray}
H_{I}& =& \sum_{ {\bf{q}} \neq 0 }\frac{ v_{ {\bf{q}} } }{2V}
\sum_{ {\bf{k}} \neq {\bf{k}}^{'} } c^{\dagger}_{
{\bf{k}}+{\bf{q}}/2 } c_{ {\bf{k}}-{\bf{q}}/2 } c^{\dagger}_{
{\bf{k}}^{'}-{\bf{q}}/2 }
c_{ {\bf{k}}^{'}+{\bf{q}}/2 } , \label{HI1} \\
& =&  -\sum_{ {\bf{q}} \neq 0 }\frac{ v_{ {\bf{q}} } }{2V} \sum_{
{\bf{k}} \neq {\bf{k}}^{'} } c^{\dagger}_{ {\bf{k}}+{\bf{q}}/2 }
c_{ {\bf{k}}^{'}+{\bf{q}}/2 } c^{\dagger}_{
{\bf{k}}^{'}-{\bf{q}}/2 } c_{ {\bf{k}}-{\bf{q}}/2 } \nonumber \\
& =& -\sum_{ {\bf{P}} \neq {\bf{P}}^{'} } \sum_{ {\bf{Q}} \neq 0 }
\frac{ v_{ {\bf{P}}-{\bf{P}}^{'} } }{2V} c^{\dagger}_{
{\bf{P}}+{\bf{Q}}/2 } c_{ {\bf{P}}-{\bf{Q}}/2 } c^{\dagger}_{
{\bf{P}}^{'}-{\bf{Q}}/2 } c_{ {\bf{P}}^{'}+{\bf{Q}}/2 } .
\label{HI2}
\end{eqnarray}
The last result is obtained simply by relabeling the indices. If
we set (see end of Appendix B) $c^{\dagger}_{ {\bf{k}}+{\bf{q}}/2
}c_{ {\bf{k}}-{\bf{q}}/2 }
 =_{RPA}
[A_{ {\bf{k}} }(-{\bf{q}}) + A^{\dagger}_{ {\bf{k}} }({\bf{q}})]$
with equality to the level of RPA,  then we may write
\begin{equation}
H_{I} = \sum_{ {\bf{q}} \neq 0 }\frac{ v_{ {\bf{q}} } }{2V} \sum_{
{\bf{k}} \neq {\bf{k}}^{'} } [A_{ {\bf{k}} }(-{\bf{q}} )+
A^{\dagger}_{ {\bf{k}} }({\bf{q}})][A_{ {\bf{k}}^{'} }({\bf{q}}) +
A^{\dagger}_{ {\bf{k}}^{'} }(-{\bf{q}})] .
\end{equation}
Alternatively, we may also write
\begin{equation}
H_{I} = -\sum_{ {\bf{q}} \neq 0 }\sum_{ {\bf{k}} \neq {\bf{k}}^{'}
} \frac{ v_{ {\bf{k}}-{\bf{k}}^{'} }}{2V} [A_{ {\bf{k}}
}(-{\bf{q}}) + A^{\dagger}_{ {\bf{k}} }({\bf{q}})][A_{
{\bf{k}}^{'} }({\bf{q}}) + A^{\dagger}_{ {\bf{k}}^{'}
}(-{\bf{q}})] .
\end{equation}
The first alternative yields results identical to the ones
presented in our earlier work \cite{Setlur}. The momentum
distributions derived from this interaction term possess sharp
discontinuities. The second alternative is more interesting. It
has a minus sign signifying an apparent change from repulsion to
attraction (only apparently so). It has
been argued in the literature\cite{Mahan} that if we assume that $
v_{ {\bf{q}} } = w_{0} $ is independent of $ {\bf{q}} $ (in other
words, a $\delta$-function interaction in real space), then we may
conclude by examining Eqs.\ (\ref{HI1}) and (\ref{HI2}), that $
H_{I} = - H_{I} $. Therefore, $H_{I} =0$. {\em The exchange
correlation energy of a system of fermions interacting via a
$\delta$-function interaction is identically zero.}  This is not
surprising since we know that two fermions cannot be at the same
point and the interaction is zero unless they are at the same
point, therefore one obtains an answer zero. This means that for
such systems the Hamiltonian is a simple functional of the number
operator.
\begin{equation}
H = \sum_{ {\bf{k}} }\epsilon_{ {\bf{k}} } n_{ {\bf{k}} }
 -\frac{ w_{0} }{2V}
\sum_{ {\bf{k}},{\bf{q}} \neq 0 } n_{ {\bf{k}} + {\bf{q}}/2 }n_{
{\bf{k}} - {\bf{q}}/2 } .
\end{equation}
Unfortunately, this Hamiltonian does not yield results too
different from the noninteracting case. However, we have found
that if one relaxes the assumption that $ v_{ {\bf{q}} } $ is
strictly independent of $ {\bf{q}} $ and instead say, puts in a
weak dependence on $ {\bf{q}} $, it is still a good approximation
to ignore the exchange correlation energy (at least in comparison
with the kinetic energy and the exchange self-energy), and we
obtain some rather interesting results for the one-particle Green
function. For instance, we could assume that the potential in real
space instead of being a strict $\delta$-function is somewhat
smeared. For example in 1D, we might  take $V(x) = e^{2} ( a/ \pi)
( x^{2} + a^{2} )^{-1}$. We merely quote the final results for the
one particle properties as this is just an aside. In order to
derive these results one has to make use of the Schwinger's
functional approach to dealing with quantum field theories
\cite{Baym}. This exercise highlights the importance of
fluctuations in the momentum distribution in determining the
salient features of the one-particle Green function in systems
that have very short range interactions and are not too strong.
Following the notation of Kadanoff and Baym \cite{Baym}, we write
down the spectral function and the spectral width as
\begin{eqnarray}
A({\bf{p}},\omega)& =& \frac{ \sqrt{ -\kappa({\bf{p}},\omega) } }
{
F({\bf{p}}) } ,  \label{SPEC} \\
 \Gamma({\bf{p}},\omega)
 & = & \sqrt{ -\kappa({\bf{p}},\omega) }
\end{eqnarray}
 where $ \kappa({\bf{p}},\omega) = (\omega - {\tilde{\epsilon}}_{ {\bf{p}} }
 + \mu)^{2} - 4 F({\bf{p}}) < 0 $.
 In the event $ \kappa({\bf{p}},\omega)  > 0 $,
 $  \sqrt{ -\kappa({\bf{p}},\omega) } = 0 $, by definition.
 Here $ F({\bf{k}}) $ is related to the number-number correlation
 function
$N({\bf{k}},{\bf{k}}^{'}) = \langle n_{ {\bf{k}} }n_{ {\bf{k}}^{'}
} \rangle
 - \langle n_{ {\bf{k}} }  \rangle
\langle n_{ {\bf{k}}^{'} } \rangle $ where $ {\tilde{\epsilon}}_{
{\bf{k}} } $ is the modified dispersion relation.
${\tilde{\epsilon}}_{ {\bf{k}} } = \epsilon_{ {\bf{k}} }
 - V^{-1}\sum_{ {\bf{q}} \neq 0 }
 v_{ {\bf{q}} }  \langle n_{ {\bf{k}}-{\bf{q}} } \rangle $.
Furthermore, $F({\bf{k}}) = V^{-2} \sum_{ {\bf{q}},{\bf{q}}^{'}
\neq 0 }  v_{ {\bf{q}} } v_{ {\bf{q}}^{'} }  N({\bf{k}}-{\bf{q}},
{\bf{k}}-{\bf{q}}^{'})$. The momentum distribution has to be
determined self-consistently.
\begin{equation}
\langle n_{ {\bf{p}} } \rangle
 = \left( \frac{2}{\pi} \right)
\int^{\pi/2}_{-\pi/2} d \theta \, \frac{\cos^{2} \theta } {
e^{\beta( {\tilde{\epsilon}}_{ {\bf{p}} } - \mu )}
 e^{2\beta\sqrt{F({\bf{p}})} \sin \theta }
 + 1 } .
\end{equation}
These formulas are incomplete unless one has a prescription for
the object $ N({\bf{k}},{\bf{k}}^{'}) $. We note here some
properties of this object. First, $ N({\bf{k}},{\bf{k}}) = \langle
n_{ {\bf{k}} } \rangle (1 - \langle n_{ {\bf{k}} } \rangle) $. It
may occur to the reader that since $ [n_{ {\bf{k}} },H] = 0 $, the
ground state of the system should be an eigenstate of $ n_{
{\bf{k}} } $, and therefore $ N({\bf{k}},{\bf{k}}^{'}) = 0 $ for $
{\bf{k}} \neq {\bf{k}}^{'} $. This quantity then signifies the
importance of terms beyond exchange energy. We have have tried an
number of ansatzs for this object. In particular, if we ignore the
dependence on the angle $ {\bf{k}}.{\bf{k}}^{'} $, we may write a
reasonable-looking formula for this object($ {\bf{k}} \neq
{\bf{k}}^{'} $).
\begin{equation}
N({\bf{k}},{\bf{k}}^{'}) = (\langle n_{ {\bf{k}} } \rangle -
n_{\beta}({\bf{k}}) ) (\langle n_{ {\bf{k}}^{'} } \rangle -
n_{\beta}({\bf{k}}^{'}) ) \label{NNCORR1}
\end{equation}
This has the attractive feature of conserving the number of
particles and also the net momentum of the electrons ( $ \sum_{
{\bf{k}} }N({\bf{k}},{\bf{k}}^{'}) \approx 0 $, $   \sum_{
{\bf{k}} } {\bf{k}} \mbox{       }N({\bf{k}},{\bf{k}}^{'}) \approx
0 $ ) and further since $ 0 < n_{ {\bf{k}} } < 1 $ we expect $ -1
< N({\bf{k}},{\bf{k}}^{'} ) < 1 $ which is also obeyed. It has
been pointed out that this object $ N({\bf{k}},{\bf{k}}^{'}) $ is
singular in the BCS case \cite{Leggett} . In particular, it
 is identically zero unless $ {\bf{k}} = \pm {\bf{k}}^{'} $.
 This means that for the BCS case which is a special case we may
 write,
\begin{equation}
 N({\bf{k}},{\bf{k}}^{'}) = \langle N \rangle \mbox{  }\Delta({\bf{k}})
 ( \delta_{ {\bf{k}},{\bf{k}}^{'} } - \delta_{ {\bf{k}}, -{\bf{k}}^{'} })
\label{NNCORRBCS}
\end{equation}
Here $ \Delta({\bf{k}}) $ is a function of order unity and
$\langle N \rangle $ is the number of particles. Ideally, we would
like a formula for this object that in the normal Fermi liquid
regime reduces to an unremarkable function of $ |{\bf{k}}|,
|{\bf{k}}^{'}| $ and $ {\bf{k}}.{\bf{k}}^{'} $ and exhibits a
phase transition to the singular appearance as one crosses over
into the superconducting regime. In fact it may be seen that the
BCS correlation functions supports a net momentum in the sense of
a quantum fluctuation $ \langle {\vec{P}}^{2} \rangle_{BCS}  =
 2\langle N \rangle \sum_{ {\bf{k}} }k^{2} \Delta({\bf{k}})\neq 0
$. It will be shown later that such a formula is indeed possible
in principle. Bare short-range interactions are not of much
interest in continuum systems so we leave it at that (however Eq.
( ~\ref{NNCORR1} ) and Eq. ( ~\ref{NNCORRBCS} ) are meant to be
universally valid ).

\subsection{Long-Range Interactions}

For long-range interactions, if one considers an approach similar
to the one used for deriving the traditional RPA dielectric
function, but this time including possible fluctuations in the
momentum distribution, we find a result that is different form the
usual RPA-dielectric function but one that reduces to it in the
weak coupling limit. This derivation is found in Appendix D. We
merely quote the final answer here:
\begin{equation}
\epsilon_{eff}({\bf{q}},\omega) =
\epsilon_{g-RPA}({\bf{q}},\omega)
 - (\frac{v_{ {\bf{q}} } }{V})^{2}\frac{ P_{2}({\bf{q}},\omega) }
{ \epsilon_{g-RPA}({\bf{q}},\omega) } .  \label{EFFDI}
\end{equation}
 Here,
\begin{equation}
P_{2}({\bf{q}},\omega) =
\end{equation}
\begin{equation}\sum_{ {\bf{k}},{\bf{k}}^{'} } \frac{
N({\bf{k}}+{\bf{q}}/2, {\bf{k}}^{'}+{\bf{q}}/2) -
N({\bf{k}}-{\bf{q}}/2, {\bf{k}}^{'}+{\bf{q}}/2) -
N({\bf{k}}+{\bf{q}}/2, {\bf{k}}^{'}-{\bf{q}}/2) +
N({\bf{k}}-{\bf{q}}/2, {\bf{k}}^{'}-{\bf{q}}/2) } { (\omega -
{\tilde{\epsilon}}_{ {\bf{k}}-{\bf{q}}/2 } + {\tilde{\epsilon}}_{
{\bf{k}}+{\bf{q}}/2 }) (\omega - {\tilde{\epsilon}}_{
{\bf{k}}^{'}-{\bf{q}}/2 } + {\tilde{\epsilon}}_{
{\bf{k}}^{'}+{\bf{q}}/2 }) } ,
\end{equation}
\begin{equation}
\epsilon_{g-RPA}({\bf{q}},\omega) = 1 + \frac{ v_{ {\bf{q}} } }{V}
\sum_{ {\bf{k}} } \frac{ \langle n_{0}({\bf{k}}+{\bf{q}}/2)
\rangle
 - \langle n_{0}({\bf{k}}-{\bf{q}}/2) \rangle }
{ \omega - {\tilde{\epsilon}}_{ {\bf{k}}+{\bf{q}}/2 } +
{\tilde{\epsilon}}_{ {\bf{k}}-{\bf{q}}/2  } }
\end{equation}
and,
\begin{equation}
{\tilde{\epsilon}}_{ {\bf{k}} } =  \epsilon_{ {\bf{k}} }
 - \sum_{ {\bf{q}} \neq 0 } \frac{ v_{ {\bf{q}} }  }{V}
\langle n_{ {\bf{k}}-{\bf{q}} } \rangle
\end{equation}
The momentum distribution in such a system with long-range
interactions may be written down as follows.
\begin{equation}
{\bar{n}}_{ {\bf{k}} } = n_{F}({\bf{k}}) \mbox{  } F_{1}({\bf{k}})
+ (1 - n_{F}({\bf{k}})) \mbox{    }F_{2}({\bf{k}})
\end{equation}
\begin{equation}
F_{1}({\bf{k}}) = \frac{1}{ 1 + \frac{ S_{B}({\bf{k}}) }{ 1 +
S_{A}({\bf{k}}) } }
\end{equation}
\begin{equation}
F_{2}({\bf{k}}) = \frac{1}{ 1 + \frac{ 1 + S_{B}({\bf{k}}) }{
S_{A}({\bf{k}}) } }
\end{equation}
\begin{equation}
S_{A}({\bf{k}}) = \sum_{ {\bf{q}}, i } \frac{ {\bar{n}}_{
{\bf{k-q}} } } { ( \omega_{i}(-{\bf{q}}) + {\bf{k.q}}/m -
\epsilon_{ {\bf{q}} } )^{2} } g^{2}_{i}(-{\bf{q}})
\end{equation}
\begin{equation}
S_{B}({\bf{k}}) = \sum_{ {\bf{q}}, i } \frac{ 1 - {\bar{n}}_{
{\bf{k+q}} } } { ( \omega_{i}(-{\bf{q}}) + {\bf{k.q}}/m +
\epsilon_{ {\bf{q}} } )^{2} } g^{2}_{i}(-{\bf{q}})
\end{equation}
\begin{equation}
g_{i}({\bf{q}}) = \left[ \sum_{ {\bf{k}} } \frac{ {\bar{n}}_{
{\bf{k-q}} } - {\bar{n}}_{ {\bf{k+q}} } } { (\omega_{i}({\bf{q}})
- \frac{ {\bf{k.q}} }{m})^{2} } \right]^{-\frac{1}{2}}
\end{equation}
This quantity $ g_{i}({\bf{q}}) $ is a residue of complex
integration\cite{Setlur} and is given as the frequency derivative
of the polarization.
\begin{equation}
g_{i}({\bf{q}}) = \frac{V}{v_{ {\bf{q}} }}
 \left( \frac{ \partial }{ \partial \omega } \right)
 _{ \omega = \omega_{i}({\bf{q}}) }
 \epsilon^{P}_{eff}({\bf{q}},\omega)
\end{equation}
here $  \epsilon^{P}_{eff} $ is the principal part of $
\epsilon_{eff} $. moreover $ \omega_{i} $ are the zeros of this
dielectric function.
\begin{equation}
\epsilon_{eff}({\bf{q}},\omega_{i}) = 0
\end{equation}
As it stands the above equation is ill-defined since the
dielectric function is complex. Interpreting $ \omega_{i} $ to be
the zero of the principal part results in capturing only the
collective mode. The particle-hole mode which is completely lost
in this approach is very important. In order not to lose this mode
we have to interpret the zeros in a special manner. We shall take
the point of view that all positive energies are allowed as zeros
but each comes with a weight corresponding to the strength of the
dynamical structure factor at that energy. Define the weight to be
\begin{equation}
W({\bf{q}},\omega) = Im \left( \frac{1}{
\epsilon_{eff}({\bf{q}},\omega - i0^{+}) } \right)
\end{equation}
Therefore, sum over modes is now interpreted as,
\begin{equation}
\sum_{ {\bf{q}}, i } f({\bf{q}},\omega_{i})
 = \frac{ \int^{\infty}_{0} d\omega \mbox{     }
 W({\bf{q}},\omega) f({\bf{q}},\omega) }
 {  \int^{\infty}_{0} d\omega \mbox{     } W({\bf{q}},\omega) }
\end{equation}
It is worth emphasizing that Eq.(~\ref{EFFDI}) is valid for long
range interactions where it is wrong to ignore the exchange
correlation term. The one-particle properties that were derived in
Eq.(~{\ref{SPEC}) are valid for short range interactions only. The
interaction strength had to be weak so that one could ignore the
exchange-correlation term in the short-range case. This was also
needed in the long-range case to ensure that one could ignore the
quadratic terms in the sea-displacements. However, in arriving at
Eq.(~\ref{EFFDI}) use was made of the form in Eq.(~\ref{HI1})
rather than Eq.(~\ref{HI2}). There really is no a-priori reason to
prefer one over the other. It is possible that one has to use a
combination of the two(in fact, this has been shown in a preprint
\cite{preprint}).
 Using the second form exclusively
(Eq.(~\ref{HI2})) leads to a dielectric function that is totally
different from the RPA-dielectric function (we have not been able
to write down an explicit formula for this), thereby casting doubt
on the sanctity of the traditional methods of deriving dielectric
functions. This issue has probably not been addressed fully by the
many-body community even though a well-known text-book mentions
some of these facts (Mahan\cite{Mahan} : end of chapter 4). One
may argue that the latter form, namely in Eq.(~\ref{HI2}) at least
for Coulomb repulsion ($ 1/r $) in three spatial dimensions is
negligible at high density. This is because in such a case we have
two independent length scales $ k^{-1}_{F} $ and $ a_{B} $ and if
$ k_{F}a_{B} >> 1 $, we may express all states as linear
combinations of low lying excited states of the noninteracting
system and then
 $ |{\bf{k}}| \approx |{\bf{k}}^{'}| \approx k_{F} $
and $ 0 < |{\bf{q}}| \approx a^{-1}_{B} << k_{F} $, this means
that $ |{\bf{k}}-{\bf{k}}^{'}|^{2} >> (  a^{-1}_{B} )^{2} \approx
q^{2} $. This means that Eq.(~\ref{HI2}) is negligible compared to
Eq.(~\ref{HI1}).
This ambiguity in operator ordering that leads to
very different looking hamiltonians as in Eq.(~\ref{HI1}) and
Eq.(~\ref{HI2}) suggests that this problem will persist in all
cases where electron-electron interactions are present, in other
words,
  in nearly all of many-body physics.
  This suggests to the authors that one must look elsewhere to find
 a more natural home for these techniques. We have some reason to believe
 that quantum electrodynamics is such a place. There, electron-electron
 interactions come about indirectly via-coupling to gauge fields where the
 operator ordering ambiguity is absent. Furthermore, a phenomenon important
 to condensed matter physics, namely magnetism, being primarily due to the
 spin of the electron, is taken into account naturally in the relativistic
theory, since spin is a consequence of the Dirac
equation\cite{Dirac}. In the next section, we study
charge-conserving electron-hole systems that is a precursor to
this more ambitious program of reworking gauge theories in the
sea-displacement language. To conclude this section, we point out
another definition of the sea-displacement that gives us the
illusion of understanding the meaning of the square root. This is
similar to Eq.(~\ref{EQNN2}) for bosons.
\begin{equation}
 \begin{small}
A_{ {\bf{k}} }({\bf{q}}) = n_{F}({\bf{k}}-{\bf{q}}/2)(1 -
n_{F}({\bf{k}}+{\bf{q}}/2)) \left( {\hat{ {\bf{1}} }} - \sum_{
{\bf{q}}_{1} \neq 0 } A^{\dagger}_{ {\bf{k}} - {\bf{q}}/2 +
{\bf{q}}_{1}/2 } ({\bf{q}}_{1}) A_{ {\bf{k}} - {\bf{q}}/2 +
{\bf{q}}_{1}/2 } ({\bf{q}}_{1}) \right)^{-\frac{1}{2}}
c^{\dagger}_{ {\bf{k}}-{\bf{q}}/2 }c_{ {\bf{k}}+{\bf{q}}/2 }
\end{small}
\label{SQRTINT}
\end{equation}
The reason why this formula is not too illuminating is because a
power series expansion around the unit operator and an iterative
solution, yields just as in the Bose case, a result that amounts
to ignoring the presence of the square root. That is,
\begin{equation}
A_{ {\bf{k}} }({\bf{q}}) =_{?} n_{F}({\bf{k}}-{\bf{q}}/2)(1 -
n_{F}({\bf{k}}+{\bf{q}}/2)) c^{\dagger}_{ {\bf{k}}-{\bf{q}}/2 }c_{
{\bf{k}}+{\bf{q}}/2 }
\end{equation}
This is not the definition that is intended\cite{Cunecrit}.
Instead, one must also consider regimes in which the square root
is close to zero rather than close to unity. The exact definition
of $ A_{ {\bf{k}} }({\bf{q}}) $ is the 'self-consistent' solution
of the nonlinear operator equation Eq.(~\ref{SQRTINT}). Readers
with superior mathematical skills may finally be able to solve
this operator equation and map out its domain and range in Fock
space and explain how this object acts on elements of its domain.

\subsection{Comparison With Other Truly Exactly Solved Models}

One of the other criticisms that has been expressed against the
sea-boson method is that it is unable to reproduce Luttinger
Liquid behavior. The authors have convinced themselves that the
Calogero-Sutherland Model(CSM) is in fact a Luttinger liquid. A
 Luttinger liquid is characterized by a lack of discontinuity in
  the momentum distribution, and the divergence(negative infinity)
 and continuity of the slope of momentum distribution across the
 Fermi surface.
 The
correlation functions of this model has been derived by Ha, Lesage
et.al. and Forrester\cite{CALO}. However, the correlation
functions of this model depend on the use of a singular gauge
transformation that forces a particular statistics ( in our case,
fermionic ) on the correlation functions. The authors who have
computed the correlation functions have used the natural
statistics for the CSM which is in general, fractional. This makes
comparison with our approach difficult. The CSM for statistical
parameter $ \beta = 3 $(in the notation of Lesage
et.al.\cite{CALO}) is the simplest model that describes
interacting fermions. However this choice makes the potential
energy of the same order as the kinetic energy making the system
strongly interacting. In our approach, the sea-bosons were written
down in the plane wave basis with the tacit assumption that this
is a good choice. It is easy to see that this choice is likely to
be good only for weakly nonideal systems. The fermionic
correlation functions for the weakly coupled CSM has yet to be
written down. According to our expectations, this should result in
a sharp Fermi surface for sufficiently weak repulsion. The Hubbard
model in 1D\cite{Lieb} is another system that we have to compare
our results with. It is exactly solvable only for half-filling.
The correlation functions are hard to deduce in a closed form, but
it is believed that this is a Luttinger liquid for arbitrarily
weak repulsion. This latter result is the  one we have the
greatest difficulty in proving. In any event, in order for the
community to accept the sea-boson technique,we have to somehow
reproduce Luttinger liquid features. Work is in progress to do
exactly this. It will be reported in a future publication.

\section{Electron-Hole Systems : Exciton-Exciton Interactions}

In this section, we study the exciton Green function using the
sea-displacement technique. In fact, the biexciton Green function
would be more interesting since it determines nonlinear optical
response. This technique is ideally suited to study this quantity
since exciton-exciton interactions are crucial in determining the
true nature of the bi-exciton. We find that exciton-exciton
interactions are not of the two-body type as it is often assumed
in the literature but rather more complicated. These interactions
are mediated by other bosons that are present in the system,
namely the intra-band particle hole excitations and the
zero-momentum inter-band particle-hole excitation. All this will
be made clear in the present section. Let us now move on to the
details. As we have said before, a generalization is possible of
these techniques to two-component charge-conserving electron-hole
systems. One may define sea-displacement operators $ A_{ {\bf{k}}
\sigma }({\bf{q}}\sigma^{'}) $ that include discrete internal
degrees of freedom that may be metaphorically called spin. These
objects are defined below. Since the definition in Eq.(~\ref{AKQ})
was valid for both $ {\bf{q}} = 0 $ and $ {\bf{q}} \neq 0 $, we
have,
\begin{equation}
A_{ {\bf{k}} \sigma }({\bf{q}}\sigma^{'})
 = n_{F}({\bf{k}}-{\bf{q}}/2\sigma)
(1 - n_{F}({\bf{k}}+{\bf{q}}/2\sigma^{'})) \frac{1}{ \sqrt{ n_{
{\bf{k}}-{\bf{q}}/2\sigma } } } c^{\dagger}_{
{\bf{k}}-{\bf{q}}/2\sigma } c_{ {\bf{k}}+{\bf{q}}/2\sigma^{'} }
\end{equation}
We now make the following identifications. Let $ c_{ {\bf{k}} } $
be the operator that annihilates an electron from the conduction
band and $ d^{\dagger}_{ -{\bf{k}} } $ be the operator that
creates a hole in valence band. Then we write,
\begin{equation}
c_{ {\bf{k}} \uparrow } = c_{ {\bf{k}} }
\end{equation}
\begin{equation}
c_{ {\bf{k}} \downarrow } = d^{\dagger}_{ -{\bf{k}} }
\end{equation}
Then we have the following formulas for the electron-hole
sea-displacement operators( again for both $ {\bf{q}} = 0 $ and $
{\bf{q}} \neq 0 $ )
\begin{equation}
A_{ {\bf{k}} \downarrow }({\bf{q}}\uparrow)
 = (1 - n^{h}_{F}(-{\bf{k}}+{\bf{q}}/2))
(1 - n^{e}_{F}({\bf{k}}+{\bf{q}}/2)) \frac{1}{\sqrt{ {\hat{
{\bf{1}} }} - d^{\dagger}_{ -{\bf{k}}+{\bf{q}}/2 } d_{
-{\bf{k}}+{\bf{q}}/2 } }} d_{ -{\bf{k}}+{\bf{q}}/2 }c_{
{\bf{k}}+{\bf{q}}/2 }
\end{equation}
\begin{equation}
A_{ {\bf{k}} \uparrow }({\bf{q}}\downarrow)
 = n^{h}_{F}({\bf{k}}+{\bf{q}}/2)
n^{e}_{F}({\bf{k}}-{\bf{q}}/2) \frac{1}{\sqrt{ c^{\dagger}_{
{\bf{k}}-{\bf{q}}/2 } c_{ {\bf{k}}-{\bf{q}}/2 } }} c^{\dagger}_{
{\bf{k}}-{\bf{q}}/2 }d^{\dagger}_{ -{\bf{k}}-{\bf{q}}/2 }
\end{equation}
\begin{equation}
A_{ {\bf{k}} \uparrow }({\bf{q}}\uparrow)
 = n^{e}_{F}({\bf{k}}-{\bf{q}}/2)
(1 - n^{e}_{F}({\bf{k}}+{\bf{q}}/2)) \frac{1}{\sqrt{ c^{\dagger}_{
{\bf{k}}-{\bf{q}}/2 } c_{ {\bf{k}}-{\bf{q}}/2 } }} c^{\dagger}_{
{\bf{k}}-{\bf{q}}/2 }c_{ {\bf{k}}+{\bf{q}}/2 }
\end{equation}
\begin{equation}
A_{ {\bf{k}} \downarrow }({\bf{q}}\downarrow)
 = -n^{h}_{F}(-{\bf{k}}-{\bf{q}}/2)
(1 - n^{h}_{F}(-{\bf{k}}+{\bf{q}}/2)) \frac{1}{ \sqrt{ {\hat{
{\bf{1}} }} - d^{\dagger}_{ -{\bf{k}}+{\bf{q}}/2 } d_{
-{\bf{k}}+{\bf{q}}/2 } }} d^{\dagger}_{ -{\bf{k}}-{\bf{q}}/2 } d_{
-{\bf{k}}+{\bf{q}}/2 }
\end{equation}
If the system is undoped, we are tempted to set $
n^{e}_{F}({\bf{k}}) = n^{h}_{F}({\bf{k}}) = 0 $. But this would be
unwise. For such a choice would make all the sea-displacement
operators but one, identically zero and this would lead to the
conclusion that the exciton-exciton interactions are of the
two-body type exclusively. We will soon argue that this
underestimates the strength and importance of exciton-exciton
interactions. The reason for this fallacy it seems is that we have
to careful with the order in which we take limits. At any nonzero
temperature, we expect that the $ {\bf{k}} = 0 $ state of the
noninteracting system is always occupied even in a fully undoped
system. Thermal fluctuations lead to a non-empty band. This in
turn facilitates exciton-exciton interaction, since we may now
contemplate intra-band particle-hole excitations competing for
prominence with excitons and scattering off them and so on. In
order to make all this more concrete, let us proceed as follows.
We know that for a noninteracting Fermi system,
\begin{equation}
 n^{e}_{F}({\bf{k}}) = \theta(k_{F} - |{\bf{k}}|)
  \hspace{0.3in} n^{h}_{F}({\bf{k}}) = \theta(k_{F} - |{\bf{k}}|)
\end{equation}
If we go to the limit of an undoped system, $ k_{F} \rightarrow 0
$ we find the following result,
\begin{equation}
 n^{e}_{F}({\bf{k}}) = n^{h}_{F}({\bf{k}}) = \delta_{ {\bf{k}}, 0 }
\end{equation}
That is, only the zero momentum state is occupied, the rest are
unoccupied. This choice enables us to have a scheme by which even
in an undoped system, exciton-exciton interactions may be present
and may contribute to the lineshape of the exciton. With this
simplification we may write ( for both $ {\bf{q}} = 0 $ and $
{\bf{q}} \neq 0 $ )
\begin{equation}
A_{ {\bf{k}} \downarrow }({\bf{q}} \uparrow)
 = (1 - \delta_{ {\bf{k}}, {\bf{q}}/2 })
(1 - \delta_{ {\bf{k}}, -{\bf{q}}/2 }) \frac{1}{ \sqrt{ {\hat{
{\bf{1}} }} - d^{\dagger}_{ -{\bf{k}}+{\bf{q}}/2 } d_{
-{\bf{k}}+{\bf{q}}/2 } } } d_{ -{\bf{k}}+{\bf{q}}/2 }c_{ {\bf{k}}
+ {\bf{q}}/2 }
\end{equation}
\begin{equation}
A_{ {\bf{k}} \uparrow }({\bf{q}} \downarrow)
 = \delta_{ {\bf{q}}, 0 }
\delta_{ {\bf{k}}, 0 } \frac{1}{ \sqrt{ c^{\dagger}_{ {\bf{0}} }
c_{ {\bf{0}} } } } c^{\dagger}_{ {\bf{0}} }d^{\dagger}_{ {\bf{0}}
}
\end{equation}
\begin{equation}
A_{ {\bf{k}} \uparrow }({\bf{q}} \uparrow)
 = \delta_{ {\bf{k}}, {\bf{q}}/2 }
(1 - \delta_{ {\bf{k}}, -{\bf{q}}/2 }) \frac{1}{ \sqrt{
c^{\dagger}_{ {\bf{0}} } c_{ {\bf{0}} } } } c^{\dagger}_{ {\bf{0}}
}c_{ {\bf{q}} }
\end{equation}
\begin{equation}
A_{ {\bf{k}} \downarrow }({\bf{q}} \downarrow)
 = -\delta_{ {\bf{k}},-{\bf{q}}/2 }
(1 - \delta_{ {\bf{k}},{\bf{q}}/2 }) \frac{1}{ \sqrt{ {\hat{
{\bf{1}} }} - d^{\dagger}_{ {\bf{q}} } d_{ {\bf{q}} } }  }
d^{\dagger}_{ {\bf{0}} } d_{ {\bf{q}} }
\end{equation}
Let us now ascertain the commutation rules obeyed by these
objects. Following the prescription outlined in Appendix B, we may
write down the RPA-like commutation rules by replacing the right
side of the approximate ones by the leading order results. If $
{\bf{k}} \neq \pm {\bf{q}}/2 $ and $ {\bf{k}}^{'} \neq \pm
{\bf{q}}^{'}/2 $ and $ {\bf{q}} \neq 0 $ then,
\begin{equation}
[A_{ {\bf{k}}\downarrow }({\bf{q}}\uparrow), A_{
{\bf{k}}^{'}\downarrow }({\bf{q}}^{'}\uparrow)]
 = 0
 \hspace{0.3in} [A_{ {\bf{k}}\downarrow }({\bf{q}}\uparrow),
A^{\dagger}_{ {\bf{k}}^{'}\downarrow }({\bf{q}}^{'}\uparrow)]
 = \delta_{ {\bf{k}}, {\bf{k}}^{'} }\delta_{ {\bf{q}}, {\bf{q}}^{'} }
\end{equation}
also,
\begin{equation}
[A_{ {\bf{k}}\downarrow }({\bf{0}}\uparrow), A_{
{\bf{k}}^{'}\downarrow }({\bf{0}}\uparrow)]
 = 0
 \hspace{0.3in} [A_{ {\bf{k}}\downarrow }({\bf{0}}\uparrow),
A^{\dagger}_{ {\bf{k}}^{'}\downarrow }({\bf{0}}\uparrow)]
 = \delta_{ {\bf{k}}, {\bf{k}}^{'} }
\end{equation}
\begin{equation}
[A_{ {\bf{0}}\uparrow }({\bf{0}}\downarrow),
 A^{\dagger}_{ {\bf{0}}\uparrow }({\bf{0}}\downarrow)]
  = 1
\end{equation}
\begin{equation}
[A_{ {\bf{q}}/2 \uparrow }({\bf{q}}\uparrow), A_{ {\bf{q}}^{'}/2
\uparrow }({\bf{q}}^{'}\uparrow)]
 = 0
\hspace{0.3in} [A_{ {\bf{q}}/2 \uparrow }({\bf{q}}\uparrow),
A^{\dagger}_{ {\bf{q}}^{'}/2 \uparrow }({\bf{q}}^{'}\uparrow)]
 = \delta_{ {\bf{q}}, {\bf{q}}^{'} }
\end{equation}
\begin{equation}
[A_{ {\bf{q}}/2 \downarrow }({\bf{q}}\downarrow), A_{
{\bf{q}}^{'}/2 \downarrow }({\bf{q}}^{'}\downarrow)]
 = 0
\hspace{0.3in} [A_{ {\bf{q}}/2 \downarrow }({\bf{q}}\downarrow),
A^{\dagger}_{ {\bf{q}}^{'}/2 \downarrow }({\bf{q}}^{'}\downarrow)]
 = \delta_{ {\bf{q}}, {\bf{q}}^{'} }
\end{equation}
Finally, $ A_{ {\bf{k}} \uparrow }({\bf{0}} \uparrow ) =  A_{
{\bf{k}} \downarrow }({\bf{ 0}}\downarrow) = 0 $ . All other
commutators involving any two of these sea-displacement operators
are zero. In deriving these commutation rules, use has been made
of the following approximate formulas. On the right hand side of
commutation rules
 we are obliged to set the off diagonal terms to be identically zero and
 $ c^{\dagger}_{0}c_{0} = d^{\dagger}_{0}d_{0} \approx 1 $.
 $ c^{\dagger}_{ {\bf{k}} } c_{ {\bf{k}} } =  c^{\dagger}_{ {\bf{k}} } c_{ {\bf{k}} }
\approx 0 $. It is worthwhile to verify some of these commutation
rules explicitly. For example,
\begin{equation}
[A_{ {\bf{0}} \uparrow }({\bf{0}}\downarrow), A^{\dagger}_{
{\bf{0}} \uparrow }({\bf{0}}\downarrow)]
 \approx [c^{\dagger}_{0}d^{\dagger}_{0}, d_{0}c_{0}]
 = c^{\dagger}_{0}c_{0} - d_{0}d^{\dagger}_{0}
 = c^{\dagger}_{0}c_{0} + d^{\dagger}_{0}d_{0} - 1
\approx 2 -1 = 1
\end{equation}
The rest are reasonably straightforward. Using these facts, we may
write the following correspondence for the number conserving Fermi
bilinears (the total number of electrons and holes commutes with
these objects. See Appendix B
 for some hints as to how to derive these formulas).
Here $ {\bf{q}} \neq 0 $, and we have singled out $ {\bf{q}} = 0 $
as a special case. If $ {\bf{k}} \neq \pm {\bf{q}}/2 $
\[
c^{\dagger}_{ {\bf{k}}+{\bf{q}}/2 } c_{ {\bf{k}}-{\bf{q}}/2 }
 = \lambda \mbox{        }
 A^{\dagger}_{ (1/2)({\bf{k}}+{\bf{q}}/2) \uparrow }
({\bf{k}}+{\bf{q}}/2 \uparrow )A_{ (1/2)({\bf{k}}-{\bf{q}}/2)
\uparrow } ({\bf{k}}-{\bf{q}}/2 \uparrow)
\]
\begin{equation}
 +\lambda \mbox{        } \sum_{ {\bf{q}}_{1} \neq {\bf{k}}+{\bf{q}}/2 }
A^{\dagger}_{ {\bf{k}}+{\bf{q}}/2-{\bf{q}}_{1}/2 \downarrow }
({\bf{q}}_{1}\uparrow) A_{ {\bf{k}} - {\bf{q}}_{1}/2 \downarrow
}(-{\bf{q}}+{\bf{q}}_{1} \uparrow)
\end{equation}
\[
d^{\dagger}_{ -{\bf{k}}+{\bf{q}}/2 } d_{ -{\bf{k}}-{\bf{q}}/2 }
 =\lambda \mbox{        } A^{\dagger}_{ (1/2)({\bf{k}}-{\bf{q}}/2) \downarrow }
(-{\bf{k}}+{\bf{q}}/2 \downarrow ) A_{ (1/2)({\bf{k}}+{\bf{q}}/2)
\downarrow } (-{\bf{k}}-{\bf{q}}/2 \downarrow)
\]
\begin{equation}
+\lambda \mbox{        } \sum_{ {\bf{q}}_{1} \neq
-{\bf{k}}+{\bf{q}}/2 } A^{\dagger}_{
{\bf{k}}-{\bf{q}}/2+{\bf{q}}_{1}/2 \downarrow }
({\bf{q}}_{1}\uparrow) A_{ {\bf{k}}+{\bf{q}}_{1}/2 \downarrow
}(-{\bf{q}}+{\bf{q}}_{1} \uparrow)
\end{equation}
\begin{equation}
c^{\dagger}_{0}c_{ -{\bf{q}} }
 = \sqrt{ c^{\dagger}_{0}c_{0} }
\mbox{         }A_{ -{\bf{q}}/2 \uparrow }(-{\bf{q}}\uparrow)
\approx A_{ -{\bf{q}}/2 \uparrow }(-{\bf{q}}\uparrow)
\end{equation}
\begin{equation}
c^{\dagger}_{ {\bf{q}} }c_{0}
 =
A^{\dagger}_{ {\bf{q}}/2 \uparrow }({\bf{q}}\uparrow) \sqrt{
c^{\dagger}_{0}c_{0} } \approx A^{\dagger}_{ {\bf{q}}/2 \uparrow
}({\bf{q}}\uparrow)
\end{equation}
\begin{equation}
d_{ -{\bf{q}} }d^{\dagger}_{0}
 = ( {\hat{ {\bf{1}} }} - d^{\dagger}_{ -{\bf{q}} }d_{ -{\bf{q}} } )^{\frac{1}{2
}} A_{ {\bf{q}}/2 \downarrow }(-{\bf{q}}\downarrow) \approx A_{
{\bf{q}}/2 \downarrow }(-{\bf{q}}\downarrow)
\end{equation}
\begin{equation}
 d_{0}
d^{\dagger}_{ {\bf{q}} } = A^{\dagger}_{ -{\bf{q}}/2 \downarrow
}({\bf{q}}\downarrow) ( {\hat{ {\bf{1}} }} - d^{\dagger}_{
{\bf{q}} }d_{ {\bf{q}} } ) ^{\frac{1}{2}} \approx A^{\dagger}_{
-{\bf{q}}/2 \downarrow }({\bf{q}}\downarrow)
\end{equation}
Also we have for the number operators ( $ {\bf{k}} \neq 0 $ ).
\begin{equation}
c^{\dagger}_{0}c_{0} = {\hat{ {\bf{1}} }}
 -\lambda \mbox{      } \sum_{ {\bf{q}}_{1} \neq 0 }
A^{\dagger}_{ {\bf{q}}_{1}/2\uparrow }({\bf{q}}_{1}\uparrow) A_{
{\bf{q}}_{1}/2\uparrow }({\bf{q}}_{1}\uparrow)
 - \lambda \mbox{        }
A^{\dagger}_{ {\bf{0}} \uparrow }({\bf{0}} \downarrow) A_{
{\bf{0}} \uparrow }({\bf{0}} \downarrow)
\end{equation}
\begin{equation}
c^{\dagger}_{ {\bf{k}} }c_{ {\bf{k}} } = \lambda \mbox{      }
A^{\dagger}_{ {\bf{k}}/2 \uparrow }({\bf{k}}\uparrow) A_{
{\bf{k}}/2 \uparrow }({\bf{k}}\uparrow)
 +\lambda \mbox{      }  \sum_{ {\bf{q}}_{1} \neq {\bf{k}} }
A^{\dagger}_{ {\bf{k}} - {\bf{q}}_{1}/2 \downarrow } ({\bf{q}}_{1}
\uparrow) A_{ {\bf{k}} - {\bf{q}}_{1}/2 \downarrow } ({\bf{q}}_{1}
\uparrow)
\end{equation}
\begin{equation}
d^{\dagger}_{0}d_{0} = {\hat{ {\bf{1}} }}
 -\lambda \mbox{      } \sum_{ {\bf{q}}_{1} \neq 0 }
A^{\dagger}_{ -{\bf{q}}_{1}/2\downarrow }({\bf{q}}_{1}\downarrow)
A_{ -{\bf{q}}_{1}/2\downarrow }({\bf{q}}_{1}\downarrow)
 - \lambda \mbox{      } A^{\dagger}_{ {\bf{0}} \uparrow }({\bf{0}} \downarrow)
A_{ {\bf{0}} \uparrow }({\bf{0}} \downarrow)
\end{equation}
\begin{equation}
d^{\dagger}_{ -{\bf{k}} }d_{ -{\bf{k}} } = \lambda \mbox{
}A^{\dagger}_{ {\bf{k}}/2 \downarrow }(-{\bf{k}}\downarrow) A_{
{\bf{k}}/2 \downarrow }(-{\bf{k}}\downarrow)
 +  \lambda \mbox{      }\sum_{ {\bf{q}}_{1} \neq -{\bf{k}} }
A^{\dagger}_{ {\bf{k}} + {\bf{q}}_{1}/2 \downarrow } ({\bf{q}}_{1}
\uparrow) A_{ {\bf{k}} + {\bf{q}}_{1}/2 \downarrow } ({\bf{q}}_{1}
\uparrow)
\end{equation}
Now for the charge-conserving Fermi bilinear(the total charge
operator commutes with these objects). If $ {\bf{k}} \neq \pm
{\bf{q}}/2 $ then, ( with $ {\bf{q}} \neq 0 $ )
\begin{equation}
d_{ -{\bf{k}}-{\bf{q}}/2 }c_{ {\bf{k}}-{\bf{q}}/2 } = ( {\hat{
{\bf{1}} }} - d^{\dagger}_{ -{\bf{k}}-{\bf{q}}/2 } d_{
-{\bf{k}}-{\bf{q}}/2 } )^{\frac{1}{2}} A_{ {\bf{k}} \downarrow
}(-{\bf{q}}\uparrow) \approx A_{ {\bf{k}} \downarrow
}(-{\bf{q}}\uparrow)
\end{equation}
if in addition we have $ {\bf{k}} \neq 0 $
\begin{equation}
d_{ -{\bf{k}} }c_{ {\bf{k}} } = ( {\hat{ {\bf{1}} }} -
d^{\dagger}_{ -{\bf{k}} } d_{ -{\bf{k}} } )^{\frac{1}{2}} A_{
{\bf{k}} \downarrow }(0\uparrow) \approx A_{ {\bf{k}} \downarrow
}(0\uparrow)
\end{equation}
if $ {\bf{k}} = 0 $
\begin{equation}
d_{0}c_{0} = A^{\dagger}_{0\uparrow}(0\downarrow) \sqrt{
c^{\dagger}_{0} c_{0} } \approx
A^{\dagger}_{0\uparrow}(0\downarrow)
\end{equation}
\begin{equation}
d_{0}c_{ -{\bf{q}} } = \lambda \mbox{      }
A^{\dagger}_{0\uparrow}(0\downarrow) A_{ -{\bf{q}}/2 \uparrow
}(-{\bf{q}} \uparrow)
 + \lambda \mbox{      }
\sum_{ {\bf{q}}_{1} \neq 0 } A^{\dagger}_{-{\bf{q}}_{1}/2,
\downarrow }({\bf{q}}_{1}\downarrow) A_{-{\bf{q}}/2
-{\bf{q}}_{1}/2, \downarrow }(-{\bf{q}}+{\bf{q}}_{1},\uparrow)
\end{equation}
\begin{equation}
d_{ -{\bf{q}} }c_{0} = -\lambda \mbox{      }
A^{\dagger}_{0\uparrow}(0\downarrow) A_{ {\bf{q}}/2 \downarrow
}(-{\bf{q}} \downarrow) -\lambda \mbox{      }
 \sum_{ {\bf{q}}_{1} \neq 0 }
A^{\dagger}_{ {\bf{q}}_{1}/2,\uparrow }({\bf{q}}_{1}\uparrow) A_{
{\bf{q}}/2+{\bf{q}}_{1}/2, \downarrow
}(-{\bf{q}}+{\bf{q}}_{1},\uparrow)
\end{equation}
Let us now verify some of these correspondences. For example we
know that in the Fermi language,
\begin{equation}
[c^{\dagger}_{0}c_{0},c^{\dagger}_{0}c_{ -{\bf{q}} }]
 = c^{\dagger}_{0}c_{ -{\bf{q}} }
\end{equation}
In the sea-displacement language we have,
\begin{equation}
[c^{\dagger}_{0}c_{0},c^{\dagger}_{0}c_{ -{\bf{q}} }]
 = \sum_{ {\bf{q}}_{1} \neq 0 }
[A_{ -{\bf{q}}/2 \uparrow }(-{\bf{q}}\uparrow), A^{\dagger}_{
{\bf{q}}_{1}/2 \uparrow }({\bf{q}}_{1} \uparrow) A_{
{\bf{q}}_{1}/2 \uparrow }({\bf{q}}_{1} \uparrow)]
 = A_{ -{\bf{q}}/2 \uparrow }(-{\bf{q}}\uparrow)
\approx c^{\dagger}_{0}c_{ -{\bf{q}} }
\end{equation}
as required. Similarly we have,
\begin{equation}
[c^{\dagger}_{0}c_{0},d_{0}c_{0}]
 = -d_{0}c_{0}
\end{equation}
In the sea-displacement language we have,
\begin{equation}
[c^{\dagger}_{0}c_{0},d_{0}c_{0}] = -[A^{\dagger}_{
{\bf{0}}\uparrow }({\bf{0}}\downarrow) A_{ {\bf{0}}\uparrow
}({\bf{0}}\downarrow) ,A^{\dagger}_{ {\bf{0}}\uparrow
}({\bf{0}}\downarrow) ]
  = -A^{\dagger}_{ {\bf{0}}\uparrow }({\bf{0}}\downarrow)
 \approx -d_{0}c_{0}
\end{equation}
Next, we write down the hamiltonian of the electron-hole system
interacting via repulsion and attractive interactions. We show how
excitons emerge naturally from the formalism. We are also able to
pin down the precise nature of exciton-exciton interactions. The
total hamiltonian may be split into several parts. First is the
kinetic energy plus the part of the potential energy
 that leads to the exciton.
\[
H_{0} = \sum_{ {\bf{k}} \neq 0 }(\frac{ k^{2} }{2m_{e}} + E_{g})
c^{\dagger}_{ {\bf{k}} }c_{ {\bf{k}} }
 +  \sum_{ {\bf{k}} \neq 0 }(\frac{ k^{2} }{2m_{h}})
d^{\dagger}_{ -{\bf{k}} }d_{ -{\bf{k}} }
 + E_{g}\mbox{  }c^{\dagger}_{0}c_{0}
\]
\begin{equation}
- \sum_{ {\bf{q}} \neq 0 } \frac{ v_{ {\bf{q}} } }{V} \sum_{
{\bf{k}} \pm {\bf{q}}/2 \neq 0 } \sum_{ -{\bf{k}}^{'} \pm
{\bf{q}}/2 \neq 0} c^{\dagger}_{ {\bf{k}}+{\bf{q}}/2 }
d^{\dagger}_{ -{\bf{k}}^{'} - {\bf{q}}/2 } d_{ -{\bf{k}}^{'} +
{\bf{q}}/2 } c_{ {\bf{k}}-{\bf{q}}/2 }
\end{equation}
\[
H_{I} =
 \sum_{ {\bf{q}} \neq 0 }\frac{ v_{ {\bf{q}} } }{2V}
\sum_{ {\bf{k}} \neq {\bf{k}}^{'} ; {\bf{k}} \pm {\bf{q}}/2 \neq 0
} \sum_{ {\bf{k}}^{'} \pm {\bf{q}}/2 \neq 0 } c^{\dagger}_{
{\bf{k}}+{\bf{q}}/2 }c^{\dagger}_{ {\bf{k}}^{'}-{\bf{q}}/2 } c_{
{\bf{k}}^{'}+{\bf{q}}/2 } c_{ {\bf{k}}-{\bf{q}}/2 }
\]
\[
+ \sum_{ {\bf{q}} \neq 0 }\frac{ v_{ {\bf{q}} } }{2V} \sum_{
{\bf{k}} \neq {\bf{k}}^{'} ; {\bf{k}} \pm {\bf{q}}/2 \neq 0 }
\sum_{ {\bf{k}}^{'} \pm {\bf{q}}/2 \neq 0 } d^{\dagger}_{
-{\bf{k}}-{\bf{q}}/2 }d^{\dagger}_{ -{\bf{k}}^{'}+{\bf{q}}/2 } d_{
-{\bf{k}}^{'}-{\bf{q}}/2 } d_{ -{\bf{k}}+{\bf{q}}/2 }
\]
 \begin{equation}
- \sum_{ {\bf{q}} \neq 0 }\frac{ v_{ {\bf{q}} } }{2V} \sum_{
{\bf{k}} \pm {\bf{q}}/2 \neq 0 } n^{(e)}_{ {\bf{k}}+{\bf{q}}/2
}n^{(e)}_{ {\bf{k}}-{\bf{q}}/2 }
 - \sum_{ {\bf{q}} \neq 0 }\frac{ v_{ {\bf{q}} } }{2V}
\sum_{ {\bf{k}} \pm {\bf{q}}/2 \neq 0 } n^{(h)}_{
-{\bf{k}}-{\bf{q}}/2 }n^{(h)}_{ -{\bf{k}}+{\bf{q}}/2 }
\end{equation}
\[
H_{I,0} = \sum_{ {\bf{q}} \neq 0 }\frac{ v_{ {\bf{q}} } }{V}
\sum_{ {\bf{k}} \pm {\bf{q}}/2  \neq 0 } (c^{\dagger}_{0}c_{
{\bf{q}} } + c^{\dagger}_{ -{\bf{q}} }c_{0}) c^{\dagger}_{
{\bf{k}}+{\bf{q}}/2 }c_{ {\bf{k}}-{\bf{q}}/2 } + \sum_{ {\bf{q}}
\neq 0 }\frac{ v_{ {\bf{q}} } }{V} \sum_{ {\bf{k}} \pm {\bf{q}}/2
\neq 0 } (d^{\dagger}_{ -{\bf{q}} }d_{0} + d^{\dagger}_{0}d_{
{\bf{q}} }) d^{\dagger}_{ -{\bf{k}} + {\bf{q}}/2 }d_{ -{\bf{k}} -
{\bf{q}}/2 }
\]
\[
- \sum_{ {\bf{q}} \neq 0 }\frac{ v_{ {\bf{q}} } }{V} \sum_{
{\bf{k}} \pm {\bf{q}}/2 \neq 0 } (c^{\dagger}_{
{\bf{k}}+{\bf{q}}/2 }d^{\dagger}_{0} d_{ {\bf{q}} }c_{
{\bf{k}}-{\bf{q}}/2 }
 + c^{\dagger}_{ {\bf{k}}+{\bf{q}}/2 }d^{\dagger}_{ -{\bf{q}} }
d_{0}c_{ {\bf{k}}-{\bf{q}}/2 })
\]
\begin{equation}
- \sum_{ {\bf{q}} \neq 0 }\frac{ v_{ {\bf{q}} } }{V} \sum_{
-{\bf{k}}^{'} \pm {\bf{q}}/2 \neq 0 }
(c^{\dagger}_{0}d^{\dagger}_{ -{\bf{k}}^{'}+{\bf{q}}/2 } d_{
-{\bf{k}}^{'}-{\bf{q}}/2 } c_{ {\bf{q}} }
 + c^{\dagger}_{ -{\bf{q}} }d^{\dagger}_{ -{\bf{k}}^{'}+{\bf{q}}/2 }
d_{ -{\bf{k}}^{'}-{\bf{q}}/2 }c_{0})
\end{equation}
 The part $ H_{I} $ when written
 out in terms of sea-displacements have four bosons in them.
 Whereas the term $ H_{I,0} $ has three bosons. The kinetic energy
 operator has only two bosons. Thus we may systematically regard
 the kinetic energy
 as being more important
 than $ H_{I,0} $ which in turn
 is more important than $ H_{I} $.
 In the sea-displacement language, we may write upto additive constants,
\[
H_{0} = \sum_{ {\bf{k}} \neq 0 }(\frac{ k^{2} }{2m_{e}})
A^{\dagger}_{ {\bf{k}}/2 \uparrow }({\bf{k}}\uparrow) A_{
{\bf{k}}/2 \uparrow }({\bf{k}}\uparrow) + \sum_{ {\bf{k}} \neq 0
}(\frac{ k^{2} }{2m_{h}}) A^{\dagger}_{ {\bf{k}}/2 \downarrow
}(-{\bf{k}}\downarrow) A_{ {\bf{k}}/2 \downarrow
}(-{\bf{k}}\downarrow)
\]
\[
 - E_{g}A^{\dagger}_{0 \uparrow }(0\downarrow)A_{ 0\uparrow }
(0\downarrow)
\]
\[
+ \sum_{ {\bf{k}} \neq 0 } \left( \frac{ k^{2} }{2\mu} + E_{g}
\right) A^{\dagger}_{ {\bf{k}} \downarrow }({\bf{0}}\uparrow) A_{
{\bf{k}} \downarrow }({\bf{0}}\uparrow) - \sum_{ {\bf{k}} \neq
{\bf{k}}^{'} } \frac{ v_{ {\bf{k}}-{\bf{k}}^{'} } }{V}
A^{\dagger}_{ {\bf{k}} \downarrow }({\bf{0}}\uparrow) A_{
{\bf{k}}^{'} \downarrow }({\bf{0}}\uparrow)
\]
\begin{equation}
 +
\sum_{ {\bf{k}},{\bf{q}} \neq 0 }
  \omega_{ {\bf{k}} }({\bf{q}})
A^{\dagger}_{ {\bf{k}} \downarrow } ({\bf{q}} \uparrow) A_{
{\bf{k}} \downarrow } ({\bf{q}} \uparrow)
 - \sum_{ {\bf{k}} \neq {\bf{k}}^{'} }
\frac{ v_{ {\bf{k}}-{\bf{k}}^{'} } }{V} \sum_{ {\bf{q}} \neq 0 }
A^{\dagger}_{ {\bf{k}} \downarrow }({\bf{q}}\uparrow) A_{
{\bf{k}}^{'} \downarrow }({\bf{q}}\uparrow)
\end{equation}
here,
\begin{equation}
\omega_{ {\bf{k}} }({\bf{q}})
 = \left( \frac{k^{2}}{2\mu} + E_{g} + \frac{ {\bf{k}}.{\bf{q}} }{2}
( \frac{1}{m_{e}} - \frac{1}{m_{h}} )
 + \frac{1}{4} \frac{ q^{2} }{2 \mu} \right)
( 1 - \delta_{ {\bf{k}}, {\bf{q}}/2 }) ( 1 - \delta_{ {\bf{k}},
-{\bf{q}}/2 })
\end{equation}
 This hamiltonian has a very appealing form. The term
 containing $ A_{ {\bf{k}} \downarrow }({\bf{0}}\uparrow) $ has
 been singled out since it highlights the exciton. This is nothing
 but the hamiltonian of the free exciton with center-of mass momentum
 equal to zero. The other hamiltonian involving
 $ A_{ {\bf{k}} \downarrow }({\bf{q}}\uparrow) $
 corresponds to an exciton with non-zero center-of mass motion.
If for example, we write,
\begin{equation}
A_{ {\bf{k}}\downarrow }({\bf{0}}\uparrow)
  =  \sum_{ I }{\tilde{\varphi}}_{I}({\bf{k}})
b_{I}(0)
\end{equation}
where $ {\tilde{\varphi}}_{I}({\bf{k}}) $ is the Fourier transform
of excitonic (hydrogenic) wavefunctions, then the hamiltonian may
be recast in the diagonal form,
\begin{equation}
\sum_{ {\bf{k}} \neq 0 } \left( \frac{ k^{2} }{2\mu} + E_{g}
\right) A^{\dagger}_{ {\bf{k}} \downarrow }({\bf{0}}\uparrow) A_{
{\bf{k}} \downarrow }({\bf{0}}\uparrow) - \sum_{ {\bf{k}} \neq
{\bf{k}}^{'} } \frac{ v_{ {\bf{k}}-{\bf{k}}^{'} } }{V}
A^{\dagger}_{ {\bf{k}} \downarrow }({\bf{0}}\uparrow) A_{
{\bf{k}}^{'} \downarrow }({\bf{0}}\uparrow)
 = \sum_{I} E_{I}(0) b^{\dagger}_{I}(0)b_{I}(0)
\end{equation}
where $ E_{I}(0) $ are the energy levels of the exciton. we shall
assume that a similar transformation has been performed for the
full  hamiltonian including center-of-mass motion.
\[
H_{0} = \sum_{I, {\bf{q}}
}E_{I}({\bf{q}})b^{\dagger}_{I}({\bf{q}}) b_{I}({\bf{q}})
 - E_{g}A^{\dagger}_{0 \uparrow }(0\downarrow)A_{ 0\uparrow }
(0\downarrow)
\]
\begin{equation}
 + \sum_{ {\bf{k}} \neq 0 }(\frac{ k^{2} }{2m_{e}})
A^{\dagger}_{ {\bf{k}}/2 \uparrow }({\bf{k}}\uparrow) A_{
{\bf{k}}/2 \uparrow }({\bf{k}}\uparrow) + \sum_{ {\bf{k}} \neq 0
}(\frac{ k^{2} }{2m_{h}}) A^{\dagger}_{ {\bf{k}}/2 \downarrow
}(-{\bf{k}}\downarrow) A_{ {\bf{k}}/2 \downarrow
}(-{\bf{k}}\downarrow)
\end{equation}
Of particular interest in the term $ - E_{g}A^{\dagger}_{0
\uparrow }(0\downarrow)A_{ 0\uparrow }(0\downarrow) $. It suggests
that the ground state of the system has one of these bosons
present. This is  consistent with the earlier observation since we
have $ {\bf{k}} = 0 $ occupied by one electron and one hole. Since
$ (A_{ 0\uparrow }(0\downarrow))^{2} = 0 $ it is clear that we can
have only one of these bosons present. Let $ | 0 \rangle $ be the
vacuum of the fermions. Thus the $ A_{ 0\uparrow }(0\downarrow) $
creates a solitron. The vacuum of the fermions may be written as,
\begin{equation}
c_{ {\bf{k}} }| 0 \rangle  = 0 \hspace{0.2in} d_{ -{\bf{k}} }| 0
\rangle  = 0
\end{equation}
It may be seen that this vacuum is the vacuum of all the bosons as
well {\it{except}} $ A_{ {\bf{0}} \uparrow }({\bf{0}}\downarrow)
$. The action of $ A_{ {\bf{0}} \uparrow }({\bf{0}}\downarrow) $
on
 $ |0\rangle $ produces
 an electron-hole pair at $ {\bf{k}} = 0 $. Since we have
 argued earlier that this corresponds to the ground state of the
 free theory, we have the following fact.
The ground state of $ H_{0} $ is given by,
\begin{equation}
| G \rangle =  A_{0 \uparrow }(0\downarrow)| 0 \rangle
\end{equation}
It may be seen quite easily that
\begin{equation}
c^{\dagger}_{0}c_{0} | G \rangle =  d^{\dagger}_{0}d_{0} | G
\rangle = 1| G \rangle
\end{equation}
The object $ A_{ {\bf{k}}/2 \uparrow }({\bf{k}}\uparrow) $
annihilates a conductron. Finally the operator
$ A_{ -{\bf{k}}/2 \downarrow }({\bf{k}}\downarrow) $  annihilates
a valeron. Now we would like
to see how this evolves under the presence of the exciton-exciton
interaction terms. A few words regarding these are in order. We
will see here perhaps for the first time, the true nature of
exciton-exciton interactions. They do not interact via simple
two-body interactions rather the interaction is mediated by other
bosons such as $ A_{ {\bf{k}}/2 \uparrow }({\bf{k}}\uparrow) $
which are intra-band particle-hole excitations. Further we have
also coupling to the object $ A^{\dagger}_{0 \uparrow
}(0\downarrow) $ which is an inter-band zero momentum
electron-hole excitation. Thus exciton-exciton interactions are
mediated by these other bosons and the interactions are rather
more complex than the simple two-body variety. Let us write down
precisely what they are.

\tiny

\[
H_{I,0} = \lambda \sum_{ {\bf{q}} \neq 0 } \frac{ v_{ {\bf{q}} }
}{V} \sum_{ {\bf{k}} \pm {\bf{q}}/2 \neq 0 } \left( A_{ {\bf{q}}/2
\uparrow }({\bf{q}} \uparrow)
 + A^{\dagger}_{ -{\bf{q}}/2 \uparrow }(-{\bf{q}} \uparrow) \right)
A^{\dagger}_{ (1/2)({\bf{k}}+{\bf{q}}/2) \uparrow }
({\bf{k}}+{\bf{q}}/2, \uparrow) A_{ (1/2)({\bf{k}}-{\bf{q}}/2)
\uparrow } ({\bf{k}}-{\bf{q}}/2,\uparrow)
\]
\[
+ \lambda \sum_{ {\bf{q}} \neq 0 } \frac{ v_{ {\bf{q}} } }{V}
\sum_{ {\bf{k}} \pm {\bf{q}}/2 \neq 0 } \left( A_{ {\bf{q}}/2
\uparrow }({\bf{q}} \uparrow)
 + A^{\dagger}_{ -{\bf{q}}/2 \uparrow }(-{\bf{q}} \uparrow) \right)
\sum_{ {\bf{q}}_{1} \neq {\bf{k}}+{\bf{q}}/2 } A^{\dagger}_{
{\bf{k}}+{\bf{q}}/2 - {\bf{q}}_{1}/2 \downarrow }
({\bf{q}}_{1},\uparrow) A_{ {\bf{k}}-{\bf{q}}_{1}/2, \downarrow
}(-{\bf{q}}+{\bf{q}}_{1},\uparrow)
\]
\[
+ \lambda \sum_{ {\bf{q}} \neq 0 } \frac{ v_{ {\bf{q}} } }{V}
\sum_{ {\bf{k}} \pm {\bf{q}}/2 \neq 0 } \left( -A^{\dagger}_{
{\bf{q}}/2 \downarrow }(-{\bf{q}} \downarrow)
 - A_{ -{\bf{q}}/2 \downarrow }({\bf{q}} \downarrow) \right)
A^{\dagger}_{ (1/2)({\bf{k}}-{\bf{q}}/2) \downarrow }
(-{\bf{k}}+{\bf{q}}/2, \downarrow) A_{ (1/2)({\bf{k}}+{\bf{q}}/2)
\downarrow } (-{\bf{k}}-{\bf{q}}/2,\downarrow)
\]
\[
+ \lambda \sum_{ {\bf{q}} \neq 0 } \frac{ v_{ {\bf{q}} } }{V}
\left( - A^{\dagger}_{ {\bf{q}}/2 \downarrow }(-{\bf{q}}
\downarrow)
 - A_{ -{\bf{q}}/2 \downarrow }({\bf{q}} \downarrow) \right)
\sum_{ {\bf{q}}_{1} \neq -{\bf{k}}+{\bf{q}}/2 } A^{\dagger}_{
{\bf{k}}-{\bf{q}}/2+{\bf{q}}_{1}/2 \downarrow }
({\bf{q}}_{1},\uparrow) A_{ {\bf{k}}+{\bf{q}}_{1}/2, \downarrow
}(-{\bf{q}}+{\bf{q}}_{1},\uparrow)
\]
\[
-\lambda \sum_{ {\bf{q}} \neq 0 } \frac{ v_{ {\bf{q}} } }{V}
\sum_{ {\bf{k}} \pm {\bf{q}}/2 \neq 0 } A^{\dagger}_{
(1/2)({\bf{k}}+{\bf{q}}/2) \uparrow } ({\bf{k}}+{\bf{q}}/2,
\uparrow) A_{ {\bf{0}}, \uparrow }( {\bf{0}}, \downarrow) A_{
(1/2)({\bf{k}}-{\bf{q}}/2 -{\bf{q}}), \downarrow }
({\bf{k}}+{\bf{q}}/2, \uparrow)
\]
\[
-\lambda \sum_{ {\bf{q}} \neq 0 } \frac{ v_{ {\bf{q}} } }{V}
\sum_{ {\bf{k}} \pm {\bf{q}}/2 \neq 0 } A^{\dagger}_{
(1/2)({\bf{k}}-{\bf{q}}/2 -{\bf{q}}), \downarrow }
({\bf{k}}+{\bf{q}}/2, \uparrow) A^{\dagger}_{ {\bf{0}}, \uparrow
}( {\bf{0}}, \downarrow) A_{ (1/2)({\bf{k}}+{\bf{q}}/2) \uparrow }
({\bf{k}}+{\bf{q}}/2, \uparrow)
\]
\[
+ \lambda \sum_{ {\bf{q}} \neq 0 } \frac{ v_{ {\bf{q}} } }{V}
\sum_{ -{\bf{k}}^{'} \pm {\bf{q}}/2 \neq 0 } A^{\dagger}_{
(1/2)({\bf{k}}^{'}-{\bf{q}}/2),\downarrow }
(-{\bf{k}}^{'}+{\bf{q}}/2, \downarrow) A_{ {\bf{0}}, \uparrow }(
{\bf{0}}, \downarrow) A_{ (1/2)({\bf{k}}^{'}+{\bf{q}}/2+{\bf{q}}),
\downarrow } (-{\bf{k}}^{'}+{\bf{q}}/2, \uparrow)
\]
\[
+\lambda \sum_{ {\bf{q}} \neq 0 } \frac{ v_{ {\bf{q}} } }{V}
\sum_{ -{\bf{k}}^{'} \pm {\bf{q}}/2 \neq 0 } A^{\dagger}_{
(1/2)({\bf{k}}^{'}+{\bf{q}}/2 +{\bf{q}}), \downarrow }
(-{\bf{k}}^{'}+{\bf{q}}/2, \uparrow) A^{\dagger}_{
{\bf{0}},\uparrow }( {\bf{0}}, \downarrow) A_{
(1/2)({\bf{k}}^{'}-{\bf{q}}/2) \downarrow }
(-{\bf{k}}^{'}+{\bf{q}}/2, \downarrow)
\]
\[
- \lambda \sum_{ {\bf{q}} \neq 0 } \frac{ v_{ {\bf{q}} } }{V}
\sum_{ {\bf{k}} \pm {\bf{q}}/2 \neq 0 } \sum_{ {\bf{q}}_{1} \neq 0
} A^{\dagger}_{ (1/2)({\bf{k}}+{\bf{q}}/2) - {\bf{q}}_{1}/2,
\downarrow } ({\bf{k}}+{\bf{q}}/2 +{\bf{q}}_{1}, \uparrow) A_{
-{\bf{q}}_{1}/2, \downarrow}({\bf{q}}_{1}\downarrow) A_{
(1/2)({\bf{k}}-{\bf{q}}/2-{\bf{q}}),\downarrow}
({\bf{k}}+{\bf{q}}/2,\uparrow)
\]
\[
- \lambda \sum_{ {\bf{q}} \neq 0 } \frac{ v_{ {\bf{q}} } }{V}
\sum_{ {\bf{k}} \pm {\bf{q}}/2 \neq 0 } \sum_{ {\bf{q}}_{1} \neq 0
} A^{\dagger}_{ (1/2)({\bf{k}}-{\bf{q}}/2-{\bf{q}}), \downarrow }
({\bf{k}}+{\bf{q}}/2, \uparrow) A^{\dagger}_{ -{\bf{q}}_{1}/2,
\downarrow}({\bf{q}}_{1}\downarrow) A_{
(1/2)({\bf{k}}+{\bf{q}}/2)-{\bf{q}}_{1}/2,\downarrow}
({\bf{k}}+{\bf{q}}/2+{\bf{q}}_{1},\uparrow)
\]
\[
 + \lambda \sum_{ {\bf{q}} \neq 0 } \frac{ v_{ {\bf{q}} } }{V}
\sum_{ -{\bf{k}}^{'} \pm {\bf{q}}/2 \neq 0 } \sum_{ {\bf{q}}_{1}
\neq 0 } A^{\dagger}_{
(1/2)({\bf{k}}^{'}-{\bf{q}}/2)+{\bf{q}}_{1}/2, \downarrow }
(-{\bf{k}}^{'}+{\bf{q}}/2+{\bf{q}}_{1}, \uparrow) A_{
{\bf{q}}_{1}/2,\uparrow}({\bf{q}}_{1}\uparrow) A_{
(1/2)({\bf{k}}^{'}+{\bf{q}}/2+{\bf{q}}),\downarrow}
(-{\bf{k}}^{'}+{\bf{q}}/2,\uparrow)
\]
\begin{equation}
+ \lambda \sum_{ {\bf{q}} \neq 0 } \frac{ v_{ {\bf{q}} } }{V}
\sum_{ -{\bf{k}}^{'} \pm {\bf{q}}/2 \neq 0 } \sum_{ {\bf{q}}_{1}
\neq 0 } A^{\dagger}_{
(1/2)({\bf{k}}^{'}+{\bf{q}}/2+{\bf{q}}),\downarrow }
(-{\bf{k}}^{'}+{\bf{q}}/2, \uparrow) A^{\dagger}_{
{\bf{q}}_{1}/2,\uparrow}({\bf{q}}_{1}\uparrow) A_{
(1/2)({\bf{k}}^{'}-{\bf{q}}/2)+{\bf{q}}_{1}/2,\downarrow}
(-{\bf{k}}^{'}+{\bf{q}}/2+{\bf{q}}_{1},\uparrow)
\end{equation}
\[
H_{I} = \lambda^{2} \sum_{ {\bf{q}} \neq 0 } \frac{ v_{ {\bf{q}} }
}{2V} \sum_{ {\bf{k}} \neq {\bf{k}}^{'} } \sum_{ {\bf{k}} \pm
{\bf{q}}/2 \neq 0 } \sum_{ {\bf{k}}^{'} \pm {\bf{q}}/2 \neq 0 }
A^{\dagger}_{ (1/2)({\bf{k}}+{\bf{q}}/2) \uparrow }
({\bf{k}}+{\bf{q}}/2,\uparrow) A_{ (1/2)({\bf{k}}-{\bf{q}}/2)
\uparrow } ({\bf{k}}-{\bf{q}}/2,\uparrow) A^{\dagger}_{
(1/2)({\bf{k}}^{'}-{\bf{q}}/2) \uparrow }
({\bf{k}}^{'}-{\bf{q}}/2,\uparrow) A_{
(1/2)({\bf{k}}^{'}+{\bf{q}}/2) \uparrow }
({\bf{k}}^{'}+{\bf{q}}/2,\uparrow)
\]
\[
 +\lambda^{2} \sum_{ {\bf{q}} \neq 0 }
\frac{ v_{ {\bf{q}} } }{2V} \sum_{ {\bf{k}} \neq {\bf{k}}^{'} }
\sum_{ {\bf{k}} \pm {\bf{q}}/2 \neq 0 } \sum_{ {\bf{k}}^{'} \pm
{\bf{q}}/2 \neq 0 }
 \sum_{ {\bf{q}}_{1} \neq {\bf{k}}+{\bf{q}}/2 }
A^{\dagger}_{ {\bf{k}}+{\bf{q}}/2 - {\bf{q}}_{1}/2, \downarrow }
({\bf{q}}_{1}\uparrow) A_{ {\bf{k}}- {\bf{q}}_{1}/2, \downarrow }
(-{\bf{q}}+{\bf{q}}_{1}, \uparrow) A^{\dagger}_{
(1/2)({\bf{k}}^{'}-{\bf{q}}/2) \uparrow }
({\bf{k}}^{'}-{\bf{q}}/2,\uparrow) A_{
(1/2)({\bf{k}}^{'}+{\bf{q}}/2) \uparrow }
({\bf{k}}^{'}+{\bf{q}}/2,\uparrow)
\]
\[
+ \lambda^{2} \sum_{ {\bf{q}} \neq 0 } \frac{ v_{ {\bf{q}} } }{2V}
\sum_{ {\bf{k}} \neq {\bf{k}}^{'} } \sum_{ {\bf{k}} \pm {\bf{q}}/2
\neq 0 } \sum_{ {\bf{k}}^{'} \pm {\bf{q}}/2 \neq 0 } A^{\dagger}_{
(1/2)({\bf{k}}+{\bf{q}}/2) \uparrow }
({\bf{k}}+{\bf{q}}/2,\uparrow) A_{ (1/2)({\bf{k}}-{\bf{q}}/2)
\uparrow } ({\bf{k}}-{\bf{q}}/2,\uparrow) A^{\dagger}_{
{\bf{k}}^{'}-{\bf{q}}/2-{\bf{q}}_{2}/2, \downarrow }
({\bf{q}}_{2}\uparrow) A_{ {\bf{k}}^{'} - {\bf{q}}_{2}/2,
\downarrow } ({\bf{q}}+{\bf{q}}_{2}, \uparrow)
\]
\[
 +\lambda^{2} \sum_{ {\bf{q}} \neq 0 }
\frac{ v_{ {\bf{q}} } }{2V} \sum_{ {\bf{k}} \neq {\bf{k}}^{'} }
\sum_{ {\bf{k}} \pm {\bf{q}}/2 \neq 0 } \sum_{ {\bf{k}}^{'} \pm
{\bf{q}}/2 \neq 0 } \sum_{ {\bf{q}}_{1} \neq {\bf{k}}+{\bf{q}}/2 }
\sum_{ {\bf{q}}_{2} \neq {\bf{k}}^{'}-{\bf{q}}/2 } A^{\dagger}_{
{\bf{k}}+{\bf{q}}/2 - {\bf{q}}_{1}/2, \downarrow }
({\bf{q}}_{1}\uparrow) A_{ {\bf{k}}- {\bf{q}}_{1}/2, \downarrow }
(-{\bf{q}}+{\bf{q}}_{1}, \uparrow)
 A^{\dagger}_{ {\bf{k}}^{'}-{\bf{q}}/2-{\bf{q}}_{2}/2, \downarrow }
({\bf{q}}_{2}\uparrow) A_{ {\bf{k}}^{'} - {\bf{q}}_{2}/2,
\downarrow } ({\bf{q}}+{\bf{q}}_{2}, \uparrow)
\]
\[
 + \lambda^{2} \sum_{ {\bf{q}} \neq 0 }
\frac{ v_{ {\bf{q}} } }{2V} \sum_{ {\bf{k}} \neq {\bf{k}}^{'} }
\sum_{ {\bf{k}} \pm {\bf{q}}/2 \neq 0 } \sum_{ {\bf{k}}^{'} \pm
{\bf{q}}/2 \neq 0 } A^{\dagger}_{ (1/2)({\bf{k}}-{\bf{q}}/2)
\downarrow } (-{\bf{k}}+{\bf{q}}/2,\downarrow) A_{
(1/2)({\bf{k}}+{\bf{q}}/2) \downarrow }
(-{\bf{k}}-{\bf{q}}/2,\downarrow) A^{\dagger}_{
(1/2)({\bf{k}}^{'}+{\bf{q}}/2) \downarrow }
(-{\bf{k}}^{'}-{\bf{q}}/2,\downarrow) A_{
(1/2)({\bf{k}}^{'}-{\bf{q}}/2) \downarrow }
(-{\bf{k}}^{'}+{\bf{q}}/2,\downarrow)
\]
\[
 +\lambda^{2} \sum_{ {\bf{q}} \neq 0 }
\frac{ v_{ {\bf{q}} } }{2V} \sum_{ {\bf{k}} \neq {\bf{k}}^{'} }
\sum_{ {\bf{k}} \pm {\bf{q}}/2 \neq 0 } \sum_{ {\bf{k}}^{'} \pm
{\bf{q}}/2 \neq 0 } \sum_{ {\bf{q}}_{1} \neq -{\bf{k}}+{\bf{q}}/2
} A^{\dagger}_{ {\bf{k}}-{\bf{q}}/2 + {\bf{q}}_{1}/2, \downarrow }
({\bf{q}}_{1}\uparrow) A_{ {\bf{k}}+ {\bf{q}}_{1}/2, \downarrow }
(-{\bf{q}}+{\bf{q}}_{1}, \uparrow) A^{\dagger}_{
(1/2)({\bf{k}}^{'}+{\bf{q}}/2) \downarrow }
(-{\bf{k}}^{'}-{\bf{q}}/2,\downarrow) A_{
(1/2)({\bf{k}}^{'}-{\bf{q}}/2) \downarrow }
(-{\bf{k}}^{'}+{\bf{q}}/2,\downarrow)
\]
\[
+ \lambda^{2} \sum_{ {\bf{q}} \neq 0 } \frac{ v_{ {\bf{q}} } }{2V}
\sum_{ {\bf{k}} \neq {\bf{k}}^{'} } \sum_{ {\bf{k}} \pm {\bf{q}}/2
\neq 0 } \sum_{ {\bf{k}}^{'} \pm {\bf{q}}/2 \neq 0 } \sum_{
{\bf{q}}_{2} \neq -{\bf{k}}^{'}-{\bf{q}}/2 } A^{\dagger}_{
(1/2)({\bf{k}}-{\bf{q}}/2) \downarrow }
(-{\bf{k}}+{\bf{q}}/2,\downarrow) A_{ (1/2)({\bf{k}}+{\bf{q}}/2)
\downarrow } (-{\bf{k}}-{\bf{q}}/2,\downarrow) A^{\dagger}_{
{\bf{k}}^{'}+{\bf{q}}/2+{\bf{q}}_{2}/2, \downarrow }
({\bf{q}}_{2}\uparrow) A_{ {\bf{k}}^{'}+{\bf{q}}_{2}/2, \downarrow
} ({\bf{q}}+{\bf{q}}_{2}, \uparrow)
\]
\[
 +\lambda^{2} \sum_{ {\bf{q}} \neq 0 }
\frac{ v_{ {\bf{q}} } }{2V} \sum_{ {\bf{k}} \neq {\bf{k}}^{'} }
\sum_{ {\bf{k}} \pm {\bf{q}}/2 \neq 0 } \sum_{ {\bf{k}}^{'} \pm
{\bf{q}}/2 \neq 0 }
 \sum_{ {\bf{q}}_{1} \neq -{\bf{k}}+{\bf{q}}/2 }
\sum_{ {\bf{q}}_{2} \neq -{\bf{k}}^{'}-{\bf{q}}/2 } A^{\dagger}_{
{\bf{k}}-{\bf{q}}/2+{\bf{q}}_{1}/2, \downarrow }
({\bf{q}}_{1}\uparrow) A_{ {\bf{k}}+{\bf{q}}_{1}/2, \downarrow }
(-{\bf{q}}+{\bf{q}}_{1}, \uparrow)
 A^{\dagger}_{ {\bf{k}}^{'}+{\bf{q}}/2+{\bf{q}}_{2}/2, \downarrow }
({\bf{q}}_{2}\uparrow) A_{ {\bf{k}}^{'}+{\bf{q}}_{2}/2, \downarrow
} ({\bf{q}}+{\bf{q}}_{2}, \uparrow)
\]
\[
 - \lambda^{2} \sum_{ {\bf{q}} \neq 0 } \frac{ v_{ {\bf{q}} } }{2V}
\sum_{ {\bf{k}} \pm {\bf{q}}/2 \neq 0 }
A^{\dagger}_{(1/2)({\bf{k}}+{\bf{q}}/2)\uparrow }
({\bf{k}}+{\bf{q}}/2 \uparrow) A_{
(1/2)({\bf{k}}+{\bf{q}}/2)\uparrow } ({\bf{k}}+{\bf{q}}/2
\uparrow) A^{\dagger}_{(1/2)({\bf{k}}-{\bf{q}}/2)\uparrow }
({\bf{k}}-{\bf{q}}/2 \uparrow) A_{
(1/2)({\bf{k}}-{\bf{q}}/2)\uparrow } ({\bf{k}}-{\bf{q}}/2
\uparrow)
\]
\[
 - \lambda^{2} \sum_{ {\bf{q}} \neq 0 } \frac{ v_{ {\bf{q}} } }{2V}
\sum_{ {\bf{k}} \pm {\bf{q}}/2 \neq 0 } \sum_{ {\bf{q}}_{2} \neq
{\bf{k}} - {\bf{q}}/2 }
A^{\dagger}_{(1/2)({\bf{k}}+{\bf{q}}/2)\uparrow }
({\bf{k}}+{\bf{q}}/2 \uparrow) A_{
(1/2)({\bf{k}}+{\bf{q}}/2)\uparrow } ({\bf{k}}+{\bf{q}}/2
\uparrow) A^{\dagger}_{ {\bf{k}}-{\bf{q}}/2 - {\bf{q}}_{2}/2
\downarrow } ({\bf{q}}_{2} \uparrow) A_{ {\bf{k}}-{\bf{q}}/2 -
{\bf{q}}_{2}/2 \downarrow } ( {\bf{q}}_{2}\uparrow)
\]
\[
-\lambda^{2} \sum_{ {\bf{q}} \neq 0 } \frac{ v_{ {\bf{q}} } }{2V}
\sum_{ {\bf{k}} \pm {\bf{q}}/2 \neq 0 } \sum_{ {\bf{q}}_{1} \neq
{\bf{k}} + {\bf{q}}/2 } A^{\dagger}_{ {\bf{k}}+{\bf{q}}/2 -
{\bf{q}}_{1}/2 \downarrow } ({\bf{q}}_{1} \uparrow) A_{
{\bf{k}}+{\bf{q}}/2 - {\bf{q}}_{1}/2 \downarrow } ({\bf{q}}_{1}
\uparrow) A^{\dagger}_{(1/2)({\bf{k}}-{\bf{q}}/2)\uparrow }
({\bf{k}}-{\bf{q}}/2,\uparrow) A_{
(1/2)({\bf{k}}-{\bf{q}}/2),\uparrow } ({\bf{k}}-{\bf{q}}/2
\uparrow)
\]
\[
 - \lambda^{2} \sum_{ {\bf{q}} \neq 0 } \frac{ v_{ {\bf{q}} } }{2V}
\sum_{ {\bf{k}} \pm {\bf{q}}/2 \neq 0 } \sum_{ {\bf{q}}_{1} \neq
{\bf{k}} + {\bf{q}}/2 } \sum_{ {\bf{q}}_{2} \neq {\bf{k}} -
{\bf{q}}/2 } A^{\dagger}_{ {\bf{k}}+{\bf{q}}/2 - {\bf{q}}_{1}/2
\downarrow } ({\bf{q}}_{1} \uparrow) A_{ {\bf{k}}+{\bf{q}}/2 -
{\bf{q}}_{1}/2 \downarrow } ({\bf{q}}_{1} \uparrow) A^{\dagger}_{
{\bf{k}}-{\bf{q}}/2 - {\bf{q}}_{2}/2 \downarrow } ({\bf{q}}_{2}
\uparrow) A_{ {\bf{k}}-{\bf{q}}/2 - {\bf{q}}_{2}/2 \downarrow } (
{\bf{q}}_{2}\uparrow)
\]
\[
 - \lambda^{2} \sum_{ {\bf{q}} \neq 0 } \frac{ v_{ {\bf{q}} } }{2V}
\sum_{ -{\bf{k}} \pm {\bf{q}}/2 \neq 0 }
A^{\dagger}_{(1/2)({\bf{k}}+{\bf{q}}/2)\downarrow }
(-{\bf{k}}-{\bf{q}}/2 \downarrow) A_{
(1/2)({\bf{k}}+{\bf{q}}/2)\downarrow } (-{\bf{k}}-{\bf{q}}/2
\downarrow) A^{\dagger}_{(1/2)({\bf{k}}-{\bf{q}}/2)\downarrow }
(-{\bf{k}}+{\bf{q}}/2 \downarrow) A_{
(1/2)({\bf{k}}-{\bf{q}}/2)\downarrow } (-{\bf{k}}+{\bf{q}}/2
\downarrow)
\]
\[
 - \lambda^{2} \sum_{ {\bf{q}} \neq 0 } \frac{ v_{ {\bf{q}} } }{2V}
\sum_{ -{\bf{k}} \pm {\bf{q}}/2 \neq 0 } \sum_{ {\bf{q}}_{2} \neq
-{\bf{k}}+{\bf{q}}/2 }
A^{\dagger}_{(1/2)({\bf{k}}+{\bf{q}}/2)\downarrow }
(-{\bf{k}}-{\bf{q}}/2 \downarrow) A_{
(1/2)({\bf{k}}+{\bf{q}}/2)\downarrow } (-{\bf{k}}-{\bf{q}}/2
\downarrow) A^{\dagger}_{ {\bf{k}}-{\bf{q}}/2 + {\bf{q}}_{2}/2
\downarrow } ({\bf{q}}_{2} \uparrow) A_{ {\bf{k}}-{\bf{q}}/2 +
{\bf{q}}_{2}/2 \downarrow } ( {\bf{q}}_{2}\uparrow)
\]
\[
-\lambda^{2} \sum_{ {\bf{q}} \neq 0 } \frac{ v_{ {\bf{q}} } }{2V}
\sum_{ -{\bf{k}} \pm {\bf{q}}/2 \neq 0 } \sum_{ {\bf{q}}_{1} \neq
-{\bf{k}}-{\bf{q}}/2 } A^{\dagger}_{ {\bf{k}}+{\bf{q}}/2 +
{\bf{q}}_{1}/2 \downarrow } ({\bf{q}}_{1} \uparrow) A_{
{\bf{k}}+{\bf{q}}/2 + {\bf{q}}_{1}/2 \downarrow } ({\bf{q}}_{1}
\uparrow) A^{\dagger}_{(1/2)({\bf{k}}-{\bf{q}}/2)\downarrow }
(-{\bf{k}}+{\bf{q}}/2,\downarrow)
A_{(1/2)({\bf{k}}-{\bf{q}}/2)\downarrow }
(-{\bf{k}}+{\bf{q}}/2,\downarrow)
\]
\begin{equation}
- \lambda^{2} \sum_{ {\bf{q}} \neq 0 } \frac{ v_{ {\bf{q}} } }{2V}
\sum_{ {\bf{k}} \pm {\bf{q}}/2 \neq 0 } \sum_{ {\bf{q}}_{1} \neq
-{\bf{k}}-{\bf{q}}/2 } \sum_{ {\bf{q}}_{2} \neq
-{\bf{k}}+{\bf{q}}/2 } A^{\dagger}_{
{\bf{k}}+{\bf{q}}/2+{\bf{q}}_{1}/2 \downarrow } ({\bf{q}}_{1}
\uparrow) A_{ {\bf{k}}+{\bf{q}}/2+{\bf{q}}_{1}/2 \downarrow }
({\bf{q}}_{1} \uparrow) A^{\dagger}_{
{\bf{k}}-{\bf{q}}/2+{\bf{q}}_{2}/2 \downarrow } ({\bf{q}}_{2}
\uparrow) A_{ {\bf{k}}-{\bf{q}}/2+{\bf{q}}_{2}/2 \downarrow } (
{\bf{q}}_{2}\uparrow)
\end{equation}

\small It is clear that this hamiltonian is rather more complex
than the simple two-body variety. In particular, the leading
contribution which is first order in $ \lambda $ has three
operators signifying a non-conservation of exciton number. In one
such process, an exciton recombines with a solitron to produce a
conductron. Only excitons couple with external fields and we are
able to infer the existence of the other excitations is inferred
indirectly by studying the binding energy of biexcitons that are
due to these exotic many-body processes. From the above form of
the interaction terms we see that the ground state of $ H_{0} $
does not evolve with time under the action of the perturbation.
However, if we first create an exciton (or two excitons) and then
evolve the state under the action of the full hamiltonian, then
the state does indeed evolve and produces all the effects that we
expect, such as a possible exciton-line width, biexciton (bound
state of two-excitons) and so on. Let us now try to verify these
expectations rigorously.

\subsection{ The Exciton Green Function }

In this section, we compute the exciton Green function as
resulting from exciton-exciton interactions and compare the
lifetime of the exciton arising from exciton-exciton interactions
with the lifetime from usual radiative recombination processes.
Consider the initial state which corresponds to an exciton in
internal state $ I  $ and whose center of mass is moving with
momentum
 $ {\bf{Q}} $. Then we may write,
\begin{equation}
| I, {\bf{Q}} \rangle =  b^{\dagger}_{I}({\bf{Q}}) | G \rangle
\end{equation}
Let us now examine how this state evolves with time.

\tiny
\begin{equation}
e^{-i\mbox{  }t\mbox{   }(H + \lambda H_{I,0} + \lambda^{2} H_{I}
)} |I,{\bf{Q}} \rangle \approx e^{-i\mbox{  }t\mbox{   }H_{0}}
\left( {\hat{ {\bf{1}} }} - i\mbox{   } \lambda \mbox{
}\int^{t}_{0} dt_{1} {\hat{H}}_{I,0}(t_{1})
 + (-i)^{2}\lambda^{2}
\mbox{ }\int^{t}_{0} dt_{1} {\hat{H}}_{I,0}(t_{1}) \mbox{
}\int^{t_{1}}_{0} dt_{2} {\hat{H}}_{I,0}(t_{2})
 -i \mbox{  }
\lambda^{2} \mbox{ }\int^{t}_{0}dt_{1} {\hat{H}}_{I}(t_{1})
\right)
 b^{\dagger}_{I}({\bf{Q}}) | G \rangle
\end{equation}
\small There are many final states possible. We would like to
compute the Green function,
\begin{equation}
G_{I}({\bf{Q}};\omega) = i\mbox{ }\int_{0}^{\infty} dt\mbox{ }e^{i
\mbox{ }\omega \mbox{    }t} \langle  I , {\bf{Q}} | e^{-i\mbox{
}t\mbox{ }H } | I , {\bf{Q}}\rangle
\end{equation}
The exciton lineshape is given by plotting,
\begin{equation}
L( \omega; {\bf{{Q}}} ) = - \sum_{I}
 Im( G_{I}({\bf{Q}};\omega - i\mbox{   }0^{+}))
\end{equation}
We may evaluate the matrix element as,
 \tiny
\[
\langle I, {\bf{Q}} | e^{-i\mbox{  }t\mbox{  }(H_{0} + \lambda
H_{I,0} + \lambda^{2} H_{I} ) } | I, {\bf{Q}} \rangle
 = e^{-i \epsilon_{I}({\bf{Q}}) t }
 + (-i)^{2} \lambda^{2}
e^{-i \epsilon_{I}({\bf{Q}}) t}
 \sum
 ( \frac{ v_{ {\bf{q}}_{1} } }{V} )^{2}
\]
\[
\times
 \varphi_{I}^{*}({\bf{k}}_{1} + {\bf{q}}_{1}/2 - {\bf{Q}}/2
; {\bf{Q}}) \varphi_{I_{2}}({\bf{k}}_{1} - {\bf{Q}}/2;
-{\bf{q}}_{1} + {\bf{Q}} ) \varphi^{*}_{I_{2}}({\bf{k}}_{2}
 - {\bf{Q}}/2; -{\bf{q}}_{1} + {\bf{Q}} )
 \varphi_{I}({\bf{k}}_{2} + {\bf{q}}_{1}/2 - {\bf{Q}}/2; {\bf{Q}})
\]
\[
\times
 \int^{t}_{0}dt_{1} \mbox{  }
 e^{-i \mbox{  }\epsilon_{e}({\bf{q}}_{1}) t_{1} }
e^{i \epsilon_{I}({\bf{Q}})t_{1}}
e^{-i\epsilon_{I_{2}}(-{\bf{q}}_{1}+{\bf{Q}}) t_{1}}
\int^{t_{1}}_{0}dt_{2} \mbox{  }
 e^{i \mbox{  }\epsilon_{e}({\bf{q}}_{1}) t_{2} }
e^{i \epsilon_{I_{2}}({\bf{Q}}-{\bf{q}}_{1})t_{2}}
e^{-i\epsilon_{I}({\bf{Q}}) t_{2}}
\]
\[
 + (-i)^{2} \lambda^{2}
e^{-i \epsilon_{I}({\bf{Q}}) t}
 \sum
 ( \frac{ v_{ {\bf{q}}_{1} } }{V} )^{2}
\varphi_{I}^{*}({\bf{k}}_{1} - {\bf{q}}_{1}/2 + {\bf{Q}}/2 ;
{\bf{Q}}) \varphi_{I_{2}}({\bf{k}}_{1} + {\bf{Q}}/2; -{\bf{q}}_{1}
+ {\bf{Q}} ) \varphi^{*}_{I_{2}}({\bf{k}}_{2}
 + {\bf{Q}}/2; -{\bf{q}}_{1} + {\bf{Q}} )
 \varphi_{I}({\bf{k}}_{2} - {\bf{q}}_{1}/2 + {\bf{Q}}/2; {\bf{Q}})
\]

\[
\times
 \int^{t}_{0}dt_{1} \mbox{  }
 e^{-i \mbox{  }\epsilon_{h}({\bf{q}}_{1}) t_{1} }
e^{i \epsilon_{I}({\bf{Q}})t_{1}}
e^{-i\epsilon_{I_{2}}(-{\bf{q}}_{1}+{\bf{Q}}) t_{1}}
\int^{t_{1}}_{0}dt_{2} \mbox{  }
 e^{i \mbox{  }\epsilon_{h}({\bf{q}}_{1}) t_{2} }
e^{i \epsilon_{I_{2}}({\bf{Q}}-{\bf{q}}_{1})t_{2}}
e^{-i\epsilon_{I}({\bf{Q}}) t_{2}}
\]

\[
 + (-i)^{2} \lambda^{2}
e^{-i \epsilon_{I}({\bf{Q}}) t}
 \sum
 ( \frac{ v_{ {\bf{q}}_{1} } }{V} )
( \frac{ v_{ {\bf{q}}_{2} } }{V} )
 \varphi_{I}^{*}((1/2){\bf{Q}}
 -{\bf{Q}}_{1} ; {\bf{Q}} )
  \varphi_{I_{2}}( (1/2)({\bf{Q}}-{\bf{Q}}_{1})-{\bf{q}}_{1};
{\bf{Q}}-{\bf{Q}}_{1})
\varphi^{*}_{I_{2}}((1/2)({\bf{Q}}-{\bf{Q}}_{1})-{\bf{q}}_{2} ;
{\bf{Q}}-{\bf{Q}}_{1} ) \varphi_{I}((1/2){\bf{Q}}-{\bf{Q}}_{1};
{\bf{Q}})
\]

\[
\times
 \int^{t}_{0}dt_{1} \mbox{  }
 e^{-i \mbox{  }\epsilon_{h}({\bf{Q}}_{1}) t_{1} }
e^{i \epsilon_{I}({\bf{Q}})t_{1}}
e^{-i\epsilon_{I_{2}}(-{\bf{Q}}_{1}+{\bf{Q}}) t_{1}}
\int^{t_{1}}_{0}dt_{2} \mbox{  }
 e^{i \mbox{  }\epsilon_{h}({\bf{Q}}_{1}) t_{2} }
e^{i \epsilon_{I_{2}}({\bf{Q}}-{\bf{Q}}_{1})t_{2}}
e^{-i\epsilon_{I}({\bf{Q}}) t_{2}}
\]

\[
 + (-i)^{2} \lambda^{2}
e^{-i \epsilon_{I}({\bf{Q}}) t}
 \sum
 ( \frac{ v_{ {\bf{q}}_{1} } }{V} )
( \frac{ v_{ {\bf{q}}_{2} } }{V} )
 \varphi_{I}^{*}({\bf{Q}}_{1} -{\bf{Q}}/2 ; {\bf{Q}} )
  \varphi_{I_{2}}( (1/2)({\bf{Q}}_{1}-{\bf{Q}}) + {\bf{q}}_{1};
{\bf{Q}}-{\bf{Q}}_{1})
\varphi^{*}_{I_{2}}((1/2)({\bf{Q}}_{1}-{\bf{Q}})+{\bf{q}}_{2} ;
{\bf{Q}}-{\bf{Q}}_{1})
 \varphi_{I}({\bf{Q}}_{1}-{\bf{Q}}/2; {\bf{Q}})
\]

\begin{equation}
\times
 \int^{t}_{0}dt_{1} \mbox{   }
 e^{-i\mbox{   }\epsilon_{e}({\bf{Q}}_{1}) t_{1} }
e^{i \epsilon_{I}({\bf{Q}})t_{1}}
e^{-i\epsilon_{I_{2}}(-{\bf{Q}}_{1}+{\bf{Q}}) t_{1}}
\int^{t_{1}}_{0}dt_{2} \mbox{  }
 e^{i \mbox{  }\epsilon_{e}({\bf{Q}}_{1}) t_{2} } e^{i
\epsilon_{I_{2}}({\bf{Q}}-{\bf{Q}}_{1})t_{2}}
e^{-i\epsilon_{I}({\bf{Q}}) t_{2}}
\end{equation}
\small We may rewrite the above equation retaining only the most
singular parts as,
\begin{equation}
\langle I {\vec{Q}}| e^{-it(H_{0} + \lambda H_{I0} + \lambda^{2}
H_{I})} | I {\vec{Q}} \rangle
 = e^{-i\mbox{     }t \mbox{     }\epsilon_{I}({\bf{Q}})}
-i\mbox{       } F_{I}({\bf{Q}}) \mbox{    } t \mbox{ }e^{-i\mbox{
} t \mbox{         } \epsilon_{I}({\bf{Q}})} \approx
 e^{ -i\mbox{    }t \mbox{     }
 (\epsilon_{I}({\bf{Q}}) + F_{I}({\bf{Q}})) }
\end{equation}
where, \tiny
\[
F_{I}({\bf{Q}}) = -\lambda^{2} \sum \left( \frac{ v_{
{\bf{q}}_{1}} }{V} \right)^{2} \frac{
\varphi^{*}_{I}({\bf{k}}_{1}+{\bf{q}}_{1}/2-{\bf{Q}}/2;{\bf{Q}})
 \varphi_{I_{2}}({\bf{k}}_{1}-{\bf{Q}}/2;-{\bf{q}}_{1}+{\bf{Q}})
 \varphi^{*}_{I_{2}}({\bf{k}}_{2}-{\bf{Q}}/2;-{\bf{q}}_{1}+{\bf{Q}})
 \varphi_{I}({\bf{k}}_{2}+{\bf{q}}_{1}/2-{\bf{Q}}/2;{\bf{Q}}) }
 { \epsilon_{e}({\bf{q}}_{1}) +
\epsilon_{I_{2}}({\bf{Q}}-{\bf{q}}_{1}) - \epsilon_{I}({\bf{Q}}) }
\]
\[
-\lambda^{2} \sum \left( \frac{ v_{ {\bf{q}}_{1}} }{V} \right)^{2}
\frac{
\varphi^{*}_{I}({\bf{k}}_{1}-{\bf{q}}_{1}/2+{\bf{Q}}/2;{\bf{Q}})
 \varphi_{I_{2}}({\bf{k}}_{1}+{\bf{Q}}/2;-{\bf{q}}_{1}+{\bf{Q}})
 \varphi^{*}_{I_{2}}({\bf{k}}_{2}+{\bf{Q}}/2;-{\bf{q}}_{1}+{\bf{Q}})
 \varphi_{I}({\bf{k}}_{2}-{\bf{q}}_{1}/2+{\bf{Q}}/2;{\bf{Q}}) }
 { \epsilon_{h}({\bf{q}}_{1}) +
\epsilon_{I_{2}}({\bf{Q}}-{\bf{q}}_{1}) - \epsilon_{I}({\bf{Q}}) }
\]
\[
 -\lambda^{2} \sum \left( \frac{ v_{
{\bf{q}}_{1}} }{V} \right) \left( \frac{ v_{ {\bf{q}}_{2}} }{V}
\right)
 \frac{ \varphi^{*}_{I}({\bf{Q}}_{1}-{\bf{Q}}/2;{\bf{Q}})
 \varphi_{I_{2}}((\frac{1}{2})
 ({\bf{Q}}_{1}-{\bf{Q}})+{\bf{q}}_{1};{\bf{Q}}-{\bf{Q}}_{1})
 \varphi^{*}_{I_{2}}((\frac{1}{2})
 ({\bf{Q}}_{1}-{\bf{Q}})+{\bf{q}}_{2};{\bf{Q}}-{\bf{Q}}_{1})
 \varphi_{I}({\bf{Q}}_{1}-{\bf{Q}}/2;{\bf{Q}}) }
 { \epsilon_{e}({\bf{Q}}_{1}) +
\epsilon_{I_{2}}({\bf{Q}}-{\bf{Q}}_{1}) - \epsilon_{I}({\bf{Q}}) }
\]
\begin{equation}
 -\lambda^{2} \sum \left( \frac{ v_{
{\bf{q}}_{1}} }{V} \right) \left( \frac{ v_{ {\bf{q}}_{2}} }{V}
\right)
 \frac{ \varphi^{*}_{I}(-{\bf{Q}}_{1}+{\bf{Q}}/2;{\bf{Q}})
 \varphi_{I_{2}}((\frac{1}{2})
 ({\bf{Q}}-{\bf{Q}}_{1})-{\bf{q}}_{1};{\bf{Q}}-{\bf{Q}}_{1})
 \varphi^{*}_{I_{2}}((\frac{1}{2})
 ({\bf{Q}}-{\bf{Q}}_{1})-{\bf{q}}_{2};{\bf{Q}}-{\bf{Q}}_{1})
 \varphi_{I}(-{\bf{Q}}_{1}+{\bf{Q}}/2;{\bf{Q}}) }
 {\epsilon_{h}({\bf{Q}}_{1}) +
\epsilon_{I_{2}}({\bf{Q}}-{\bf{Q}}_{1}) - \epsilon_{I}({\bf{Q}}) }
\end{equation}
\small Here $  \epsilon_{e}({\bf{k}}) = k^{2}/2m_{e} $ and
 $ \epsilon_{h}({\bf{k}}) = k^{2}/2m_{h} $.
 The lineshape of the exciton may be written as,
\begin{equation}
L({\bf{Q}},\omega) = \sum_{I} \frac{ Im( F_{I}({\bf{Q}}) ) }
 { (\omega - \epsilon_{I}({\bf{Q}}) - Re( F_{I}({\bf{Q}}) ))^{2}
  +  (Im( F_{I}({\bf{Q}}) ))^{2} }
\end{equation}
In this article, we merely point out the feasibility of these
computations. In our next article we intend to explore the
practical consequences more thoroughly, specifically, the
biexciton Green functions and nonlinear optical susceptibilities.

\section{ Conclusions }

In this article, we have laid down the ground work for a very
promising new approach for understanding excitons in
semiconductors and charge-conserving systems such as electrons and
positrons. We have shown how to compute the exciton-lineshape that
(if broadened) is due to also to nonradiative many-body processes.
We have identified the elementary excitations in the system and
precisely pointed out the role each of them play in the theory.
The elementary entities in the two-component Fermi system may be
thought of in two equivalent ways. One may consider them to be
electrons and holes interacting via two-body attractive and
repulsive interactions or we may consider the system to be made of
excitons, solitrons, valerons and conductrons all interacting with
each other by somewhat complicated, but purely local hamiltonians.
The formidable technical challenges have been overcome and now a
clear path has been mapped out for a systematic exploration of
nonperturbative phenomena in many-body physics.

\section{Appendix A}

In this section, we prove the claims made in the first section
namely that $ X_{0} $ is canonically conjugate to $ {\hat{N}}_{0}
$. Let $ {\hat{O}} = (b_{0} - \sqrt{ {\hat{N}}_{0} })(1/\sqrt{
{\hat{N}}_{0} }) $.
\begin{equation}
X_{0} = \frac{i}{2}\sum^{\infty}_{ n = 1 }\frac{ (-1)^{n+1} }{n}
{\hat{O}}^{n} - \frac{i}{2}\sum^{\infty}_{ n = 1 }\frac{
(-1)^{n+1} }{n} {\hat{O}}^{\dagger n}
\end{equation}
\begin{equation}
[X_{0},{\hat{N}}_{0}]
 = \frac{i}{2}\sum^{\infty}_{ n = 1 }\frac{ (-1)^{n+1} }{n}
[{\hat{O}}^{n}, {\hat{N}}_{0}] - \frac{i}{2}\sum^{\infty}_{ n = 1
}\frac{ (-1)^{n+1} }{n} [{\hat{O}}^{\dagger n}, {\hat{N}}_{0}]
\end{equation}
\begin{equation}
[{\hat{O}}^{n}, {\hat{N}}_{0}] = \sum_{m=1}^{n} {\hat{O}}^{n-m}
[{\hat{O}},{\hat{N}}_{0}] {\hat{O}}^{m-1}
\end{equation}
But,
\begin{equation}
[{\hat{O}},{\hat{N}}_{0}] =  {\hat{ {\bf{1}} }} + {\hat{O}}
\end{equation}
Therefore,
\begin{equation}
[{\hat{O}}^{n}, {\hat{N}}_{0}] = n \mbox{   }{\hat{O}}^{n-1}(
{\hat{ {\bf{1}} }} + {\hat{O}} )
\end{equation}
It is a relatively simple matter to sum this series and we may
write,
\[
[X_{0}, {\hat{N}}_{0}] =
 \frac{i}{2}\sum^{\infty}_{ n = 1 }\frac{ (-1)^{n+1} }{n}
n \mbox{   }{\hat{O}}^{n-1}( {\hat{ {\bf{1}} }} + {\hat{O}} ) +
\frac{i}{2}\sum^{\infty}_{ n = 1 }\frac{ (-1)^{n+1} }{n} n \mbox{
}{\hat{O}}^{\dagger n-1} ( {\hat{ {\bf{1}} }} +
{\hat{O}}^{\dagger} )
\]
\begin{equation}
 = \frac{i}{2}( {\hat{ {\bf{1}} }} + {\hat{O}} )
\frac{ {\hat{ {\bf{1}} }}  }{ {\hat{ {\bf{1}} }} + {\hat{O}}  }
 + \frac{i}{2}( {\hat{ {\bf{1}} }} + {\hat{O}}^{\dagger} )
\frac{ {\hat{ {\bf{1}} }}  }{ {\hat{ {\bf{1}} }} +
{\hat{O}}^{\dagger}  }
 = i\mbox{    } {\hat{ {\bf{1}} }}
\end{equation}
Therefore we find much to our relief,
\begin{equation}
[X_{0}, {\hat{N}}_{0}] = i \mbox{    }  {\hat{ {\bf{1}} }}
\end{equation}
Now we move on to fermions.

\section{ Appendix B }

Here we would like to prove some facts about fermions that were
claimed in the main text. Writing down the commutation rules
obeyed by $ A_{ {\bf{k}} }({\bf{q}}) $ involves knowing the
precise meaning of the square root in the denominator. Since we
have not been successful in pinning down its true meaning, we
shall have to take the point of view that the meaning is uniquely
fixed by {\it{demanding}} that the following commutation rules be
obeyed. This is not very satisfying but the authors have exhausted
their meagre capabilities. Let us consider the various
possibilities.

(a) $ n_{F}({\bf{k}}+{\bf{q}}/2) = 0 $,
 $ n_{F}({\bf{k}}-{\bf{q}}/2) = 1 $.

\begin{equation}
\sqrt{ n_{ {\bf{k}}-{\bf{q}}/2 } } A_{ {\bf{k}} }({\bf{q}})
 =
 \left( {\bf{1}} - \sum_{ {\bf{q}}_{1} \neq 0 }
 A^{\dagger}_{ {\bf{k}}-{\bf{q}}/2 + {\bf{q}}_{1}/2 }
 ({\bf{q}}_{1})
 A_{ {\bf{k}}-{\bf{q}}/2 + {\bf{q}}_{1}/2 }
 ({\bf{q}}_{1}) \right)^{\frac{1}{2}}
A_{ {\bf{k}} }({\bf{q}}) =
 c^{\dagger}_{ {\bf{k}}-{\bf{q}}/2 }c_{ {\bf{k}}+{\bf{q}}/2 }
\end{equation}
We have to convince ourselves that the left hand side obeys the
same commutation rules as the right hand side.

(b) $ n_{F}({\bf{k}}+{\bf{q}}/2) = 1 $,
 $ n_{F}({\bf{k}}-{\bf{q}}/2) = 0 $.

\begin{equation}
A^{\dagger}_{ {\bf{k}} }(-{\bf{q}})
 \sqrt{ n_{ {\bf{k}}+{\bf{q}}/2 } }
  = A^{\dagger}_{ {\bf{k}} }(-{\bf{q}})
   \left( {\bf{1}} - \sum_{ {\bf{q}}_{1} \neq 0 }
    A^{\dagger}_{ {\bf{k}}+{\bf{q}}/2 +
{\bf{q}}_{1}/2 }({\bf{q}}_{1}) A_{ {\bf{k}}+{\bf{q}}/2 +
{\bf{q}}_{1}/2 }({\bf{q}}_{1}) \right)^{\frac{1}{2}}
 =
c^{\dagger}_{ {\bf{k}}-{\bf{q}}/2 }c_{ {\bf{k}}+{\bf{q}}/2 }
\end{equation}

(c) $ n_{F}({\bf{k}}+{\bf{q}}/2) = 0 $, $
n_{F}({\bf{k}}-{\bf{q}}/2) = 0 $.

\begin{equation}
c^{\dagger}_{ {\bf{k}} + {\bf{q}}/2 }c_{ {\bf{k}} - {\bf{q}}/2 }
 = \sum_{ {\bf{q}}_{1} \neq {\bf{q}}, 0 }
A^{\dagger}_{ {\bf{k}}+{\bf{q}}/2-{\bf{q}}_{1}/2 } ({\bf{q}}_{1})
A_{ {\bf{k}}-{\bf{q}}_{1}/2 }(-{\bf{q}}+ {\bf{q}}_{1})
\label{EQNC}
\end{equation}

(d) $ n_{F}({\bf{k}}+{\bf{q}}/2) = 1 $, $
n_{F}({\bf{k}}-{\bf{q}}/2) = 1 $.

\begin{equation}
c^{\dagger}_{ {\bf{k}} + {\bf{q}}/2 }c_{ {\bf{k}} - {\bf{q}}/2 }
 = -\sum_{ {\bf{q}}_{1} \neq {\bf{q}}, 0 }
A^{\dagger}_{ {\bf{k}}-{\bf{q}}/2+{\bf{q}}_{1}/2 } ({\bf{q}}_{1})
A_{ {\bf{k}}+{\bf{q}}_{1}/2 }(-{\bf{q}}+ {\bf{q}}_{1})
\label{EQND}
\end{equation}
Similarly, we have

$ (a^{'}) $ $ n_{F}({\bf{k}}^{'}+{\bf{q}}^{'}/2) = 0 $,
 $ n_{F}({\bf{k}}^{'}-{\bf{q}}^{'}/2) = 1 $.
\begin{equation}
\sqrt{ n_{ {\bf{k}}^{'}-{\bf{q}}^{'}/2 } } A_{ {\bf{k}}^{'}
}({\bf{q}}^{'})
 = c^{\dagger}_{ {\bf{k}}^{'}-{\bf{q}}^{'}/2 }
c_{ {\bf{k}}^{'}+{\bf{q}}^{'}/2 }
\end{equation}
and so on for the other cases. We would now like to write down
some statements that would be analogous to commutation rules. Let
us first define the following statements( it goes without saying
that $ {\bf{q}} \neq 0 $ and  $ {\bf{q}}^{'} \neq 0 $ ).

S1 : $ {\bf{k}} + {\bf{q}}/2 = {\bf{k}}^{'} + {\bf{q}}^{'}/2 $

S2 : $ {\bf{k}} + {\bf{q}}/2 = {\bf{k}}^{'} - {\bf{q}}^{'}/2 $

S3 : $ {\bf{k}} - {\bf{q}}/2 = {\bf{k}}^{'} + {\bf{q}}^{'}/2 $

S4 : $ {\bf{k}} - {\bf{q}}/2 = {\bf{k}}^{'} - {\bf{q}}^{'}/2 $

SS1 : $ {\bf{k}} = {\bf{k}}^{'} \mbox{ } and \mbox{ } {\bf{q}} =
{\bf{q}}^{'} $

SS2 : $ {\bf{k}} = {\bf{k}}^{'} \mbox{ } and \mbox{ } {\bf{q}} =
-{\bf{q}}^{'} $

\noindent
 Consider the object :
\begin{equation}
 \sqrt{n_{ {\bf{k}}-{\bf{q}}/2 } }
 A_{ {\bf{k}} }({\bf{q}})
 \sqrt{n_{ {\bf{k}}^{'}-{\bf{q}}^{'}/2 } }
 A_{ {\bf{k}}^{'} }({\bf{q}}^{'})
\end{equation}
We would now like to ascertain the meaning of this object when
say, S1 is true but S2, S3 and S4 are false. This state of affairs
in symbolic logic is written as
 $ S1 \wedge \neg S2 \wedge \neg S3\wedge \neg S4 $. Let us define,
\[
AS1 = S1 \wedge \neg S2 \wedge \neg S3 \wedge \neg S4
 \hspace{0.3in}
AS2 = S2 \wedge \neg S1 \wedge \neg S3 \wedge \neg S4
\]

\[
AS3 = S3 \wedge \neg S1 \wedge \neg S2 \wedge \neg S4
 \hspace{0.3in}
AS4 = S4 \wedge \neg S1 \wedge \neg S2 \wedge \neg S3
\]

\[
AA0 = \neg S1 \wedge \neg S2 \wedge \neg S3 \wedge \neg S4
\]

(1) If AS1 is true then we have,

\begin{equation}
 \sqrt{n_{ {\bf{k}}-{\bf{q}}/2 } }
 A_{ {\bf{k}} }({\bf{q}})
 \sqrt{n_{ {\bf{k}}^{'}-{\bf{q}}^{'}/2 } }
 A_{ {\bf{k}}^{'} }({\bf{q}}^{'})
  =
 \sqrt{n_{ {\bf{k}}^{'}-{\bf{q}}^{'}/2 } }
 A_{ {\bf{k}}^{'} }({\bf{q}}^{'})
\sqrt{n_{ {\bf{k}}-{\bf{q}}/2 } }
 A_{ {\bf{k}} }({\bf{q}})
 = 0
\end{equation}

\vspace{0.3in}

\begin{equation}
 \sqrt{n_{ {\bf{k}}-{\bf{q}}/2 } }
 A_{ {\bf{k}} }({\bf{q}})
 A^{\dagger}_{ {\bf{k}}^{'} }(-{\bf{q}}^{'})
 \sqrt{ n_{ {\bf{k}}^{'}+{\bf{q}}^{'}/2 } } =
A^{\dagger}_{ {\bf{k}}^{'} }(-{\bf{q}}^{'})
 \sqrt{ n_{ {\bf{k}}^{'}+{\bf{q}}^{'}/2 } }
 \sqrt{n_{ {\bf{k}}-{\bf{q}}/2 } }
 A_{ {\bf{k}} }({\bf{q}})
 = 0
\end{equation}

\vspace{0.3in}

\[
 \sqrt{ n_{ {\bf{k}}-{\bf{q}}/2 } }
 A_{ {\bf{k}} }({\bf{q}})
\left(  \sum_{ {\bf{q}}_{1} \neq {\bf{q}}^{'},  0 }
 A^{\dagger}_{ {\bf{k}}^{'} + {\bf{q}}^{'}/2 - {\bf{q}}_{1}/2 }
 ({\bf{q}}_{1})
 A_{ {\bf{k}}^{'} - {\bf{q}}_{1}/2 }(-{\bf{q}}^{'} + {\bf{q}}_{1})
 \right)
\]
\begin{equation}
= c^{\dagger}_{ {\bf{k}}-{\bf{q}}/2 }c_{ {\bf{k}}+{\bf{q}}/2 }
c^{\dagger}_{ {\bf{k}}+{\bf{q}}/2 }
 c_{ {\bf{k}}^{'} - {\bf{q}}^{'}/2 }
  = \sqrt{ n_{ {\bf{k}} - {\bf{q}}/2 } }
  A_{ {\bf{k}} - {\bf{q}}^{'}/2 }({\bf{q}}-{\bf{q}}^{'})
   \left( {\hat{ {\bf{1}} }}
    - \sum_{ {\bf{q}}_{1} \neq 0 }
    A^{\dagger}_{ {\bf{k}}+{\bf{q}}/2 - {\bf{q}}_{1}/2 }
    ({\bf{q}}_{1})A_{ {\bf{k}}+{\bf{q}}/2 - {\bf{q}}_{1}/2 }
    ({\bf{q}}_{1}) \right)
\end{equation}

\vspace{0.3in}

\[
 \left( \sum_{ {\bf{q}}_{1} \neq {\bf{q}}^{'}, 0 }
 A^{\dagger}_{ {\bf{k}}^{'} + {\bf{q}}^{'}/2 - {\bf{q}}_{1}/2 }
 ({\bf{q}}_{1})
 A_{ {\bf{k}}^{'} - {\bf{q}}_{1}/2 }(-{\bf{q}}^{'} + {\bf{q}}_{1})
 \right)
\sqrt{ n_{ {\bf{k}}-{\bf{q}}/2 } }
 A_{ {\bf{k}} }({\bf{q}})
\]
\begin{equation}
= c^{\dagger}_{ {\bf{k}}+{\bf{q}}/2 }
 c_{ {\bf{k}}^{'} - {\bf{q}}^{'}/2 }
c^{\dagger}_{ {\bf{k}}-{\bf{q}}/2 }c_{ {\bf{k}}+{\bf{q}}/2 } =
-\sqrt{ n_{ {\bf{k}} - {\bf{q}}/2 } }
  A_{ {\bf{k}} - {\bf{q}}^{'}/2 }({\bf{q}}-{\bf{q}}^{'})
   \left( \sum_{ {\bf{q}}_{1} \neq 0 }
    A^{\dagger}_{ {\bf{k}}+{\bf{q}}/2 - {\bf{q}}_{1}/2 }
    ({\bf{q}}_{1})A_{ {\bf{k}}+{\bf{q}}/2 - {\bf{q}}_{1}/2 }
    ({\bf{q}}_{1}) \right)
\end{equation}

\vspace{0.3in}

\begin{equation}
-\sqrt{ n_{ {\bf{k}}-{\bf{q}}/2 } }
 A_{ {\bf{k}} }({\bf{q}})
\left(  \sum_{ {\bf{q}}_{1} \neq {\bf{q}}^{'},  0 }
 A^{\dagger}_{ {\bf{k}}^{'} - {\bf{q}}^{'}/2 + {\bf{q}}_{1}/2 }
 ({\bf{q}}_{1})
 A_{ {\bf{k}}^{'} + {\bf{q}}_{1}/2 }(-{\bf{q}}^{'} + {\bf{q}}_{1})
 \right) = 0
\end{equation}

\begin{equation}
-\left( \sum_{ {\bf{q}}_{1} \neq {\bf{q}}^{'}, 0 }
 A^{\dagger}_{ {\bf{k}}^{'} - {\bf{q}}^{'}/2 + {\bf{q}}_{1}/2 }
 ({\bf{q}}_{1})
 A_{ {\bf{k}}^{'} + {\bf{q}}_{1}/2 }(-{\bf{q}}^{'} + {\bf{q}}_{1})
 \right)
\sqrt{ n_{ {\bf{k}}-{\bf{q}}/2 } }
 A_{ {\bf{k}} }({\bf{q}})
 = 0
\end{equation}

\begin{equation}
A^{\dagger}_{ {\bf{k}} }(-{\bf{q}})
 \sqrt{ n_{ {\bf{k}} + {\bf{q}}/2 } }
A^{\dagger}_{ {\bf{k}}^{'} }(-{\bf{q}}^{'})
 \sqrt{ n_{ {\bf{k}}^{'} + {\bf{q}}^{'}/2 } }
 =
A^{\dagger}_{ {\bf{k}}^{'} }(-{\bf{q}}^{'})
 \sqrt{ n_{ {\bf{k}}^{'} + {\bf{q}}^{'}/2 } }
A^{\dagger}_{ {\bf{k}}^{'} }(-{\bf{q}}^{'})
 \sqrt{ n_{ {\bf{k}}^{'} + {\bf{q}}^{'}/2 } }
  = 0
\end{equation}

\begin{equation}
A^{\dagger}_{ {\bf{k}} }(-{\bf{q}})
 \sqrt{ n_{ {\bf{k}} + {\bf{q}}/2 } }
\left( \sum_{ {\bf{q}}_{1} \neq {\bf{q}}^{'}, 0 }
 A^{\dagger}_{ {\bf{k}}^{'}+{\bf{q}}^{'}/2 - {\bf{q}}_{1}/2 }
 ({\bf{q}}_{1})A_{ {\bf{k}}^{'} - {\bf{q}}_{1}/2 }
 ( - {\bf{q}}^{'} + {\bf{q}}_{1}) \right)
 = 0
\end{equation}

\begin{equation}
\left( \sum_{ {\bf{q}}_{1} \neq {\bf{q}}^{'}, 0 }
 A^{\dagger}_{ {\bf{k}}^{'}+{\bf{q}}^{'}/2 - {\bf{q}}_{1}/2 }
 ({\bf{q}}_{1})A_{ {\bf{k}}^{'} - {\bf{q}}_{1}/2 }
 ( - {\bf{q}}^{'} + {\bf{q}}_{1}) \right)
A^{\dagger}_{ {\bf{k}} }(-{\bf{q}})
 \sqrt{ n_{ {\bf{k}} + {\bf{q}}/2 } }
 = 0
\end{equation}

\tiny
\begin{equation}
-A^{\dagger}_{ {\bf{k}} }(-{\bf{q}})
 \sqrt{ n_{ {\bf{k}} + {\bf{q}}/2 } }
\left( \sum_{ {\bf{q}}_{1} \neq {\bf{q}}^{'}, 0 }
 A^{\dagger}_{ {\bf{k}}^{'}-{\bf{q}}^{'}/2 + {\bf{q}}_{1}/2 }
 ({\bf{q}}_{1})A_{ {\bf{k}}^{'} + {\bf{q}}_{1}/2 }
 ( - {\bf{q}}^{'} + {\bf{q}}_{1})
  \right)
 = \left( \sum_{ {\bf{q}}_{1} \neq {\bf{q}}, 0 }
 A^{\dagger}_{ {\bf{k}}+{\bf{q}}/2 + {\bf{q}}_{1}/2 }
 ({\bf{q}}_{1})A_{ {\bf{k}}+{\bf{q}}/2 + {\bf{q}}_{1}/2 }
 ({\bf{q}}_{1})
  \right)
\sqrt{ n_{ {\bf{k}} - {\bf{q}}/2 } } A_{ {\bf{k}} -
{\bf{q}}^{'}/2}({\bf{q}}-{\bf{q}}^{'})
\end{equation}

\begin{equation}
 -\left( \sum_{ {\bf{q}}_{1} \neq {\bf{q}}^{'}, 0 }
 A^{\dagger}_{ {\bf{k}}^{'}-{\bf{q}}^{'}/2 + {\bf{q}}_{1}/2 }
 ({\bf{q}}_{1})A_{ {\bf{k}}^{'} + {\bf{q}}_{1}/2 }
 ( - {\bf{q}}^{'} + {\bf{q}}_{1})
  \right)
A^{\dagger}_{ {\bf{k}} }(-{\bf{q}})
 \sqrt{ n_{ {\bf{k}} + {\bf{q}}/2 } }
 = -\left( {\hat{ {\bf{1}} }} -  \sum_{ {\bf{q}}_{1} \neq {\bf{q}}, 0 }
 A^{\dagger}_{ {\bf{k}}+{\bf{q}}/2 + {\bf{q}}_{1}/2 }
 ({\bf{q}}_{1})A_{ {\bf{k}}+{\bf{q}}/2 + {\bf{q}}_{1}/2 }
 ({\bf{q}}_{1})
  \right)
\sqrt{ n_{ {\bf{k}} - {\bf{q}}/2 } } A_{ {\bf{k}} -
{\bf{q}}^{'}/2}({\bf{q}}-{\bf{q}}^{'})
\end{equation}

\small
\begin{equation}
\left( \sum_{ {\bf{q}}_{1} \neq {\bf{q}}, 0 } A^{\dagger}_{
{\bf{k}} + {\bf{q}}/2 - {\bf{q}}_{1}/2 }({\bf{q}}_{1}) A_{
{\bf{k}} - {\bf{q}}_{1}/2 }(-{\bf{q}}+{\bf{q}}_{1})
 \right)
\left( \sum_{ {\bf{q}}_{2} \neq {\bf{q}}, 0 } A^{\dagger}_{
{\bf{k}}^{'} + {\bf{q}}^{'}/2 - {\bf{q}}_{2}/2 }({\bf{q}}_{2}) A_{
{\bf{k}}^{'} - {\bf{q}}_{2}/2 }(-{\bf{q}}^{'}+{\bf{q}}_{2})
 \right) = 0
\end{equation}

\begin{equation}
\left( \sum_{ {\bf{q}}_{2} \neq {\bf{q}}, 0 } A^{\dagger}_{
{\bf{k}}^{'} + {\bf{q}}^{'}/2 - {\bf{q}}_{2}/2 }({\bf{q}}_{2}) A_{
{\bf{k}}^{'} - {\bf{q}}_{2}/2 }(-{\bf{q}}^{'}+{\bf{q}}_{2})
 \right)
\left( \sum_{ {\bf{q}}_{1} \neq {\bf{q}}, 0 } A^{\dagger}_{
{\bf{k}} + {\bf{q}}/2 - {\bf{q}}_{1}/2 }({\bf{q}}_{1}) A_{
{\bf{k}} - {\bf{q}}_{1}/2 }(-{\bf{q}}+{\bf{q}}_{1})
 \right)  = 0
\end{equation}
\begin{equation}
\left( \sum_{ {\bf{q}}_{1} \neq {\bf{q}}, 0 } A^{\dagger}_{
{\bf{k}} - {\bf{q}}/2 + {\bf{q}}_{1}/2 }({\bf{q}}_{1}) A_{
{\bf{k}} + {\bf{q}}_{1}/2 }(-{\bf{q}}+{\bf{q}}_{1})
 \right)
\left( \sum_{ {\bf{q}}_{2} \neq {\bf{q}}^{'}, 0 } A^{\dagger}_{
{\bf{k}}^{'} - {\bf{q}}^{'}/2 + {\bf{q}}_{2}/2 }({\bf{q}}_{2}) A_{
{\bf{k}}^{'} + {\bf{q}}_{2}/2 }(-{\bf{q}}^{'}+{\bf{q}}_{2})
 \right) = 0
\end{equation}

\begin{equation}
\left( \sum_{ {\bf{q}}_{2} \neq {\bf{q}}^{'}, 0 } A^{\dagger}_{
{\bf{k}}^{'}-{\bf{q}}^{'}/2+{\bf{q}}_{2}/2 }({\bf{q}}_{2}) A_{
{\bf{k}}^{'} + {\bf{q}}_{2}/2 }(-{\bf{q}}^{'}+{\bf{q}}_{2})
 \right)
\left( \sum_{ {\bf{q}}_{1} \neq {\bf{q}}, 0 } A^{\dagger}_{
{\bf{k}} - {\bf{q}}/2 + {\bf{q}}_{1}/2 }({\bf{q}}_{1}) A_{
{\bf{k}} + {\bf{q}}_{1}/2 }(-{\bf{q}}+{\bf{q}}_{1})
 \right)  = 0
\end{equation}
\small Similarly, the reader may write down the corresponding
relations when (2) AS2 is true (3) AS3 is true and (4) when AS4 is
true. If AA0 is true then we have,
\begin{equation}
[A_{ {\bf{k}} }({\bf{q}}),A_{ {\bf{k}}^{'} }({\bf{q}}^{'})] =
 [A_{ {\bf{k}} }({\bf{q}}),A^{\dagger}_{ {\bf{k}}^{'} }({\bf{q}}^{'})]
 = 0
\end{equation}
If SS1 is true then we have,
\begin{equation}
\left[ \left( {\bf{1}} - \sum_{ {\bf{q}}_{1} \neq 0 }
 A^{\dagger}_{ {\bf{k}}-{\bf{q}}/2 + {\bf{q}}_{1}/2
 }({\bf{q}}_{1}) A_{ {\bf{k}}-{\bf{q}}/2 + {\bf{q}}_{1}/2 }
 ({\bf{q}}_{1}) \right)^{\frac{1}{2}}
A_{ {\bf{k}} }({\bf{q}})  \right]^{2} = 0
\end{equation}
and for example,
\[
\left( \sum_{ {\bf{q}}_{1} \neq {\bf{q}}, 0 } A^{\dagger}_{
{\bf{k}} + {\bf{q}}/2 - {\bf{q}}_{1}/2 }({\bf{q}}_{1})
 A_{ {\bf{k}}-{\bf{q}}_{1}/2 }(-{\bf{q}}+{\bf{q}}_{1}) \right)
\left( \sum_{ {\bf{q}}_{2} \neq {\bf{q}}, 0 }
 A^{\dagger}_{ {\bf{k}}-{\bf{q}}_{2}/2 }(-{\bf{q}}+{\bf{q}}_{2})
  A_{ {\bf{k}}+{\bf{q}}/2-{\bf{q}}_{2}/2 }({\bf{q}}_{2}) \right)
\]
\begin{equation}
 =
\left( \sum_{ {\bf{q}}_{1} \neq 0 }
 A^{\dagger}_{ {\bf{k}}+{\bf{q}}/2 - {\bf{q}}_{1}/2 }
 ({\bf{q}}_{1})
A_{ {\bf{k}}+{\bf{q}}/2 -{\bf{q}}_{1}/2 }({\bf{q}}_{1}) \right)
\left( {\bf{1}} -  \sum_{ {\bf{q}}_{2} \neq 0 }
 A^{\dagger}_{ {\bf{k}}-{\bf{q}}/2 -{\bf{q}}_{2}/2 }
 ({\bf{q}}_{2})
A_{ {\bf{k}}-{\bf{q}}/2 - {\bf{q}}_{2}/2 }({\bf{q}}_{2}) \right)
\end{equation}
Similarly, the reader can fill in the rest of the rules once the
main techniques for deducing these rules have been laid down as we
have done here. We see here that the objects $ A_{ {\bf{k}}
}({\bf{q}}) $ and $ A^{\dagger}_{ {\bf{k}} }({\bf{q}}) $ obey
{\it{exact}}, {\it{closed}} commutation rules. This enables us to
treat any theory involving fermions obeying simple fermion
commutation rules in terms of a theory involving sea-displacements
obeying these rather complicated-looking commutation rules.
However, in all cases of practical interest we shall adopt an
approximation that in the one-component Fermi system is equivalent
to the RPA or its generalizations. In the multicomponent case, we
may write down the following formula for the Fermi bilinears. This
has been used to derive the corresponding formulas for the charge
conserving electron-hole systems in the main text. If $ {\bf{q}}
\neq 0 $ then,
\[
c^{\dagger}_{ {\bf{k}}+{\bf{q}}/2 \sigma } c_{ {\bf{k}}-{\bf{q}}/2
\sigma^{'} }
 = \sqrt{ n_{ {\bf{k}}+{\bf{q}}/2 \sigma}  }
A_{ {\bf{k}} \sigma} (-{\bf{q}}\sigma^{'})
 + A^{\dagger}_{ {\bf{k}} \sigma^{'} } ({\bf{q}}\sigma)
\sqrt{ n_{ {\bf{k}}-{\bf{q}}/2 \sigma^{'}  }}
\]
\begin{equation}
+ \sum_{ {\bf{q}}_{1} \neq {\bf{q}}, 0 \sigma_{1} } A^{\dagger}_{
{\bf{k}}+{\bf{q}}/2 - {\bf{q}}_{1}/2 \sigma_{1} }
({\bf{q}}_{1}\sigma) A_{ {\bf{k}}- {\bf{q}}_{1}/2 \sigma_{1} }
(-{\bf{q}}+{\bf{q}}_{1} \sigma^{'})
 - \sum_{ {\bf{q}}_{1} \neq {\bf{q}}, 0 \sigma_{1} }
A^{\dagger}_{ {\bf{k}}-{\bf{q}}/2+{\bf{q}}_{1}/2 \sigma^{'} }
({\bf{q}}_{1}\sigma_{1}) A_{ {\bf{k}}+{\bf{q}}_{1}/2 \sigma }
(-{\bf{q}}+{\bf{q}}_{1} \sigma_{1})
\end{equation}
\begin{equation}
c^{\dagger}_{ {\bf{k}} \sigma } c_{ {\bf{k}}\sigma }
 = n_{F}({\bf{k}}\sigma) {\hat{ {\bf{1}} }}
 +  \sum_{ {\bf{q}}_{1} \neq  0, \sigma_{1} }
A^{\dagger}_{ {\bf{k}} - {\bf{q}}_{1}/2 \sigma_{1} }
({\bf{q}}_{1}\sigma) A_{ {\bf{k}}- {\bf{q}}_{1}/2 \sigma_{1} }
({\bf{q}}_{1} \sigma)
 - \sum_{ {\bf{q}}_{1} \neq 0, \sigma_{1} }
A^{\dagger}_{ {\bf{k}}+{\bf{q}}_{1}/2 \sigma }
({\bf{q}}_{1}\sigma_{1}) A_{ {\bf{k}}+{\bf{q}}_{1}/2 \sigma }
({\bf{q}}_{1} \sigma_{1})
\end{equation}

\subsection{ Approximate Commutation Rules : The Random Phase Approximation }

The commutation rules presented above together with prescriptions
such as Eq.(~\ref{AADAG}) and Eq.(~\ref{ACOMMN}) form a complete
and closed system of rules. These rules however are far from
simple to use. It is desirable to make some approximations in a
systematic manner that enables us to simplify these rules. One
such natural approximation is the RPA of Bohm and Pines
\cite{Bohm}. The way to do this is to make the following
assertion.
 {\bf{Defn}} :
       The RPA
       is obtained by retaining only the lowest
       order sea-displacement terms in Eq.(~\ref{OFFDIAG}) and
       setting $ n_{ {\bf{k}} } = n_{F}({\bf{k}}) {\hat{ {\bf{1}} }} $
       in Eq.(~\ref{DIAG}).
\noindent This means that in the exact definition in
Eq.(~\ref{AKQ}) we have to replace the number operator by the unit
operator. When this is done, we have the RPA result for $ A_{
{\bf{k}} }({\bf{q}}) $.
\begin{equation}
A_{ {\bf{k}} }({\bf{q}}) = n_{F}({\bf{k}}-{\bf{q}}/2) (1 -
n_{F}({\bf{k}}+{\bf{q}}/2)) c^{\dagger}_{ {\bf{k}}-{\bf{q}}/2 }c_{
{\bf{k}}+{\bf{q}}/2 } \label{RPARES}
\end{equation}
The ignoring of higher order terms amounts to the following (from
Eq.(~\ref{EQNC}) and Eq.(~\ref{EQND}))
\begin{equation}
 c^{\dagger}_{ {\bf{k}}+{\bf{q}}/2 }c_{ {\bf{k}} - {\bf{q}}/2 } = 0
\label{SETZERO}
\end{equation}
when $ n_{F}({\bf{k}}+{\bf{q}}/2) = 0 $ and $
n_{F}({\bf{k}}-{\bf{q}}/2) = 0 $
 \mbox{   }OR \mbox{   }
when $ n_{F}({\bf{k}}+{\bf{q}}/2) = 1 $ and $
n_{F}({\bf{k}}-{\bf{q}}/2) = 1 $. Let us now use these
simplifications to obtain a set of closed commutation rules valid
in the RPA sense for the object $ A_{ {\bf{k}} }({\bf{q}}) $. From
Eq.(~\ref{RPARES}) we have,
\[
[A_{ {\bf{k}} }({\bf{q}}),A_{ {\bf{k}}^{'} }({\bf{q}}^{'})]
 = n_{F}({\bf{k}}-{\bf{q}}/2)(1-n_{F}({\bf{k}}+{\bf{q}}/2))
n_{F}({\bf{k}}^{'}-{\bf{q}}^{'}/2)(1-n_{F}({\bf{k}}^{'}+{\bf{q}}^{'}/2))
\]
\[
\times [c^{\dagger}_{ {\bf{k}}-{\bf{q}}/2 }c_{ {\bf{k}}+{\bf{q}}/2
}, c^{\dagger}_{ {\bf{k}}^{'}-{\bf{q}}^{'}/2 }c_{
{\bf{k}}^{'}+{\bf{q}}^{'}/2 }]
\]
\[
 = n_{F}({\bf{k}}-{\bf{q}}/2)(1-n_{F}({\bf{k}}+{\bf{q}}/2))
n_{F}({\bf{k}}^{'}-{\bf{q}}^{'}/2)(1-n_{F}({\bf{k}}^{'}+{\bf{q}}^{'}/2))
\]
\begin{equation}
\times ( c^{\dagger}_{ {\bf{k}}-{\bf{q}}/2 }c_{
{\bf{k}}^{'}+{\bf{q}}^{'}/2 } \delta_{ {\bf{k}}+{\bf{q}}/2,
{\bf{k}}^{'}-{\bf{q}}^{'}/2 }
 - c^{\dagger}_{ {\bf{k}}^{'}-{\bf{q}}^{'}/2 }c_{ {\bf{k}}+{\bf{q}}/2 }
\delta_{  {\bf{k}}-{\bf{q}}/2, {\bf{k}}^{'}+{\bf{q}}^{'}/2 } )
 = 0
\end{equation}
Now the right hand side is identically zero as we see from this
argument. When $ {\bf{k}}+{\bf{q}}/2 = {\bf{k}}^{'}-{\bf{q}}^{'}/2
$ we have the factor $
(1-n_{F}({\bf{k}}+{\bf{q}}/2))n_{F}({\bf{k}}^{'}-{\bf{q}}^{'}/2) =
0 $. When $ {\bf{k}}-{\bf{q}}/2 = {\bf{k}}^{'}+{\bf{q}}^{'}/2 $ we
have $
n_{F}({\bf{k}}-{\bf{q}}/2)(1-n_{F}({\bf{k}}^{'}+{\bf{q}}^{'}/2)) =
0 $. Therefore,
\begin{equation}
[A_{ {\bf{k}} }({\bf{q}}),A_{ {\bf{k}}^{'} }({\bf{q}}^{'})]_{RPA}
= 0
\end{equation}
Now let us compute,
\[
[A_{ {\bf{k}} }({\bf{q}}), A^{\dagger}_{ {\bf{k}}^{'}
}({\bf{q}}^{'})]
 = n_{F}({\bf{k}}-{\bf{q}}/2)(1 - n_{F}({\bf{k}}+{\bf{q}}/2))
 n_{F}({\bf{k}}^{'}-{\bf{q}}^{'}/2)(1 - n_{F}({\bf{k}}^{'}+{\bf{q}}^{'}/2))
\]
\[
\times [c^{\dagger}_{ {\bf{k}} - {\bf{q}}/2 }c_{ {\bf{k}} +
{\bf{q}}/2 }, c^{\dagger}_{ {\bf{k}}^{'} + {\bf{q}}^{'}/2 } c_{
{\bf{k}}^{'} - {\bf{q}}^{'}/2 }]
\]
\[
 = n_{F}({\bf{k}}-{\bf{q}}/2)(1 - n_{F}({\bf{k}}+{\bf{q}}/2))
 n_{F}({\bf{k}}^{'}-{\bf{q}}^{'}/2)(1 - n_{F}({\bf{k}}^{'}+{\bf{q}}^{'}/2))
\]
\begin{equation}
\times (c^{\dagger}_{ {\bf{k}} - {\bf{q}}/2 } c_{ {\bf{k}}^{'} -
{\bf{q}}^{'}/2 }
 \delta_{ {\bf{k}} + {\bf{q}}/2, {\bf{k}}^{'} + {\bf{q}}^{'}/2 }
 - c^{\dagger}_{ {\bf{k}}^{'} + {\bf{q}}^{'}/2 }c_{ {\bf{k}} + {\bf{q}}/2 }
\delta_{ {\bf{k}} - {\bf{q}}/2, {\bf{k}}^{'} - {\bf{q}}^{'}/2 }
 )
\label{EQNAADAG}
\end{equation}
Because of Eq.(~\ref{SETZERO}), the right hand side of the above
equation ( Eq.(~\ref{EQNAADAG}) ) is identically zero unless $
{\bf{k}} = {\bf{k}}^{'} $ and $ {\bf{q}} = {\bf{q}}^{'} $.
Therefore we may write,
\begin{equation}
[A_{ {\bf{k}} }({\bf{q}}), A^{\dagger}_{ {\bf{k}}^{'}
}({\bf{q}}^{'})]
 = n_{F}({\bf{k}}-{\bf{q}}/2)(1 - n_{F}({\bf{k}}+{\bf{q}}/2))
\delta_{ {\bf{k}},{\bf{k}}^{'} }\delta_{ {\bf{q}}, {\bf{q}}^{'} }
( n_{ {\bf{k}} - {\bf{q}}/2 } - n_{ {\bf{k}} + {\bf{q}}/2 } )
\label{NUMFLU}
\end{equation}
If we retain the number operator as it is in the right hand side
of Eq.( ~\ref{NUMFLU}) we are dealing with the generalized RPA.
The generalized  RPA pays attention to possible fluctuations in
the momentum distribution around a nonideal mean \cite{Setlur}.
However in the simple case we are obliged to set $ n_{ {\bf{k}} }
= n_{F}({\bf{k}}){\hat{ {\bf{1}} }} $. When this is done we find
the following simple answer.
\begin{equation}
[A_{ {\bf{k}} }({\bf{q}}), A^{\dagger}_{ {\bf{k}}^{'}
}({\bf{q}}^{'})]_{RPA}
 = n_{F}({\bf{k}}-{\bf{q}}/2)(1 - n_{F}({\bf{k}}+{\bf{q}}/2))
\delta_{ {\bf{k}},{\bf{k}}^{'} }\delta_{ {\bf{q}}, {\bf{q}}^{'} }
{\hat{ {\bf{1}} }}
\end{equation}
Finally, we would like to address a rather important question,
namely is the RPA as described above a controlled approximation ?
Indeed, what really is a 'controlled approximation' ? And finally,
does it matter whether some approximation is controlled or not ?
In order to answer these questions it is important to first define
what is meant by a controlled approximation. A definition that
seems reasonable is as follows : An expansion in powers of a
dimensionless parameter small compared to unity that is obtained
by combining the various dimensionful parameters of the theory, is
a controlled approximation. By this definition it is clear that
the RPA is not a controlled approximation. Rather than expanding
in powers of a dimensionless {\it{parameter}} it seems that we are
expanding in powers of a dimensionless {\it{operator}} namely $
A_{ {\bf{k}} }({\bf{q}}) $ . For this to be justifiable, we have
to show that this object is in some sense small compared to unity.
This means that the matrix elements of this object have to be
small compared to unity\cite{smallness}.  This is possible only if
we restrict our Hilbert space to be one that contains low lying
excited stated of the noninteracting system. But this is really
begging the question. We would like to know {\it{beforehand}},
given a type of interaction and its strength whether or not such
an assumption is justified. In fact, in our earlier article
\cite{Setlur} we made some rather unfortunate remarks that may be
forgiven since it was the first article in the series and we were
going to fix the technical aspects later anyway. The formula for
the 'sea-boson' given there in terms of the Fermi fields and
justly criticized Cune and Apostol\cite{Cune} is totally wrong
despite its appealing form. The other remarks in the
appendix\cite{Setlur} justifying the controlled nature of the
RPA-approximation provided one restricts the functional form of
the potential to obey the constraints outlined
earlier\cite{Setlur} so far seem alright. In any event, the moral
of this discussion is that the RPA is a strange kind of
approximation. It is not possible to make useful statements about
when such an approximation breaks down. To say that RPA is valid
if all states of the interacting system are expressible as linear
combination of low lying states is as illuminating as saying : the
best strategy to win a game of chess is to play it so that the
opponent's king is placed under check and cannot move and the
opponent cannot block or eliminate the check.

\section{Appendix C}

Consider the hamiltonian
\begin{equation}
H = \sum_{ {\bf{k}} }{\tilde{\epsilon}}_{ {\bf{k}} } n_{ {\bf{k}}
}
 + \sum_{ {\bf{q}} \neq 0 }\frac{v_{ {\bf{q}} } }{2V}
\sum_{ {\bf{k}} \neq {\bf{k}}^{'} } c^{\dagger}_{
{\bf{k}}+{\bf{q}}/2 }c^{\dagger}_{ {\bf{k}}^{'}-{\bf{q}}/2 } c_{
{\bf{k}}^{'}+{\bf{q}}/2 } c_{ {\bf{k}}-{\bf{q}}/2 }
\end{equation}
here the exchange energy (minus the term quartic in the
sea-displacements) has been absorbed into the kinetic energy.
\begin{equation}
{\tilde{\epsilon}}_{ {\bf{k}} } = \epsilon_{ {\bf{k}} } - \sum_{
{\bf{q}} \neq 0 }\frac{ v_{ {\bf{q}} } }{V} \langle n_{
{\bf{k}}-{\bf{q}} } \rangle
\end{equation}
Let $ n_{ {\bf{q}} }({\bf{k}}) = c^{\dagger}_{ {\bf{k}} +
{\bf{q}}/2 }c_{ {\bf{k}}-{\bf{q}}/2 } $ and $ n_{ 0 }({\bf{k}}) =
c^{\dagger}_{ {\bf{k}} }c_{ {\bf{k}} } $ Let us now write down the
equation of motion for $n_{{\bf{q}} }({\bf{k}}) $. Introduce a
source,
\begin{equation}
H_{ext}(t) = \sum_{ {\bf{q}} }(U_{ext}({\bf{q}}t) +
U^{*}_{ext}(-{\bf{q}}t)) \sum_{ {\bf{k}} }n_{ {\bf{q}} }({\bf{k}})
\end{equation}
where $ U_{ext}({\bf{q}}t) = e^{-i\omega t} U_{0}({\bf{q}}) $.
\[
i\frac{ \partial }{ \partial t} n^{t}_{ {\bf{q}} }({\bf{k}})
 = ({\tilde{\epsilon}}_{ {\bf{k}}-{\bf{q}}/2 }
- {\tilde{\epsilon}}_{ {\bf{k}}+{\bf{q}}/2 })
 n^{t}_{ {\bf{q}} }({\bf{k}})
+ \sum_{ {\bf{q}}^{'} \neq 0 }\frac{ v_{ {\bf{q}}^{'} } }{2V} [n_{
{\bf{q}} }({\bf{k}}), \rho_{ {\bf{q}}^{'} }] \rho^{t}_{
-{\bf{q}}^{'} } + \sum_{ {\bf{q}}^{'} \neq 0 }\frac{ v_{
{\bf{q}}^{'} } }{2V} \rho^{t}_{ {\bf{q}}^{'} }[n_{ {\bf{q}}
}({\bf{k}}), \rho_{ -{\bf{q}}^{'} }]
\]
\begin{equation}
 + \sum_{ {\bf{q}}^{'} \neq 0 }(U_{ext}({\bf{q}}^{'}t)
+ U^{*}_{ext}(-{\bf{q}}^{'}t)) [n_{ {\bf{q}} }({\bf{k}}),\rho_{
{\bf{q}}^{'} }]
\end{equation}
\[
[n_{ {\bf{q}} }({\bf{k}}),\rho_{ {\bf{q}}^{'} }]
 = \sum_{ {\bf{k}}^{'} }
[c^{\dagger}_{ {\bf{k}} + {\bf{q}}/2 } c_{  {\bf{k}} - {\bf{q}}/2
}, c^{\dagger}_{ {\bf{k}}^{'} + {\bf{q}}^{'}/2 }
 c_{  {\bf{k}}^{'} - {\bf{q}}^{'}/2 }]
\]
\[
 = \sum_{ {\bf{k}}^{'} }
c^{\dagger}_{ {\bf{k}} + {\bf{q}}/2 }c_{  {\bf{k}}^{'} -
{\bf{q}}^{'}/2 } \delta_{ {\bf{k}} - {\bf{q}}/2, {\bf{k}}^{'} +
{\bf{q}}^{'}/2 }
 - \sum_{ {\bf{k}}^{'} }
c^{\dagger}_{ {\bf{k}}^{'} + {\bf{q}}^{'}/2 }c_{  {\bf{k}} -
{\bf{q}}/2 } \delta_{ {\bf{k}} + {\bf{q}}/2, {\bf{k}}^{'} -
{\bf{q}}^{'}/2 }
\]
\begin{equation}
\approx \delta_{ {\bf{q}}, -{\bf{q}}^{'} } [n_{0}({\bf{k}} +
{\bf{q}}/2) - n_{0}({\bf{k}} - {\bf{q}}/2)]
\end{equation}
\[
i\frac{ \partial }{ \partial t} n^{t}_{ {\bf{q}} }({\bf{k}})
 = ({\tilde{\epsilon}}_{ {\bf{k}}-{\bf{q}}/2 }
- {\tilde{\epsilon}}_{ {\bf{k}}+{\bf{q}}/2 })
 n^{t}_{ {\bf{q}} }({\bf{k}})
+ \frac{ v_{ {\bf{q}} } }{V} (n_{0}({\bf{k}} + {\bf{q}}/2) -
n_{0}({\bf{k}} - {\bf{q}}/2))
 \rho^{t}_{ {\bf{q}} }
\]
\begin{equation}
 + (U_{ext}(-{\bf{q}}t)
+ U^{*}_{ext}({\bf{q}}t)) (n_{0}({\bf{k}} + {\bf{q}}/2) -
n_{0}({\bf{k}} - {\bf{q}}/2))
\end{equation}
Let us make a first pass at the computation of the dielectric
function. Here, we make use of mean-field theory, that is, replace
 $ \langle n_{0}( {\bf{k}}^{'} )\rho_{ {\bf{q}} } \rangle = \langle n_{0}( {\bf{
k}}^{'} ) \rangle\langle \rho_{ {\bf{q}} } \rangle $
\[
\omega \mbox{      }\langle  n^{t}_{ {\bf{q}} }({\bf{k}}) \rangle
 = ({\tilde{\epsilon}}_{ {\bf{k}}-{\bf{q}}/2 } - {\tilde{\epsilon}}_{
{\bf{k}}+{\bf{q}}/2 }) \langle  n^{t}_{ {\bf{q}} }({\bf{k}})
\rangle + \frac{ v_{ {\bf{q}} } }{V} (\langle  n_{0}({\bf{k}} +
{\bf{q}}/2)  \rangle
 -  \langle  n_{0}({\bf{k}} - {\bf{q}}/2)  \rangle)
\langle  \rho^{t}_{ {\bf{q}} } \rangle
\]
\begin{equation}
 + (U_{ext}(-{\bf{q}}t)
+ U^{*}_{ext}({\bf{q}}t)) (\langle  n_{0}({\bf{k}} + {\bf{q}}/2)
\rangle - \langle  n_{0}({\bf{k}} - {\bf{q}}/2) \rangle)
\end{equation}
\[
\langle  n^{t}_{ {\bf{q}} }({\bf{k}}) \rangle
 = \frac{ v_{ {\bf{q}} } }{V}
\frac{\langle  n_{0}({\bf{k}} + {\bf{q}}/2)  \rangle
 -  \langle  n_{0}({\bf{k}} - {\bf{q}}/2)  \rangle }
{ \omega  - {\tilde{\epsilon}}_{ {\bf{k}}-{\bf{q}}/2 } +
 {\tilde{\epsilon}}_{ {\bf{k}}+{\bf{q}}/2 }  }\langle  \rho^{t}_{ {\bf{q}} } \rangle
\]
\begin{equation}
 + (U_{ext}(-{\bf{q}}t)
+ U^{*}_{ext}({\bf{q}}t)) \frac{\langle  n_{0}({\bf{k}} +
{\bf{q}}/2)  \rangle
 -  \langle  n_{0}({\bf{k}} - {\bf{q}}/2)  \rangle }
{ \omega  - {\tilde{\epsilon}}_{ {\bf{k}}-{\bf{q}}/2 }+
{\tilde{\epsilon}}_{ {\bf{k}}+{\bf{q}}/2 }
 }
\end{equation}
This means,
\begin{equation}
\langle \rho_{ -{\bf{q}} } \rangle = U_{ext}({\bf{q}}t) \frac{
P_{0}({\bf{q}},\omega) }{ \epsilon({\bf{q}}, \omega) }
\end{equation}
\begin{equation}
 P_{0}({\bf{q}},\omega) = \sum_{ {\bf{k}} }
\frac{ \langle  n_{0}({\bf{k}} - {\bf{q}}/2)  \rangle
 -  \langle  n_{0}({\bf{k}} + {\bf{q}}/2)  \rangle }
{\omega - {\tilde{\epsilon}}_{ {\bf{k}} + {\bf{q}}/2 } +
{\tilde{\epsilon}}_{ {\bf{k}} - {\bf{q}}/2 } }
\end{equation}
\begin{equation}
\epsilon({\bf{q}}, \omega) = 1 - \frac{v_{ {\bf{q}} }}{V}
P_{0}({\bf{q}},\omega)
\end{equation}
>From this and the fact that
\begin{equation}
\epsilon_{g-RPA}({\bf{q}},\omega) = \frac{ U_{ext}({\bf{q}}t) } {
U_{eff}({\bf{q}}t) } = \epsilon({\bf{q}},\omega)
\end{equation}
Next we would like to include fluctuations. Let us do this
differently this time via the use of the BBGKY hierarchy.
\[
i\frac{ \partial }{ \partial t} \langle n^{t}_{ {\bf{q}}
}({\bf{k}}) \rangle
 = ({\tilde{\epsilon}}_{ {\bf{k}}-{\bf{q}}/2 }
- {\tilde{\epsilon}}_{ {\bf{k}}+{\bf{q}}/2 })
 \langle n^{t}_{ {\bf{q}} }({\bf{k}})  \rangle
+ \frac{ v_{ {\bf{q}} } }{V} (\langle n_{0}({\bf{k}} + {\bf{q}}/2)
\rangle -
 \langle n_{0}({\bf{k}} - {\bf{q}}/2)  \rangle)
\langle  \rho^{t}_{ {\bf{q}} } \rangle
\]
\[
+  \frac{ v_{ {\bf{q}} } }{V} (F_{2A}({\bf{k}} + {\bf{q}}/2,
{\bf{q}}) - F_{2A}({\bf{k}} - {\bf{q}}/2, {\bf{q}}))
\]
\begin{equation}
 + (U_{ext}(-{\bf{q}}t)
+ U^{*}_{ext}({\bf{q}}t)) (n_{0}({\bf{k}} + {\bf{q}}/2) -
n_{0}({\bf{k}} - {\bf{q}}/2))
\end{equation}
Here,
\begin{equation}
F_{2A}({\bf{k}}^{'},{\bf{q}};t) = \langle n_{0}({\bf{k}}^{'})
\rho^{t}_{ {\bf{q}} } \rangle - \langle n_{0}({\bf{k}}^{'})\rangle
 \langle \rho^{t}_{ {\bf{q}} } \rangle
\end{equation}
\begin{equation}
F_{2}({\bf{k}}^{'};{\bf{k}},{\bf{q}};t)
 = \langle n_{0}({\bf{k}}^{'})n^{t}_{ {\bf{q}} }({\bf{k}}) \rangle
 - \langle n_{0}({\bf{k}}^{'}) \rangle
 \langle n^{t}_{ {\bf{q}} }({\bf{k}}) \rangle
\end{equation}
\begin{equation}
F_{2A}({\bf{k}}^{'},{\bf{q}};t) = \sum_{ {\bf{k}} }
F_{2}({\bf{k}}^{'};{\bf{k}},{\bf{q}};t)
\end{equation}
\[
i\frac{ \partial }{ \partial t }F_{2}({\bf{k}}^{'};
{\bf{k}},{\bf{q}};t)
 = ({\tilde{\epsilon}}_{ {\bf{k}}-{\bf{q}}/2 }
- {\tilde{\epsilon}}_{ {\bf{k}}+{\bf{q}}/2 }) F_{2}({\bf{k}}^{'};
{\bf{k}},{\bf{q}};t)
 + \frac{ v_{ {\bf{q}} } }{V}(N({\bf{k}}^{'},{\bf{k}}+{\bf{q}}/2)
 - N({\bf{k}}^{'},{\bf{k}}-{\bf{q}}/2))
\langle \rho^{t}_{ {\bf{q}} } \rangle
\]
\begin{equation}
+  \frac{ v_{ {\bf{q}} } }{V}(\langle n_{0}({\bf{k}}+{\bf{q}}/2)
\rangle
 - \langle n_{0}({\bf{k}}-{\bf{q}}/2) \rangle)
F_{2A}({\bf{k}}^{'},{\bf{q}};t)
 + (U_{ext}(-{\bf{q}}t) + U^{*}_{ext}({\bf{q}}t))
(N({\bf{k}}^{'}, {\bf{k}}+{\bf{q}}/2) - N({\bf{k}}^{'},
{\bf{k}}-{\bf{q}}/2))
\end{equation}
Let us write,
\begin{equation}
F_{2}({\bf{k}}^{'}; {\bf{k}},{\bf{q}};t) = U_{ext}(-{\bf{q}}t)
F_{2,a}({\bf{k}}^{'}; {\bf{k}},{\bf{q}})
 + U^{*}_{ext}({\bf{q}}t) F_{2,b}({\bf{k}}^{'}; {\bf{k}},{\bf{q}})
\end{equation}
\begin{equation}
\langle \rho^{t}_{ {\bf{q}} } \rangle = U_{ext}(-{\bf{q}}t)
\langle \rho^{'}_{ {\bf{q}} } \rangle +  U^{*}_{ext}({\bf{q}}t)
\langle \rho^{''}_{ {\bf{q}} } \rangle
\end{equation}
Also define,
\begin{equation}
N({\bf{k}},{\bf{k}}^{'}) = \langle
n_{0}({\bf{k}})n_{0}({\bf{k}}^{'}) \rangle
 - \langle n_{0}({\bf{k}}) \rangle \langle  n_{0}({\bf{k}}^{'}) \rangle
\end{equation}
\[
\omega \mbox{      }F_{2,a}({\bf{k}}^{'}; {\bf{k}},{\bf{q}})
 = ({\tilde{\epsilon}}_{ {\bf{k}}-{\bf{q}}/2 }
- {\tilde{\epsilon}}_{ {\bf{k}}+{\bf{q}}/2 })
F_{2,a}({\bf{k}}^{'}; {\bf{k}},{\bf{q}})
 + \frac{ v_{ {\bf{q}} } }{V}(N({\bf{k}}^{'},{\bf{k}}+{\bf{q}}/2)
 - N({\bf{k}}^{'},{\bf{k}}-{\bf{q}}/2))
\langle \rho^{'}_{ {\bf{q}} } \rangle
\]
\begin{equation}
+  \frac{ v_{ {\bf{q}} } }{V}(\langle n_{0}({\bf{k}}+{\bf{q}}/2)
\rangle
 - \langle n_{0}({\bf{k}}-{\bf{q}}/2) \rangle)
F^{a}_{2A}({\bf{k}}^{'},{\bf{q}})
 +
(N({\bf{k}}^{'}, {\bf{k}}+{\bf{q}}/2) - N({\bf{k}}^{'},
{\bf{k}}-{\bf{q}}/2))
\end{equation}
\[
-\omega \mbox{      }F_{2,b}({\bf{k}}^{'}; {\bf{k}},{\bf{q}})
 = ({\tilde{\epsilon}}_{ {\bf{k}}-{\bf{q}}/2 }
- {\tilde{\epsilon}}_{ {\bf{k}}+{\bf{q}}/2 })
F_{2,b}({\bf{k}}^{'}; {\bf{k}},{\bf{q}})
 + \frac{ v_{ {\bf{q}} } }{V}(N({\bf{k}}^{'},{\bf{k}}+{\bf{q}}/2)
 - N({\bf{k}}^{'},{\bf{k}}-{\bf{q}}/2))
\langle \rho^{''}_{ {\bf{q}} } \rangle
\]
\begin{equation}
+  \frac{ v_{ {\bf{q}} } }{V}(\langle n_{0}({\bf{k}}+{\bf{q}}/2)
\rangle
 - \langle n_{0}({\bf{k}}-{\bf{q}}/2) \rangle)
F^{b}_{2A}({\bf{k}}^{'},{\bf{q}})
 +
(N({\bf{k}}^{'}, {\bf{k}}+{\bf{q}}/2) - N({\bf{k}}^{'},
{\bf{k}}-{\bf{q}}/2))
\end{equation}
\[
\epsilon({\bf{q}},\omega) \mbox{
}F_{2A}^{a}({\bf{k}}^{'},{\bf{q}})
 = \frac{ v_{ {\bf{q}} } }{V}\sum_{ {\bf{k}} }
\frac{ N({\bf{k}}^{'}, {\bf{k}}+{\bf{q}}/2)
 -  N({\bf{k}}^{'}, {\bf{k}}-{\bf{q}}/2) }
{\omega - {\tilde{\epsilon}}_{ {\bf{k}}-{\bf{q}}/2 } +
{\tilde{\epsilon}}_{ {\bf{k}}+{\bf{q}}/2 }} \langle \rho^{'}_{
{\bf{q}} } \rangle
\]
\begin{equation}
+ \sum_{ {\bf{k}} }\frac{ N({\bf{k}}^{'}, {\bf{k}}+{\bf{q}}/2)
 -  N({\bf{k}}^{'}, {\bf{k}}-{\bf{q}}/2) }
{\omega - {\tilde{\epsilon}}_{ {\bf{k}}-{\bf{q}}/2 } +
{\tilde{\epsilon}}_{ {\bf{k}}+{\bf{q}}/2 }}
\end{equation}
\[
\omega\mbox{   }C_{ {\bf{q}} }({\bf{k}})
 = ({\tilde{\epsilon}}_{ {\bf{k}}-{\bf{q}}/2 }
- {\tilde{\epsilon}}_{ {\bf{k}}+{\bf{q}}/2 }) C_{ {\bf{q}}
}({\bf{k}}) + \frac{ v_{ {\bf{q}} } }{V}(\langle
n_{0}({\bf{k}}+{\bf{q}}/2) \rangle
 - \langle n_{0}({\bf{k}}-{\bf{q}}/2) \rangle)
\langle \rho^{'}_{ {\bf{q}} } \rangle +  \frac{ v_{ {\bf{q}} }
}{V}(F^{a}_{2A}({\bf{k}}+{\bf{q}}/2)
 - F^{a}_{2A}({\bf{k}}-{\bf{q}}/2))
\]
\begin{equation}
+ (\langle n_{0}({\bf{k}}+{\bf{q}}/2) \rangle - \langle
n_{0}({\bf{k}}-{\bf{q}}/2) \rangle)
\end{equation}
\[
C_{ {\bf{q}} }({\bf{k}}) = \frac{ v_{ {\bf{q}} } }{V} \frac{
\langle n_{0}({\bf{k}}+{\bf{q}}/2) \rangle
 - \langle n_{0}({\bf{k}}-{\bf{q}}/2) \rangle }
{ \omega - {\tilde{\epsilon}}_{ {\bf{k}}-{\bf{q}}/2 } +
{\tilde{\epsilon}}_{ {\bf{k}}+{\bf{q}}/2 } } \langle \rho^{'}_{
{\bf{q}} } \rangle
\]
\[
+ \frac{ \langle n_{0}({\bf{k}}+{\bf{q}}/2) \rangle
 - \langle n_{0}({\bf{k}}-{\bf{q}}/2) \rangle }
{ \omega - {\tilde{\epsilon}}_{ {\bf{k}}-{\bf{q}}/2 } +
{\tilde{\epsilon}}_{ {\bf{k}}+{\bf{q}}/2 } }
\]
\begin{equation}
+  \frac{ v_{ {\bf{q}} } }{V} \frac{
F^{a}_{2A}({\bf{k}}+{\bf{q}}/2,{\bf{q}})
 - F^{a}_{2A}({\bf{k}}-{\bf{q}}/2,{\bf{q}}) }
{ \omega - \epsilon_{ {\bf{k}}-{\bf{q}}/2 } + \epsilon_{
{\bf{k}}+{\bf{q}}/2 } }
\end{equation}
After all this, it may be shown that the overall dielectric
function including
 possible fluctuations in the momentum distribution is given by,
\begin{equation}
\epsilon_{eff}({\bf{q}},\omega) =
\epsilon_{g-RPA}({\bf{q}},\omega)
 - (\frac{v_{ {\bf{q}} } }{V})^{2}\frac{ P_{2}({\bf{q}},\omega) }
{ \epsilon_{g-RPA}({\bf{q}},\omega) }
\end{equation}
Here,
\begin{equation}
P_{2}({\bf{q}},\omega) = \sum_{ {\bf{k}},{\bf{k}}^{'} } \frac{
N({\bf{k}}+{\bf{q}}/2, {\bf{k}}^{'}+{\bf{q}}/2) -
N({\bf{k}}-{\bf{q}}/2, {\bf{k}}^{'}+{\bf{q}}/2) -
N({\bf{k}}+{\bf{q}}/2, {\bf{k}}^{'}-{\bf{q}}/2) +
N({\bf{k}}-{\bf{q}}/2, {\bf{k}}^{'}-{\bf{q}}/2) } { (\omega -
{\tilde{\epsilon}}_{ {\bf{k}}-{\bf{q}}/2 } + {\tilde{\epsilon}}_{
{\bf{k}}+{\bf{q}}/2 }) (\omega - {\tilde{\epsilon}}_{
{\bf{k}}^{'}-{\bf{q}}/2 } + {\tilde{\epsilon}}_{
{\bf{k}}^{'}+{\bf{q}}/2 }) }
\end{equation}
\begin{equation}
\epsilon_{g-RPA}({\bf{q}},\omega) = 1 + \frac{ v_{ {\bf{q}} } }{V}
\sum_{ {\bf{k}} } \frac{ \langle n_{0}({\bf{k}}+{\bf{q}}/2)
\rangle
 - \langle n_{0}({\bf{k}}-{\bf{q}}/2) \rangle }
{ \omega - \epsilon_{ {\bf{k}}+{\bf{q}}/2 } +\epsilon_{
{\bf{k}}-{\bf{q}}/2  } }
\end{equation}

{\bf Aknowledgment} This work was supported by the Office of Naval
Research and the Jawaharlal Nehru Centre.


\begin{thebibliography}{25}
\bibitem[11]{footnote} While these arguments are compelling to the authors
                      it is likely that a more elegant justification
                      exists, perhaps the reader can find it.
\bibitem[13]{Cunecrit} The reason why replacing the square root
of the number
                      operator by the unit operator does not work is because
                      for example, Eq.(~\ref{EQCUNE}) is inconsistent with
                      Eq.(~\ref{DIAG}) except when
                      $ n_{ {\bf{k}} } = n_{F}({\bf{k}}) {\hat{ {\bf{1}} }} $.
                      Therefore, Eq.(~\ref{EQCUNE}) is correct in the sense
                      of the RPA.
\bibitem[14]{idem} The authors are not
clear as to the role played by idempotence
               in this framework. It is likely that idempotence is
 violated by the sea-boson approach. It is likely to be obeyed only in some
 restricted sense. In any event the reader is advised
 not to invoke idempotence anywhere else in this article, or else all assertions are
inconsistent.
\bibitem[18]{smallness} This issue is also not so simple. If one considers
 the restricted Hilbert space to be so small as to contain only the ground
 state of the noninteracting system, then trivially
 $ A_{ {\bf{k}} }({\bf{q}}) $
 is small since $ \langle FS | A_{ {\bf{k}} }({\bf{q}}) | FS \rangle = 0 $.
 If one starts adding states that are low lying excitations
  to this restricted Hilbert space,
 then $ \langle I | A_{ {\bf{k}} }({\bf{q}}) | J \rangle = 0 $
 unless $ {\bf{k}}-{\bf{q}}/2 $  is just below the Fermi surface and
 $ {\bf{k}}+{\bf{q}}/2 $  is just above the Fermi surface. This means
 that $ |{\bf{k}}| \approx k_{F} $ and $ |{\bf{q}}| << k_{F} $.
 Since the matrix elements of $ A_{ {\bf{k}} }({\bf{q}}) $
  are zero 'most of the time', the matrix
 representation of $ A_{ {\bf{k}} }({\bf{q}}) $ is sparse and in this
 sense the  operator $ A_{ {\bf{k}} }({\bf{q}}) $
 may be regarded as being small even though some elements
 of this matrix may be of order unity.
\end{thebibliography}
\end{document}